
%
\expandafter\ifx\csname phyzzx\endcsname\relax\else
 \errhelp{Hit <CR> and go ahead.}
 \errmessage{PHYZZX macros are already loaded or input. }
 \endinput \fi
\catcode`\@=11 
%
%
%
\font\seventeenrm=cmr17
\font\fourteenrm=cmr12 scaled\magstep1
\font\twelverm=cmr12
\font\ninerm=cmr9            \font\sixrm=cmr6
%
\font\fourteenbf=cmbx10 scaled\magstep2
\font\twelvebf=cmbx12
\font\ninebf=cmbx9            \font\sixbf=cmbx6
%
\font\fourteeni=cmmi10 scaled\magstep2      \skewchar\fourteeni='177
\font\twelvei=cmmi12			        \skewchar\twelvei='177
\font\ninei=cmmi9                           \skewchar\ninei='177
\font\sixi=cmmi6                            \skewchar\sixi='177
%
\font\fourteensy=cmsy10 scaled\magstep2     \skewchar\fourteensy='60
\font\twelvesy=cmsy10 scaled\magstep1	    \skewchar\twelvesy='60
\font\ninesy=cmsy9                          \skewchar\ninesy='60
\font\sixsy=cmsy6                           \skewchar\sixsy='60
%
\font\fourteenex=cmex10 scaled\magstep2
\font\twelveex=cmex10 scaled\magstep1
%
\font\fourteensl=cmsl12 scaled\magstep1
\font\twelvesl=cmsl12
\font\ninesl=cmsl9
%
\font\fourteenit=cmti12 scaled\magstep1
\font\twelveit=cmti12
\font\nineit=cmti9
\font\fourteentt=cmtt10 scaled\magstep2
\font\twelvett=cmtt12
\font\fourteencp=cmcsc10 scaled\magstep2
\font\twelvecp=cmcsc10 scaled\magstep1
\font\tencp=cmcsc10
\newfam\cpfam
\newdimen\b@gheight		\b@gheight=12pt
\newcount\f@ntkey		\f@ntkey=0
\def\f@m{\afterassignment\samef@nt\f@ntkey=}
\def\samef@nt{\fam=\f@ntkey \the\textfont\f@ntkey\relax}
\def\rm{\f@m0 }
\def\mit{\f@m1 }         
\def\cal{\f@m2 }
\def\it{\f@m\itfam}
\def\sl{\f@m\slfam}
\def\bf{\f@m\bffam}
\def\tt{\f@m\ttfam}
\def\caps{\f@m\cpfam}
\def\fourteenpoint{\relax
    \textfont0=\fourteenrm          \scriptfont0=\tenrm
      \scriptscriptfont0=\sevenrm
    \textfont1=\fourteeni           \scriptfont1=\teni
      \scriptscriptfont1=\seveni
    \textfont2=\fourteensy          \scriptfont2=\tensy
      \scriptscriptfont2=\sevensy
    \textfont3=\fourteenex          \scriptfont3=\twelveex
      \scriptscriptfont3=\tenex
    \textfont\itfam=\fourteenit     \scriptfont\itfam=\tenit
    \textfont\slfam=\fourteensl     \scriptfont\slfam=\tensl
    \textfont\bffam=\fourteenbf     \scriptfont\bffam=\tenbf
      \scriptscriptfont\bffam=\sevenbf
    \textfont\ttfam=\fourteentt
    \textfont\cpfam=\fourteencp
    \samef@nt
    \b@gheight=14pt
    \setbox\strutbox=\hbox{\vrule height 0.85\b@gheight
				depth 0.35\b@gheight width\z@ }}
\def\twelvepoint{\relax
    \textfont0=\twelverm          \scriptfont0=\ninerm
      \scriptscriptfont0=\sixrm
    \textfont1=\twelvei           \scriptfont1=\ninei
      \scriptscriptfont1=\sixi
    \textfont2=\twelvesy           \scriptfont2=\ninesy
      \scriptscriptfont2=\sixsy
    \textfont3=\twelveex          \scriptfont3=\tenex
      \scriptscriptfont3=\tenex
    \textfont\itfam=\twelveit     \scriptfont\itfam=\nineit
    \textfont\slfam=\twelvesl     \scriptfont\slfam=\ninesl
    \textfont\bffam=\twelvebf     \scriptfont\bffam=\ninebf
      \scriptscriptfont\bffam=\sixbf
    \textfont\ttfam=\twelvett
    \textfont\cpfam=\twelvecp
    \samef@nt
    \b@gheight=12pt
    \setbox\strutbox=\hbox{\vrule height 0.85\b@gheight
				depth 0.35\b@gheight width\z@ }}
\def\tenpoint{\relax
    \textfont0=\tenrm          \scriptfont0=\sevenrm
      \scriptscriptfont0=\fiverm
    \textfont1=\teni           \scriptfont1=\seveni
      \scriptscriptfont1=\fivei
    \textfont2=\tensy          \scriptfont2=\sevensy
      \scriptscriptfont2=\fivesy
    \textfont3=\tenex          \scriptfont3=\tenex
      \scriptscriptfont3=\tenex
    \textfont\itfam=\tenit     \scriptfont\itfam=\seveni
    \textfont\slfam=\tensl     \scriptfont\slfam=\sevenrm
    \textfont\bffam=\tenbf     \scriptfont\bffam=\sevenbf
      \scriptscriptfont\bffam=\fivebf
    \textfont\ttfam=\tentt
    \textfont\cpfam=\tencp
    \samef@nt
    \b@gheight=10pt
    \setbox\strutbox=\hbox{\vrule height 0.85\b@gheight
				depth 0.35\b@gheight width\z@ }}
%
%
%
\normalbaselineskip = 20pt plus 0.2pt minus 0.1pt
\normallineskip = 1.5pt plus 0.1pt minus 0.1pt
\normallineskiplimit = 1.5pt
\newskip\normaldisplayskip
\normaldisplayskip = 20pt plus 5pt minus 10pt
\newskip\normaldispshortskip
\normaldispshortskip = 6pt plus 5pt
\newskip\normalparskip
\normalparskip = 6pt plus 2pt minus 1pt
\newskip\skipregister
\skipregister = 5pt plus 2pt minus 1.5pt
\newif\ifsingl@    \newif\ifdoubl@
\newif\iftwelv@    \twelv@true
\def\singlespace{\singl@true\doubl@false\spaces@t}
\def\doublespace{\singl@false\doubl@true\spaces@t}
\def\normalspace{\singl@false\doubl@false\spaces@t}
\def\Tenpoint{\tenpoint\twelv@false\spaces@t}
\def\Twelvepoint{\twelvepoint\twelv@true\spaces@t}
\def\spaces@t{\relax
      \iftwelv@ \ifsingl@\subspaces@t3:4;\else\subspaces@t1:1;\fi
       \else \ifsingl@\subspaces@t3:5;\else\subspaces@t4:5;\fi \fi
      \ifdoubl@ \multiply\baselineskip by 5
         \divide\baselineskip by 4 \fi }
\def\subspaces@t#1:#2;{
      \baselineskip = \normalbaselineskip
      \multiply\baselineskip by #1 \divide\baselineskip by #2
      \lineskip = \normallineskip
      \multiply\lineskip by #1 \divide\lineskip by #2
      \lineskiplimit = \normallineskiplimit
      \multiply\lineskiplimit by #1 \divide\lineskiplimit by #2
      \parskip = \normalparskip
      \multiply\parskip by #1 \divide\parskip by #2
      \abovedisplayskip = \normaldisplayskip
      \multiply\abovedisplayskip by #1 \divide\abovedisplayskip by #2
      \belowdisplayskip = \abovedisplayskip
      \abovedisplayshortskip = \normaldispshortskip
      \multiply\abovedisplayshortskip by #1
        \divide\abovedisplayshortskip by #2
      \belowdisplayshortskip = \abovedisplayshortskip
      \advance\belowdisplayshortskip by \belowdisplayskip
      \divide\belowdisplayshortskip by 2
      \smallskipamount = \skipregister
      \multiply\smallskipamount by #1 \divide\smallskipamount by #2
      \medskipamount = \smallskipamount \multiply\medskipamount by 2
      \bigskipamount = \smallskipamount \multiply\bigskipamount by 4 }
\def\normalbaselines{ \baselineskip=\normalbaselineskip
   \lineskip=\normallineskip \lineskiplimit=\normallineskip
   \iftwelv@\else \multiply\baselineskip by 4 \divide\baselineskip by 5
     \multiply\lineskiplimit by 4 \divide\lineskiplimit by 5
     \multiply\lineskip by 4 \divide\lineskip by 5 \fi }
\Twelvepoint  
\interlinepenalty=50
\interfootnotelinepenalty=5000
\predisplaypenalty=9000
\postdisplaypenalty=500
\hfuzz=1pt
\vfuzz=0.2pt
\voffset=0pt
\dimen\footins=8 truein
%
%
%
\def\pagecontents{
   \ifvoid\topins\else\unvbox\topins\vskip\skip\topins\fi
   \dimen@ = \dp255 \unvbox255
   \ifvoid\footins\else\vskip\skip\footins\footrule\unvbox\footins\fi
   \ifr@ggedbottom \kern-\dimen@ \vfil \fi }
\def\makeheadline{\vbox to 0pt{ \skip@=\topskip
      \advance\skip@ by -12pt \advance\skip@ by -2\normalbaselineskip
      \vskip\skip@ \line{\vbox to 12pt{}\the\headline} \vss
      }\nointerlineskip}
\def\makefootline{\baselineskip = 1.5\normalbaselineskip
                 \line{\the\footline}}
\newif\iffrontpage
\newif\ifletterstyle
\newif\ifp@genum
\def\nopagenumbers{\p@genumfalse}
\def\pagenumbers{\p@genumtrue}
\pagenumbers
\newtoks\paperheadline
\newtoks\letterheadline
\newtoks\paperfootline
\newtoks\letterfootline
\newtoks\letterinfo
\newtoks\Letterinfo
\newtoks\date
\footline={\ifletterstyle\the\letterfootline\else\the\paperfootline\fi}
\paperfootline={\hss\iffrontpage\else\ifp@genum\tenrm\folio\hss\fi\fi}
\letterfootline={\iffrontpage\LETTERFOOT\else\hfil\fi}
\Letterinfo={\hfil}
\letterinfo={\hfil}
\def\LETTERFOOT{\hfil} 
%
\def\LETTERHEAD{\vtop{\baselineskip=20pt\hbox to
\hsize{\hfil\seventeenrm\strut
CALIFORNIA INSTITUTE OF TECHNOLOGY \hfil}
\hbox to \hsize{\hfil\ninerm\strut
CHARLES C. LAURITSEN LABORATORY OF HIGH ENERGY PHYSICS \hfil}
\hbox to \hsize{\hfil\ninerm\strut
PASADENA, CALIFORNIA 91125 \hfil}}}
\headline={\ifletterstyle\the\letterheadline\else\the\paperheadline\fi}
\paperheadline={\hfil}
\letterheadline{\iffrontpage \LETTERHEAD\else
    \rm \ifp@genum \hfil \folio\hfil\fi\fi}
\def\monthname{\relax\ifcase\month 0/\or January\or February\or
   March\or April\or May\or June\or July\or August\or September\or
   October\or November\or December\else\number\month/\fi}
\def\today{\monthname\ \number\day, \number\year}
\date={\today}
\countdef\pageno=1      \countdef\pagen@=0
\countdef\pagenumber=1  \pagenumber=1
\def\advancepageno{\global\advance\pagen@ by 1
   \ifnum\pagenumber<0 \global\advance\pagenumber by -1
    \else\global\advance\pagenumber by 1 \fi \global\frontpagefalse }
\def\folio{\ifnum\pagenumber<0 \romannumeral-\pagenumber
           \else \number\pagenumber \fi }
\def\footrule{\dimen@=\prevdepth\nointerlineskip
   \vbox to 0pt{\vskip -0.25\baselineskip \hrule width 0.35\hsize \vss}
   \prevdepth=\dimen@ }
\newtoks\foottokens
\foottokens={}
\newdimen\footindent
\footindent=24pt
\def\vfootnote#1{\insert\footins\bgroup
   \interlinepenalty=\interfootnotelinepenalty \floatingpenalty=20000
   \singl@true\doubl@false\Tenpoint
   \splittopskip=\ht\strutbox \boxmaxdepth=\dp\strutbox
   \leftskip=\footindent \rightskip=\z@skip
   \parindent=0.5\footindent \parfillskip=0pt plus 1fil
   \spaceskip=\z@skip \xspaceskip=\z@skip
   \the\foottokens
   \Textindent{$ #1 $}\footstrut\futurelet\next\fo@t}
\def\Textindent#1{\noindent\llap{#1\enspace}\ignorespaces}
\def\footnote#1{\attach{#1}\vfootnote{#1}}

\def\foot{\attach\footsymbolgen\vfootnote{\footsymbol}}
\let\footsymbol=\star
\newcount\lastf@@t           \lastf@@t=-1
\newcount\footsymbolcount    \footsymbolcount=0
\newif\ifPhysRev
\def\footsymbolgen{\bumpfootsymbolcount \generatefootsymbol \footsymbol }
\def\bumpfootsymbolcount{\relax
   \iffrontpage \bumpfootsymbolNP \else \advance\lastf@@t by 1
     \ifPhysRev \bumpfootsymbolPR \else \bumpfootsymbolNP \fi \fi
   \global\lastf@@t=\pagen@ }
\def\bumpfootsymbolNP{\ifnum\footsymbolcount <0 \global\footsymbolcount =0 \fi
    \ifnum\lastf@@t<\pagen@ \global\footsymbolcount=0
     \else \global\advance\footsymbolcount by 1 \fi }
\def\bumpfootsymbolPR{\ifnum\footsymbolcount >0 \global\footsymbolcount =0 \fi
      \global\advance\footsymbolcount by -1 }
\def\fd@f#1 {\xdef\footsymbol{\mathchar"#1 }}
\def\generatefootsymbol{\ifcase\footsymbolcount \fd@f 13F \or \fd@f 279
	\or \fd@f 27A \or \fd@f 278 \or \fd@f 27B \else
	\ifnum\footsymbolcount <0 \fd@f{023 \number-\footsymbolcount }
	 \else \fd@f 203 {\loop \ifnum\footsymbolcount >5
		\fd@f{203 \footsymbol } \advance\footsymbolcount by -1
		\repeat }\fi \fi }

\def\nonfrenchspacing{\sfcode`\.=3001 \sfcode`\!=3000 \sfcode`\?=3000
	\sfcode`\:=2000 \sfcode`\;=1500 \sfcode`\,=1251 }
\nonfrenchspacing
\newdimen\d@twidth
 {\setbox0=\hbox{s.} \global\d@twidth=\wd0 \setbox0=\hbox{s}
	\global\advance\d@twidth by -\wd0 }
\def\removehglue{\loop \unskip \ifdim\lastskip >\z@ \repeat }
\def\roll@ver#1{\removehglue \nobreak \count255 =\spacefactor \dimen@=\z@
	\ifnum\count255 =3001 \dimen@=\d@twidth \fi
	\ifnum\count255 =1251 \dimen@=\d@twidth \fi
    \iftwelv@ \kern-\dimen@ \else \kern-0.83\dimen@ \fi
   #1\spacefactor=\count255 }
\def\step@ver#1{\relax \ifmmode #1\else \ifhmode
	\roll@ver{${}#1$}\else {\setbox0=\hbox{${}#1$}}\fi\fi }
\def\attach#1{\step@ver{\strut^{\mkern 2mu #1} }}
%
%
%
\newcount\chapternumber      \chapternumber=0
\newcount\sectionnumber      \sectionnumber=0
\newcount\equanumber         \equanumber=0
\let\chapterlabel=\relax
\let\sectionlabel=\relax
\newtoks\chapterstyle        \chapterstyle={\Number}
\newtoks\sectionstyle        \sectionstyle={\chapterlabel\Number}
\newskip\chapterskip         \chapterskip=\bigskipamount
\newskip\sectionskip         \sectionskip=\medskipamount
\newskip\headskip            \headskip=8pt plus 3pt minus 3pt
\newdimen\chapterminspace    \chapterminspace=15pc
\newdimen\sectionminspace    \sectionminspace=10pc
\newdimen\referenceminspace  \referenceminspace=25pc
\def\chapterreset{\global\advance\chapternumber by 1
   \ifnum\equanumber<0 \else\global\equanumber=0\fi
   \sectionnumber=0 \makechapterlabel}
\def\makechapterlabel{\let\sectionlabel=\relax
   \xdef\chapterlabel{\the\chapterstyle{\the\chapternumber}.}}
\def\alphabetic#1{\count255='140 \advance\count255 by #1\char\count255}
\def\Alphabetic#1{\count255='100 \advance\count255 by #1\char\count255}
\def\Roman#1{\uppercase\expandafter{\romannumeral #1}}
\def\roman#1{\romannumeral #1}
\def\Number#1{\number #1}
\def\BLANC#1{}
\def\titlestyle#1{\par\begingroup \interlinepenalty=9999
     \leftskip=0.02\hsize plus 0.23\hsize minus 0.02\hsize
     \rightskip=\leftskip \parfillskip=0pt
     \hyphenpenalty=9000 \exhyphenpenalty=9000
     \tolerance=9999 \pretolerance=9000
     \spaceskip=0.333em \xspaceskip=0.5em
     \iftwelv@\fourteenpoint\else\twelvepoint\fi
   \noindent #1\par\endgroup }
\def\spacecheck#1{\dimen@=\pagegoal\advance\dimen@ by -\pagetotal
   \ifdim\dimen@<#1 \ifdim\dimen@>0pt \vfil\break \fi\fi}
\def\TableOfContentEntry#1#2#3{\relax}
\def\chapter#1{\par \penalty-300 \vskip\chapterskip
   \spacecheck\chapterminspace
   \chapterreset \titlestyle{\chapterlabel\ #1}
   \TableOfContentEntry c\chapterlabel{#1}
   \nobreak\vskip\headskip \penalty 30000
   \wlog{\string\chapter\space \chapterlabel} }

\def\section#1{\par \ifnum\the\lastpenalty=30000\else
   \penalty-200\vskip\sectionskip \spacecheck\sectionminspace\fi
   \global\advance\sectionnumber by 1
   \xdef\sectionlabel{\the\sectionstyle\the\sectionnumber}
   \wlog{\string\section\space \sectionlabel}
   \TableOfContentEntry s\sectionlabel{#1}
   \noindent {\caps\enspace\sectionlabel\quad #1}\par
   \nobreak\vskip\headskip \penalty 30000 }
\def\subsection#1{\par
   \ifnum\the\lastpenalty=30000\else \penalty-100\smallskip \fi
   \noindent\undertext{#1}\enspace \vadjust{\penalty5000}}

\def\undertext#1{\vtop{\hbox{#1}\kern 1pt \hrule}}
\def\APPENDIX#1#2{\par\penalty-300\vskip\chapterskip
   \spacecheck\chapterminspace \chapterreset \xdef\chapterlabel{#1}
   \titlestyle{APPENDIX #2} \nobreak\vskip\headskip \penalty 30000
   \TableOfContentEntry a{#1}{#2}
   \wlog{\string\Appendix\ \chapterlabel} }
\def\Appendix#1{\APPENDIX{#1}{#1}}
\def\appendix{\APPENDIX{A}{}}
\def\unnumberedchapters{\let\makechapterlabel=\relax \let\chapterlabel=\relax
   \sectionstyle={\BLANC}\let\sectionlabel=\relax \sequentialequations }
%
%
%
\def\eqname#1{\relax \ifnum\equanumber<0
     \xdef#1{{\noexpand\rm(\number-\equanumber)}}%
       \global\advance\equanumber by -1
    \else \global\advance\equanumber by 1
      \xdef#1{{\noexpand\rm(\chapterlabel\number\equanumber)}} \fi #1}
\def\eqinsert#1{\noalign{\dimen@=\prevdepth \nointerlineskip
   \setbox0=\hbox to\displaywidth{\hfil #1}
   \vbox to 0pt{\kern 0.5\baselineskip\hbox{$\!\box0\!$}\vss}
   \prevdepth=\dimen@}}
%

%
%
\def\GENITEM#1;#2{\par \hangafter=0 \hangindent=#1
    \Textindent{$ #2 $}\ignorespaces}
\outer\def\newitem#1=#2;{\gdef#1{\GENITEM #2;}}
\newdimen\itemsize                \itemsize=30pt
\newitem\item=1\itemsize;
\newitem\sitem=1.75\itemsize;     
\newitem\ssitem=2.5\itemsize;     
\outer\def\newlist#1=#2&#3&#4;{\toks0={#2}\toks1={#3}%
   \count255=\escapechar \escapechar=-1
   \alloc@0\list\countdef\insc@unt\listcount     \listcount=0
   \edef#1{\par
      \countdef\listcount=\the\allocationnumber
      \advance\listcount by 1
      \hangafter=0 \hangindent=#4
      \Textindent{\the\toks0{\listcount}\the\toks1}}
   \expandafter\expandafter\expandafter
    \edef\c@t#1{begin}{\par
      \countdef\listcount=\the\allocationnumber \listcount=1
      \hangafter=0 \hangindent=#4
      \Textindent{\the\toks0{\listcount}\the\toks1}}
   \expandafter\expandafter\expandafter
    \edef\c@t#1{con}{\par \hangafter=0 \hangindent=#4 \noindent}
   \escapechar=\count255}
\def\c@t#1#2{\csname\string#1#2\endcsname}
\newlist\point=\Number&.&1.0\itemsize;
\newlist\subpoint=(\alphabetic&)&1.75\itemsize;
\newlist\subsubpoint=(\roman&)&2.5\itemsize;
%

%
%
%
%
\newcount\referencecount     \referencecount=0
\newcount\lastrefsbegincount \lastrefsbegincount=0
\newif\ifreferenceopen       \newwrite\referencewrite
\newif\ifrw@trailer
\newdimen\refindent     \refindent=30pt
\def\NPrefmark#1{\attach{\scriptscriptstyle [ #1 ] }}
\let\PRrefmark=\attach
\def\refmark#1{\relax\ifPhysRev\PRrefmark{#1}\else\NPrefmark{#1}\fi}
\def\refend@{\refmark{\number\referencecount}}
\def\refend{\refend@{}\space }
\def\refsend{\refmark{\count255=\referencecount
   \advance\count255 by-\lastrefsbegincount
   \ifcase\count255 \number\referencecount
   \or \number\lastrefsbegincount,\number\referencecount
   \else \number\lastrefsbegincount-\number\referencecount \fi}\space }
\def\refitem#1{\par \hangafter=0 \hangindent=\refindent \Textindent{#1}}
\def\Ref{\rw@trailertrue\REF}
\def\ref{\Ref\?}

\def\REF#1{\r@fstart{#1}%
   \rw@begin{\the\referencecount.}\rw@end}
\def\REFS#1{\r@fstart{#1}%
   \lastrefsbegincount=\referencecount
   \rw@begin{\the\referencecount.}\rw@end}
\def\r@fstart#1{\chardef\rw@write=\referencewrite \let\rw@ending=\refend@
   \ifreferenceopen \else \global\referenceopentrue
   \immediate\openout\referencewrite=referenc.txa
   \toks0={\catcode`\^^M=10}\immediate\write\rw@write{\the\toks0} \fi
   \global\advance\referencecount by 1 \xdef#1{\the\referencecount}}
 {\catcode`\^^M=\active %
 \gdef\rw@begin#1{\immediate\write\rw@write{\noexpand\refitem{#1}}%
   \begingroup \catcode`\^^M=\active \let^^M=\relax}%
 \gdef\rw@end#1{\rw@@end #1^^M\rw@terminate \endgroup%
   \ifrw@trailer\rw@ending\global\rw@trailerfalse\fi }%
 \gdef\rw@@end#1^^M{\toks0={#1}\immediate\write\rw@write{\the\toks0}%
   \futurelet\n@xt\rw@test}%
 \gdef\rw@test{\ifx\n@xt\rw@terminate \let\n@xt=\relax%
       \else \let\n@xt=\rw@@end \fi \n@xt}%
}
\let\rw@ending=\relax
\let\rw@terminate=\relax
\let\splitout=\relax
\def\Closeout#1{\toks0={\catcode`\^^M=5}\immediate\write#1{\the\toks0}%
   \immediate\closeout#1}
%
%
\newcount\figurecount     \figurecount=0
\newcount\tablecount      \tablecount=0
\newif\iffigureopen       \newwrite\figurewrite
\newif\iftableopen        \newwrite\tablewrite
\def\FIG#1{\f@gstart{#1}%
   \rw@begin{\the\figurecount)}\rw@end}

\def\Fig{\rw@trailertrue\def\rw@ending{Fig.~\?}\FIG\?}
\def\fig{\rw@trailertrue\def\rw@ending{fig.~\?}\FIG\?}
\def\TABLE#1{\T@Bstart{#1}%
   \rw@begin{\the\tableecount:}\rw@end}
\def\Table{\rw@trailertrue\def\rw@ending{Table~\?}\TABLE\?}
\def\f@gstart#1{\chardef\rw@write=\figurewrite
   \iffigureopen \else \global\figureopentrue
   \immediate\openout\figurewrite=figures.txa
   \toks0={\catcode`\^^M=10}\immediate\write\rw@write{\the\toks0} \fi
   \global\advance\figurecount by 1 \xdef#1{\the\figurecount}}
\def\T@Bstart#1{\chardef\rw@write=\tablewrite
   \iftableopen \else \global\tableopentrue
   \immediate\openout\tablewrite=tables.txa
   \toks0={\catcode`\^^M=10}\immediate\write\rw@write{\the\toks0} \fi
   \global\advance\tablecount by 1 \xdef#1{\the\tablecount}}
%

%
%
%
\def\getfigure#1{\global\advance\figurecount by 1
   \xdef#1{\the\figurecount}\count255=\escapechar \escapechar=-1
   \edef\n@xt{\noexpand\g@tfigure\csname\string#1Body\endcsname}%
   \escapechar=\count255 \n@xt }
\def\g@tfigure#1#2 {\errhelp=\disabledfigures \let#1=\relax
   \errmessage{\string\getfigure\space disabled}}
\newhelp\disabledfigures{ Empty figure of zero size assumed.}
\def\figinsert#1{\midinsert\Tenpoint\medskip
   \count255=\escapechar \escapechar=-1
   \edef\n@xt{\csname\string#1Body\endcsname}
   \escapechar=\count255 \centerline{\n@xt}
   \bigskip\narrower\narrower
   \noindent{\it Figure}~#1.\quad }
%
%
%
\def\masterreset{\global\pagenumber=1 \global\chapternumber=0
   \global\equanumber=0 \global\sectionnumber=0
   \global\referencecount=0 \global\figurecount=0 \global\tablecount=0 }
\def\FRONTPAGE{\ifvoid255\else\vfill\penalty-20000\fi
      \masterreset\global\frontpagetrue
      \global\lastf@@t=0 \global\footsymbolcount=0}

\def\paperstyle{\letterstylefalse\normalspace\papersize}
\def\letterstyle{\letterstyletrue\singlespace\lettersize}
\def\papersize{\hsize=35 truepc\vsize=50 truepc\hoffset=-2.51688 truepc
               \skip\footins=\bigskipamount}
\def\lettersize{\hsize=5.5 truein\vsize=8.25 truein\hoffset=.4875 truein
	\voffset=.3125 truein
   \skip\footins=\smallskipamount \multiply\skip\footins by 3 }
\paperstyle   
%
%
\def\MEMO{\letterstyle \letterinfo={\hfil } \let\rule=\memorule
	\FRONTPAGE \memohead }
\let\memohead=\relax

\def\memit@m#1{\smallskip \hangafter=0 \hangindent=1in
      \Textindent{\caps #1}}
\def\subject{\memit@m{Subject:}}
\def\topic{\memit@m{Topic:}}
\def\from{\memit@m{From:}}
\def\to{\relax \ifmmode \rightarrow \else \memit@m{To:}\fi }
\def\memorule{\medskip\hrule height 1pt\bigskip}
\newwrite\labelswrite
\newtoks\rw@toks

\def\addressee#1{\null\vskip .5truein\line{
\hskip 0.5\hsize minus 0.5\hsize\the\date\hfil}\bigskip
   \ialign to\hsize{\strut ##\hfil\tabskip 0pt plus \hsize \cr #1\crcr}
   \writelabel{#1}\medskip\par\noindent}
\def\rwl@begin#1\cr{\rw@toks={#1\crcr}\relax
   \immediate\write\labelswrite{\the\rw@toks}\futurelet\n@xt\rwl@next}
\def\rwl@next{\ifx\n@xt\rwl@end \let\n@xt=\relax
      \else \let\n@xt=\rwl@begin \fi \n@xt}
\let\rwl@end=\relax
\def\writelabel#1{\immediate\write\labelswrite{\noexpand\labelbegin}
     \rwl@begin #1\cr\rwl@end
     \immediate\write\labelswrite{\noexpand\labelend}}
\newbox\FromLabelBox

\let\labelend=\relax
\def\labelbegin#1\labelend{\setbox0=\vbox{\ialign{##\hfil\cr #1\crcr}}
     \MakeALabel }
\newtoks\FromAddress
\FromAddress={}
\def\MakeFromBox#1{\global\setbox\FromLabelBox=\vbox{\Tenpoint
     \ialign{##\hfil\cr #1\the\FromAddress\crcr}}}
\newdimen\labelwidth		\labelwidth=6in
\def\MakeALabel{\vskip 1pt \hbox{\vrule \vbox{
	\hsize=\labelwidth \hrule\bigskip
	\leftline{\hskip 1\parindent \copy\FromLabelBox}\bigskip
	\centerline{\hfil \box0 } \bigskip \hrule
	}\vrule } \vskip 1pt plus 1fil }
\newskip\signatureskip       \signatureskip=30pt
\def\signed#1{\par \penalty 9000 \medskip \dt@pfalse
  \everycr={\noalign{\ifdt@p\vskip\signatureskip\global\dt@pfalse\fi}}
  \setbox0=\vbox{\singlespace \ialign{\strut ##\hfil\crcr
   \noalign{\global\dt@ptrue}#1\crcr}}
  \line{\hskip 0.5\hsize minus 0.5\hsize \box0\hfil} \medskip }
\newbox\letterb@x
\def\lettertext{\par\unvcopy\letterb@x\par}
\def\multiletter{\setbox\letterb@x=\vbox\bgroup
      \everypar{\vrule height 1\baselineskip depth 0pt width 0pt }
      \singlespace \topskip=\baselineskip }
\def\letterend{\par\egroup}
%
%
%
\newskip\frontpageskip
\newtoks\Pubnum
\newtoks\pubtype
\newif\ifp@bblock  \p@bblocktrue
\def\PH@SR@V{\doubl@true \baselineskip=24.1pt plus 0.2pt minus 0.1pt
             \parskip= 3pt plus 2pt minus 1pt }
\def\PHYSREV{\paperstyle\PhysRevtrue\PH@SR@V}
\def\titlepage{\FRONTPAGE\paperstyle\ifPhysRev\PH@SR@V\fi
   \ifp@bblock\p@bblock \else\hrule height\z@ \relax \fi }
\def\nopubblock{\p@bblockfalse}

\frontpageskip=12pt plus .5fil minus 2pt
\pubtype={\tensl Preliminary Version}
\Pubnum={}
\def\p@bblock{\begingroup \tabskip=\hsize minus \hsize
   \baselineskip=1.5\ht\strutbox \topspace-2\baselineskip
   \halign to\hsize{\strut ##\hfil\tabskip=0pt\crcr
       \the\Pubnum\crcr\the\date\crcr\the\pubtype\crcr}\endgroup}
\def\title#1{\vskip\frontpageskip \titlestyle{#1} \vskip\headskip }
\def\author#1{\vskip\frontpageskip\titlestyle{\twelvecp #1}\nobreak}

\def\address#1{\par\kern 5pt\titlestyle{\twelvepoint\it #1}}
\def\andaddress{\par\kern 5pt \centerline{\sl and} \address}

\def\abstract{\par\dimen@=\prevdepth \hrule height\z@ \prevdepth=\dimen@
   \vskip\frontpageskip\centerline{\fourteenrm ABSTRACT}\vskip\headskip }

%
%
%

\def\\{\relax \ifmmode \backslash \else {\tt\char`\\}\fi }
\def\sequentialequations{\relax\if\equanumber<0\else\global\equanumber=-1\fi}

\def\journal#1&#2(#3){\unskip, \sl #1\unskip~\bf\ignorespaces #2\rm (19#3),}

\def\topspace{\hrule height 0pt depth 0pt \vskip}

\def\Buildrel#1\under#2{\mathrel{\mathop{#2}\limits_{#1}}}
\def\becomes#1{\mathchoice{\becomes@\scriptstyle{#1}}{\becomes@\scriptstyle
   {#1}}{\becomes@\scriptscriptstyle{#1}}{\becomes@\scriptscriptstyle{#1}}}
\def\becomes@#1#2{\mathrel{\setbox0=\hbox{$\m@th #1{\,#2\,}$}%
	\mathop{\hbox to \wd0 {\rightarrowfill}}\limits_{#2}}}

\let\int=\intop         \let\oint=\ointop
\def\lsim{\mathrel{\mathpalette\@versim<}}
\def\gsim{\mathrel{\mathpalette\@versim>}}
\def\@versim#1#2{\vcenter{\offinterlineskip
	\ialign{$\m@th#1\hfil##\hfil$\crcr#2\crcr\sim\crcr } }}
\def\big#1{{\hbox{$\left#1\vbox to 0.85\b@gheight{}\right.\n@space$}}}
\def\Big#1{{\hbox{$\left#1\vbox to 1.15\b@gheight{}\right.\n@space$}}}
\def\bigg#1{{\hbox{$\left#1\vbox to 1.45\b@gheight{}\right.\n@space$}}}
\def\Bigg#1{{\hbox{$\left#1\vbox to 1.75\b@gheight{}\right.\n@space$}}}
%
%
%
\let\sec@nt=\sec
\def\sec{\relax\ifmmode\let\n@xt=\sec@nt\else\let\n@xt\section\fi\n@xt}
\def\obsolete#1{\message{Macro \string #1 is obsolete.}}
\def\firstsec#1{\obsolete\firstsec \section{#1}}
\def\firstsubsec#1{\obsolete\firstsubsec \subsection{#1}}
\def\thispage#1{\obsolete\thispage \global\pagenumber=#1\frontpagefalse}
\def\thischapter#1{\obsolete\thischapter \global\chapternumber=#1}
\def\REFSCON{\obsolete\REFSCON\REF}
\def\splitout{\obsolete\splitout\relax}
\def\prop{\obsolete\prop \propto }
\def\nextequation#1{\obsolete\nextequation \global\equanumber=#1
   \ifnum\the\equanumber>0 \global\advance\equanumber by 1 \fi}
\def\BOXITEM{\afterassigment\B@XITEM\setbox0=}
\def\B@XITEM{\par\hangindent\wd0 \noindent\box0 }
\def\phyzzx{PHY\setbox0=\hbox{Z}\copy0 \kern-0.5\wd0 \box0 X}
%
%
\everyjob{\xdef\today{\monthname\ \number\day, \number\year}}
        
%


\hoffset=0.2truein
\voffset=0.1truein
\hsize=6truein

\def\CALT#1{\hbox to\hsize{\tenpoint \baselineskip=12pt
	\hfil\vtop{\hbox{\strut CALT-68-#1}
	\hbox{\strut DOE RESEARCH AND}
	\hbox{\strut DEVELOPMENT REPORT}}}}

\def\sqr#1#2{{\vcenter{\hrule height.#2pt
      \hbox{\vrule width.#2pt height#1pt \kern#1pt
        \vrule width.#2pt}
      \hrule height.#2pt}}}

\def\section#1#2{
\noindent\hbox{\hbox{\bf #1}\hskip 10pt\vtop{\hsize=5in
\baselineskip=12pt \noindent \bf #2 \hfil}\hfil}
\medskip}

\def\underwig#1{	
	\setbox0=\hbox{\rm \strut}
	\hbox to 0pt{$#1$\hss} \lower \ht0 \hbox{\rm \char'176}}

\def\bunderwig#1{	
	\setbox0=\hbox{\rm \strut}
	\hbox to 1.5pt{$#1$\hss} \lower 12.8pt
	 \hbox{\seventeenrm \char'176}\hbox to 2pt{\hfil}}

\def\MEMO#1#2#3#4#5{
\frontpagetrue
\centerline{\tencp INTEROFFICE MEMORANDUM}
\smallskip
\centerline{\bf CALIFORNIA INSTITUTE OF TECHNOLOGY}
\bigskip
\vtop{\tenpoint \hbox to\hsize{\strut \hbox to .75in{\caps to:\hfil}
\hbox to3in{#1\hfil}
\hbox to .75in{\caps date:\hfil}\quad \the\date\hfil}
\hbox to\hsize{\strut \hbox to.75in{\caps from:\hfil}\hbox to 2in{#2\hfil}
\hbox{{\caps extension:}\quad#3\qquad{\caps mail code:\quad}#4}\hfil}
\hbox{\hbox to.75in{\caps subject:\hfil}\vtop{\parindent=0pt
\hsize=3.5in #5\hfil}}
\hbox{\strut\hfil}}}

{}

\def\bx{{\hbox{ $\sqcup$}\llap{\hbox{$\sqcap$}}}}

\def\sqhat{{\sqrt{\hat g}}}
\def\sqphi{{\sqrt{\hat g}} e^\phi}

\def\sh{{\sqrt {\hat g}}}
\def\g{{{\hat g}e^\phi}}
\def\box{{\hbox{ $\sqcup$}\llap{\hbox{$\sqcap$}}}}
\def\hatbox{{\hbox{ $\sqcup$}\llap{\hbox{$\hat\sqcap$}}}}

{}

\Ref\kada{V. Kazakov, Phys. Lett. 150B, 282 (1985);
F. David, Nucl. Phys. B257, 45 (1985)}
\Ref\polov{A.M. Polyakov, ``Gauge Fields and Strings,'' Harwood
Academic Publishers 1987}
\Ref\poy{A.M. Polyakov, Phys. Lett. 103B, 207 (1981)}
\Ref\d{F. David, Mod. Phys. Lett. A3, 1651 (1988)}
\Ref\dk{J. Distler and H. Kawai, Nucl. Phys. B321, 509 (1988)}
\Ref\polch{J. Polchinski, Nucl. Phys. B346, 253 (1990)}
\Ref\gsw{E.g., M.B. Green, J.H. Schwarz and E. Witten, ``Superstring
Theory," Cambridge University Press 1987}
\Ref\awi{L. Alvarez-Gaum\'e and E. Witten, Nucl. Phys. B234, 269 (1983)}
\Ref\bd{E.g., N. D. Birrell and P. C. W. Davies,
``Quantum Fields in Curved Space," Cambridge University Press 1982}
\Ref\bpz{A.A. Belavin, A.M. Polyakov and A.B. Zamolodchikov,
Nucl. Phys. B241 (1984)}
\Ref\fri{D. Friedan, in Proceedings of the Les Houches session XXXIX,
``Recent Advances in Field Theory and Statistical
Mechanics." (1982)}
\Ref\kpz{V. Knizhnik, A. M. Polyakov and A. B. Zamolodchikov,
Mod. Phys. Lett. A3, 819 (1988)}
\Ref\car{E.g., J. H. Cardy, Les Houches session XLIX, 1988}
\Ref\gn{J.L. Gervais and A. Neveu, Nucl. Phys. B199, 59 (1982);
B209, 125 (1982); B238, 125; 396 (1984); Phys. Lett. 151B, 271 (1985)}
\Ref\triv{A. Gupta, S. Trivedi and M.B. Wise, Nucl. Phys. B340,
475 (1990)}
\Ref\gou{M. Goulian and M. Li, Phys. Rev. Lett. 66, 2051 (1991)}
\Ref\mxm{E.g., D. J. Gross and A. A. Migdal,
Nucl. Phys. B340, 333 (1990) and references therein.}
\Ref\sei{N. Seiberg, Prog. Theor. Phys. S102, 319 (1990)}
\Ref\dkaw{J. Distler, Z. Hlousek and H. Kawai, Int. J. of Mod.
Phys. A5, 1093 (1990)}
\Ref\pch{J. Polchinski, talk given at strings '90, Coll. Station,
Texas; publ. in Coll. Station
Wkshp. 1990, 62 (1990)}
\Ref\mss{G. Moore, N. Seiberg and M. Staudacher,
Nucl. Phys. B362, 665 (1991)}
\Ref\dis{J. Distler, Nucl. Phys. B342, 523 (1990)}
\Ref\wiw{E. Witten, Phys. Lett. 206B, 601 (1988);
J.M.F. Labastida and M. Pernici, Phys. Lett. 212B, 56 (1988)}
\Ref\wtt{E. Witten, Nucl. Phys. B373, 187 (1991)}
\Ref\pkl{I. Klebanov and A. M. Polyakov, Mod. Phys. Lett. A6, 3273 (1991)}
\Ref\kle{I. Klebanov, Lectures at ICTP Spring School,
Trieste (1991)}
\Ref\itz{E. Br\'ezin, C. Itzykson, G. Parisi and J.B. Zuber,
Comm. Math. Phys. 59, 35 (1978)}
\Ref\bre{C. Itzykson and J.B. Zuber, J. Math. Phys. 21(3), 411 (1980)}
\Ref\kmi{V. Kazakov and A. Migdal, Nucl. Phys. B311, 171 (1989)}
\Ref\double{M. Douglas and S. Shenker, Nucl. Phys. B335, 635 (1990);
E. Br\'ezin and V. Kazakov, Phys. Lett. 236B, 144 (1990);
D.J. Gross and A.A. Migdal, Phys. Rev. Lett 64, 717 (1990)}
\Ref\moo{G. Moore, Yale preprint YCTP-P1-92}
\Ref\kkv{V.A. Kazakov, Phys. Lett. 119A, 140 (1986)}

\Ref\poly{A. M. Polyakov, Mod. Phys. Lett. A6, 3273 (1991)}
\Ref\seg{G. Segal, Comm. Math. Phys. 80, 301 (1981)}
\Ref\suss{A. Cooper, L. Susskind and L. Thorlacius,
Nucl. Phys. B363, 132 (1991)}
\Ref\kut{D. Kutasov, Princeton preprint PUPT-1334 (1992)}
\Ref\pkov{A. M. Polyakov, Lectures in Les Houches, July 1992}
\Ref\gkl{D. Gross and I. Klebanov, Nucl. Phys. B344, 475 (1990)}
\Ref\jhs{J. H. Schwarz,
Phys. Lett. 272B, 239 (1991)}
\Ref\ft{E.S. Fradkin and A.A. Tseytlin, Nucl. Phys. B201, 469 (1982)}
\Ref\schm{C. Schmidhuber, Nucl. Phys. B390, 188 (1993)}
\Ref\tsey{A.A. Tseytlin, Phys. Lett. B264, 311 (1991)}
\Ref\ban{T. Banks, Nucl. Phys. B361, 166 (1991)}

\Ref\mag{M. E. Agishtein and A. A. Migdal,
Nucl. Phys. B350, 690 (1991)}
\Ref\ma{I. Antoniadis and E. Mottola, Los Alamos preprint LA-UR-91-1653 (1991)}
\Ref\dho{E. D' Hoker, Mod. Phys. Lett. A6, 745 (1991)}
\Ref\wv{E. Witten, Nucl. Phys. B340, 281 (1990); E. and H. Verlinde,
Nucl. Phys. B348, 457 (1991)}
\Ref\egu{In T. Eguchi, P. B. Gilkey and
A. J. Hanson, Phys. Rep. 66, 213 (1988), after I. M. Singer}
\Ref\bvi{A.O. Barvinsky and G.A. Vilkovisky, Phys. Rep. 119, 1 (1985)}
\Ref\amm{I. Antoniadis, P.O. Mazur and E. Mottola, preprint CPTH-A173.0492
/ LA-UR-92-1483 (1992)}
\Ref\ati{E.g., M.F. Atiyah, N.J. Hitchin and I.M. Singer,
Proc. Roy. Soc. London A362, 425 (1978)}

\hoffset .4in

\headline={\hfil}
\footline={\hfil}
\hoffset .52in
\voffset .4in

\line{}\vskip1cm
\centerline{\bf{EXTENDING THE THEORY OF RANDOM SURFACES}}
\vskip 3cm
\centerline{Thesis by}\vskip 3mm
\centerline{\bf{Christof Schmidhuber}}\vskip 1cm
\centerline{{In Partial Fulfillment of the Requirements}}\vskip 2mm
\centerline{{for the Degree of}}\vskip 2mm
\centerline{{Doctor of Philosophy}}\vskip 5cm
\centerline{{California Institute of Technology}}\vskip 3mm
\centerline{{Pasadena, California}}\vskip 10mm
\centerline{1993}\vskip3mm
\centerline{(submitted May 14, 1993)}
\vfill\eject

\headline={\hss\tenrm\folio\hss}

\pageno=-2

\line{}\vskip 15cm
\centerline{{\copyright\ \ 1993}}\vskip 3mm
\centerline{{by Christof Schmidhuber}}\vskip 3mm
\centerline{{All Rights Reserved}}
\vfill\eject

\line{}\vskip2cm
\centerline{\bf{Acknowledgements}}\vskip1cm

Being around people who share my enthusiasm for doing physics
has made me feel very much at home on the fourth floor of Lauritsen.
I thank Steve Frautschi,
John Preskill, John Schwarz, Mark Wise,
Anna, David and David, Dimitris, Helen, Jerry, Keke,
Malik, Martin, Ming, Phil, Pil, Rob, Sandip and Wolfgang
for the very friendly
atmosphere that has made it a pleasure to come to work.
\vskip3mm

I am particularly
grateful to my advisor John Schwarz, for never hesitating to support me
with help and advice when I had plans or ideas that I wanted to
carry out,
and for his encouragement that gave me confidence in
times when nothing seemed to work.
I also thank Keke Li for
listening to my ideas and my
confusions
and for his efforts to
keep my feet on the ground.
\vskip3mm

Needless to say, my Caltech years would not have been as much fun
without
occasionally extending campus to anywhere between Mount Denali
and Lake Titicaca, together with Zhaoping, Ojvind, Thomas and
the others.
And last but not least,
I am happy that the distance to
my family and my friends back home has remained purely
geographical.

\vfill\eject

{}\line{}\vskip2cm
\centerline{\bf{Abstract}}\vskip1cm

The theory of embedded random surfaces, equivalent to
two--dimensional
quantum gravity coupled to matter, is reviewed, further developed
and partly generalized to four dimensions.
It is shown that
the action of the Liouville field theory that describes random surfaces
contains terms that have not been noticed previously.
These terms are used to explain
the phase diagram
of the Sine--Gordon model coupled to gravity, in agreement with recent
results from lattice computations.
It is also demonstrated how the methods of two--dimensional quantum gravity
can be applied to four--dimensional
Euclidean gravity in the limit
of infinite Weyl coupling.
Critical exponents are predicted and an analog of the ``$c=1$ barrier''
of two--dimensional gravity is derived.

\vfill\eject

{}

\singlespace

\centerline{{\bf CONTENTS}}
\vskip 15mm

\line{{\bf Introduction}\hfill {\bf1}}
\vskip 1cm
\line{\bf Part I : Review of Previous Work on Random Surfaces
\hfill 3}
\vskip 4mm
1. Random Surfaces and Random Walks\hfill 3
\vskip0mm
2. 2D Gravity in Conformal Gauge\hfill 8
\vskip0mm
3. The Trace Anomaly and DDK\hfill 12
\vskip0mm
4. Applied Liouville Theory\hfill 20
\vskip0mm
Appendix: Random Lattices and Matrix Models\hfill 28
\vskip 1cm

\line{\bf{Part II : Running Coupling Constants in 2D Gravity
\hfill 34}}
\vskip 4mm
1. Introduction\hfill 34
\vskip0mm
2. Exactly Marginal Operators \hfill 36
\vskip0mm
3. Running Coupling Constants\hfill 42
\vskip0mm
4. Outlook\hfill 47
\vskip0mm
Appendix A: Boundary Conditions\hfill 49
\vskip0mm
Appendix B: The Cosmological Constant\hfill 52
\vskip 1cm

\line{\bf{Part III : A 4D Analog of 2D Gravity
\hfill 54}}
\vskip 4mm
1. Introduction\hfill 54
\vskip0mm
2. Conformal Gauge\hfill 57
\vskip0mm
3. Liouville in 4D\hfill 59
\vskip0mm
4. DDK in 4D\hfill 61
\vskip0mm
5. Results\hfill 67
\vskip0mm
6. Weyl Gravity at Short Distances\hfill 70
\vskip 1cm

\line{{\bf References\hfill 73}}

\vfill\eject

\normalspace
\pageno=1

{}\centerline{\bf{Introduction}}\vskip.5cm

Like random walks, random surfaces appear in many physical
systems -- in statistical mechanics, in QCD, in string theory
and in other fields. But while
random walks embedded in any number of dimensions
have long been well--understood, only in
recent years has there been much progress on
random surfaces. The
most striking development has been
the ``matrix model'' -- a gedanken experiment
that has yielded numerical values for
critical coefficients and correlation functions.

Unfortunately, this method
is restricted to random surfaces embedded in $D\le1$ dimensions.
{}From the point of view of physics,
the models with $D\le1$ are
not interesting
by themselves. The physically interesting models involve either
higher embedding dimensions -- $D=3$ for the theory of phase transitions,
$D=4$ for QCD, $D=26$ and $D=10$ for string theory and superstring theory
-- or world--sheet dimension four instead of two, if one
thinks of the theory of random geometries as quantum gravity.

It is therefore necessary to develop a field theory
of random surfaces that can be generalized to these cases.
What makes the models with $D\le1$ interesting is their role as
ideal laboratories
for testing such a theory -- ideal precisely because the answers
are known from the matrix models.

Although major progress
in this direction has recently been made
by David, Distler and Kawaii, this theory is still
incomplete even for $D\le1$.
Thus there are presently two challenges:
on the one hand, our understanding
of the matrix model results from the continuum approach must be
completed. On the other hand, this continuum approach
must be extended
to the physically interesting cases mentioned above.

This double challenge is reflected in this work.
Part II fills a gap
in the continuum theory of random surfaces in $D\le1$
dimensions,
while part III begins generalizing this theory to four dimensions.
\eject

To provide the necessary background,
previous developments in the theory of random surfaces are
summarized in part I. Based on its formulation
as two--dimensional quantum gravity coupled to matter, the theory
is discussed in conformal gauge. The conformal
anomaly, the Liouville action, the proposal of David, Distler
and Kawai, the computation of critical coefficients and the spectrum
of states are reviewed.
A brief introduction to random lattices and
matrix models is given in the appendix.

In part II, it is shown that the action for two--dimensional
quantum gravity coupled to interacting matter contains
certain terms that have not
been noticed previously. They are crucial for understanding the
renormalization group flow, and can be observed in recent matrix model results
for the phase diagram of the Sine--Gordon model coupled to gravity.
These terms ensure, order by order in the coupling constant of the
interaction, that
the theory
is scale invariant.
They are discussed up to second
order.

In part III, it is asked in how far the methods of two--dimensional
quantum gravity can be applied to four--dimensional gravity. It is found
that they can be applied to Weyl gravity at its ultraviolet fixed point
of infinite Weyl coupling.
There, the path integral over geometries
reduces to integrals over the conformal factor and over the moduli
space of conformally self--dual metrics. The conformal anomaly induces
an analog of the Liouville action. The proposal of David, Distler and Kawai
is generalized to four dimensions. Critical exponents are
predicted and the analog
of the $c=1$ barrier of two--dimensional gravity is derived.

\vfill\eject

{}




\leftline{{\bf PART I:\
REVIEW OF PREVIOUS WORK ON RANDOM SURFACES}}
\vskip1.5cm

{\leftline{\bf 1. Random Surfaces and Random Walks}}
\vskip.5cm

 {$\underline{\hbox{1.1. The Problem}}$}

The topic of Part I is the sum over embeddings of closed,
compact, euclidean two--dimensional
surfaces in D--dimensional space:
$$\int {{{\cal D}x^i(\sigma)}\over{\hbox{Diff}}}\ e^{-S},
\ \ \ S\sim\ \hbox{area +
other geometrical quantities.}
\eqno(1.1)$$
Here, $\sigma\equiv(\sigma_1,\sigma_2)$ parametrizes the surface and
the $x^i$ parametrize the embedding space.
``Diff'' in the denominator
indicates that the sum is over embeddings modulo diffeomorphisms,
\foot{coordinate transformations on the surface}
i.e., over ``geometries''.
Since such geometries
are the two--dimensional analogues of continuous random walks,
they are called ``random surfaces''. For random walks, $\sigma$ in (1.1)
would be a single parameter, and the leading term in
the action $S$ would be proportional
to the length of the walk.

Instead of summing over closed surfaces in (1.1),
corresponding to closed random walks,
we could sum over surfaces with some boundary cycles
that are fixed in the embedding
space. This would be the analog of random walks going from some point $A$
to some point $B$. Unfortunately, the theory of random surfaces
with boundaries is presently not well--developed,
so we will mostly concentrate on the sum over closed surfaces below.

Unlike closed paths, closed surfaces can have
different topologies. This is one of the difficulties one encounters
when studying random surfaces -- it is more like studying
interacting random walks (see fig. 1).
For the most part,
we will concentrate on surfaces of spherical
topology below, although some things will be said about the sum over
topologies.
\eject

 {$\underline{\hbox{1.2. The Motivation}}$}

Why do we want to study the sum (1.1) in the first place?
Like random walks, random surfaces have many
interesting applications and a better understanding
of them will benefit diverse areas of physics.
Here are some examples of where random surfaces occur:

They appear in the low--temperature
expansion of
three--dimensional statistical mechanical
systems, like the Ising model, as boundaries between
regions of different phases.
Or, the perturbation expansion of large--N QCD
can be summed in terms of surfaces
of different topology.
The theory of random surfaces embedded
in $D$ dimensions is also equivalent to two--dimensional quantum
gravity coupled to $D$ bosons and thus provides a simple toy
model for problems in quantum gravity. Surely there is little hope
of understanding quantum gravity in four dimensions,
whatever it may be, before one
understands the much simpler two--dimensional case.

Perhaps most interestingly, the
theory of random surfaces is
string theory in first--quantized formulation, just like
the theory of interacting random walks is first--quantized
$\phi^4$ theory (fig. 1). Summing random surfaces
is equivalent to summing the string perturbation expansion and could
even lead to nonperturbative predictions of string theory as
a theory of the fundamental interactions including gravity.

\vskip 6.5cm
\line{\hskip .8cm Fig. 1a: A closed path, or\hskip 3cm
Fig. 1b: A closed surface, or\hfill}
\line{\hskip 0cm 4--loop vacuum diagram of $\phi^4$ theory\hskip .7cm
4--loop vacuum diagram of string theory\hfill}
\eject

 {$\underline{\hbox{1.3. Outline}}$}

We begin with an outline of the following introduction,
and of what makes random surfaces more difficult than
random walks.
There are two ways to perform the sum (1.1):
Either one discretizes the random surfaces as random triangulations
and tries to sum all distinct triangulations (fig. 2), or one attempts to do
the path integral using field theory.
\vskip 7cm
\line{\hskip.5cm Fig. 2a: A discretized random walk\hskip 1cm Fig. 2b:
A discretized random surface\hfill}\vskip.5cm

The first way is actually the more powerful one. Amazingly,
random triangulations can be summed with the help of the matrix
model trick,${}^{[\kada]}$
explained in the appendix: they are in one--to--one
correspondence with the Feynman diagrams of
a theory of $N\times N$--matrices
in the large $N$ limit. For embedding dimensions $D\le1$
\foot{The meaning of noninteger or negative $D$ will be explained below.}
this ``matrix model'' can be solved exactly, yielding critical
exponents and correlation functions. In this way one can even sum over
all possible topologies of the surfaces.

However, while the matrix model yields results, it offers little
understanding of how they arise, and it is restricted
to unphysical embedding dimensions. One would like to have a field
theory describing random surfaces, that can be generalized
to the more physical cases
in which no matrix models are available.
These include surfaces embedded in 3, 4 and 26 dimensions,
super--surfaces, and four--dimensional ``surfaces.'' For this reason
the emphasis in this review will be on the continuum approach.
The matrix model will be viewed
as a numerical ``experiment'' whose results allow us to check
the field theory description.

To develop such a description, it is best to rewrite (1.1) as two--dimensional
quantum gravity coupled to $D$ scalar fields,${}^{[\polov]}$
as will be explained in section 2.
Likewise, the random walk can be interpreted as one-dimensional quantum
gravity coupled to $D$ scalar fields, but is much more trivial:
one--dimensional geometries are labeled by only one diffeomorphism
invariant parameter -- their total length. They have no dynamics
and the resulting theory is
just $D$--dimensional quantum mechanics.

Two--dimensional geometries are best parametrized in conformal gauge,
by the conformal factor $\phi(\sigma)$, some moduli parameters and
the genus. This is also done in section 2 and leads to a theory
of $D+1$ two--dimensional fields, the $x$'s and $\phi$. As a field
theory, it has features that have no analogy in the
case of the random walk. The
most important one is the conformal anomaly: one might think that
two--dimensional gravity is also trivial, in the sense
that the Hilbert--Einstein action
is a topological invariant and the cosmological constant provides no
dynamics for the geometries. But as will be seen in section 3, the
conformal anomaly induces dynamics for $\phi$, forcing us to study
the Liouville action,${}^{[\poy]}$
$$\int d^2\sigma\ ((\partial\phi)^2+\mu\ e^{\alpha\phi}).$$
The reformulation of two--dimensional quantum gravity as an
ordinary field theory
involving the Liouville action was a big step forward, due to
David,${}^{[\d]}$ Distler and Kawai${}^{[\dk]}$ (DDK) and also reviewed
in section 3. The most important feature of this theory
is its background independence, reflecting general
covariance of the original theory. However, since
Liouville theory is notoriously difficult to deal with,
this approach is not yet as powerful as the
matrix model methods mentioned above. In particular, it is not known how to
sum over topologies. Some aspects of Liouville theory are reviewed
in section 4 and some results are extracted.

The continuum approach reveals a transition to a branched polymer
phase when the embedding dimension exceeds 1. The interesting
case $D=1$, a main topic of part II, is also briefly discussed
in section 4.
The review closes with a brief exposition of matrix model ideas in
the appendix.

 {$\underline{\hbox{1.4. Further Problems}}$}

As emphasized, random surfaces embedded in $D\le1$ dimensions are
not physically interesting by
themselves. Rather, these models should be used as testing grounds in which a
continuum theory of random sufaces can be developed, checked
with the help of the matrix model results and then generalized to the physical
cases.

For example,
after understanding quantum gravity in one and two dimensions
(random walks and random surfaces), one would
like to go on and understand Euclidean quantum gravity in four dimensions.
A first step towards this is taken in Part III. There, it is shown
how to generalize the above methods to four dimensions in the limit
of infinite Weyl coupling.

Even the theory of random surfaces in $D\le1$ dimensions
is not yet complete.
For example, in DDK's approach, background
independence has been imposed only to lowest order, in a sense that
will be explained.
Imposing it to next order also has important consequences, as will
be seen in Part II.

Another important gap in the present theory is that we do not know how to sum
over topologies in the continuum approach.
We know from the matrix models
that the result is simple and beautiful, given by the
KdV hierarchy. This might be a hint that there is a
simple method to perform this sum.
If such a method exists and can be generalized to the
physically interesting cases, the consequences could be
far--reaching: one might be able to do nonperturbative string theory
or nonperturbative QCD in the framework of first
quantized random surfaces. Or one might be able to study more rigorously
the effects of wormholes in four dimensions and their
relevance, e.g., for the cosmological constant problem.
Many other applications can be thought of. This must be left for future
research.

\vfill\eject

{\leftline{\bf 2. 2D Gravity in Conformal Gauge}}
\vskip.5cm

 {$\underline{\hbox{2.1. Random Surfaces and 2D Gravity}}$}

It is well--known that,
instead of the integral (1.1), one may study the equivalent path integral
for quantum gravity in two
dimensions coupled to $D$ scalar fields $x^i$.${}^{[\polov]}$
The partition function is
$$\eqalign{Z &=\int {{\cal D}g_{\alpha\beta}\over\hbox{Diff}}
\ {\cal D}x^i\ e^{-S[g,x]}\ \ \ \hbox{with}\cr
S &= \int d^2\sigma{\sqrt g}\{\mu + \gamma R + g^{\alpha\beta}
\partial_\alpha x^k\partial_\beta x_k +
\hbox{other covariant terms}\}. }  \eqno(2.1)$$
Here, $g_{\alpha\beta}$
is the two--dimensional metric, ``Diff'' again indicates that
we divide the diffeomorphism group out, $R$ is the Ricci scalar, and
$\mu$ and $\gamma$ are the cosmological and inverse Newtonian constants.
To see that (1.1) and (2.1) are equivalent is not trivial.
One first considers the case $\mu=\gamma=0$ and notes that
(1.1) and (2.1) are equivalent at the classical level: The equations of motion
for $g_{\alpha\beta}$ are
$$\partial_\alpha \vec x\cdot
\partial_\beta \vec x =
 {1\over2}g_{\alpha\beta}\ \partial_\alpha \vec x\cdot
\partial^\alpha \vec x.$$
The solution is
$$\partial_\alpha \vec x\cdot
\partial_\beta \vec x = g_{\alpha\beta},\ \ \ \hbox{so}\ \ \
S =\int d^2\sigma {\sqrt g}=\int d^2\sigma\ \vert\det\ \partial_\alpha
\vec x\cdot\partial_\beta \vec x\ \vert^{1/2}\eqno(2.2)$$
is a saddle point of the action.
$g_{\alpha\beta}$ at the saddle point is the embedding space metric
induced on the surface, since
$$ds^2=dx^k dx_k=\partial_\alpha x^kd\sigma^\alpha\ \partial_\beta x_k
 d\sigma^\beta
=g_{\alpha\beta}\ d\sigma^\alpha d\sigma^\beta
,\eqno(2.3)$$
so the saddle point action (2.2) is the area as in (1.1).
It is known as the Nambu--Goto action.
It can then be shown that quantum corrections are also of the form (2.2),
so that integrating out $g$ in (2.1) yields (1.1). For $\mu\neq0$,
(2.1) has no saddle point. Nevertheless the equivalence of
(2.1) and (1.1) can be seen to hold.
We refer to ref. [\polov] for details.

The
notion of embedded random surfaces can now be generalized to noninteger
embedding dimensions by coupling any conformally
invariant matter theory to gravity,
not only scalars $x^i$. The natural definition of $D$ is then the
``central charge'' $c$ of the matter theory, which will be
introduced below. It is 1 for each free
scalar, $1/2$ for each free fermion and can be negative
for nonunitary theories.

In (2.1) we could include other renormalizable terms in the action, like
$$g_{kl}(x)\partial_\alpha x^k\partial^\alpha x^l,\ \ \ \ T(x)$$
with arbitrary analytic functions $g_{kl}(x),\ T(x)$. The
first term corresponds to random surfaces embedded in curved space.
But let us start with the terms written out in (2.1)
and add interactions of $x$ later.
\foot{We will not discuss the extrinsic curvature${}^{[\polov]}$
here.}

The Hilbert--Einstein action in (2.1) is
a topological invariant,
the Euler characteristic $\chi$:
$$\chi={1\over{4\pi}}\int d^2\sigma\ {\sqrt g}R=2-2g,\eqno(2.4)$$
where $g$ is the genus, or number of handles, of the surface.
Thus, the expansion of (2.1) in terms of the ``string coupling constant''
$\lambda=\exp\{4\pi\gamma\}$ is a topological expansion: surfaces
with $g$ handles are weighted with a relative factor $\lambda^{2g}$.
\vskip.5cm

 {$\underline{\hbox{2.2. Conformal Gauge}}$}

The form (2.1) of the integral is much more convenient than the form (1.1),
because the area expressed in terms of $x$, as in (2.2),
is difficult to handle. Also dividing out the
diffeomorphism group is difficult in (1.1). So
we will simplify
(2.1) in the next two sections. First, consider the sum over
metrics
modulo diffeomorphisms. It is most convenient to
parametrize them in conformal gauge. To this end, let us recall
the following well--known facts:${}^{[\gsw]}$

1. The topology of a closed, oriented surface is completely specified
by the genus $g$ in (2.4).

2. Two metrics are said to be in the same conformal equivalence class,
if they differ only by a rescaling and a diffeomorphism.
For given genus, there is a finite dimensional moduli space of conformal
equivalence classes. Its dimension is zero for the sphere ($g=0$),
two for the torus ($g=1$) (ratio
of the radii of the torus and its twist) and $6g-6$ for $g>1$. Denoting
the moduli as $m_i$ and fixing a reference metric $\hat g(m_i)$ in each
class, any metric can be written as
$$g_{\alpha\beta}=\hat g_{\alpha\beta}(m_i)\ e^{\phi}
\circ\hbox{Diffeomorphism $\xi$}.\eqno(2.5)$$

3. The decomposition (2.5) is not unique.
At genus 0 and 1 there are globally well--defined
diffeomorphisms that are equivalent to Weyl rescalings. They form a
two--dimensional group for genus 1 (translations) and the six--dimensional
group $SL(2,C)$ for genus 0 (translations, rotation, global scale
transformation
and two special conformal transformations).

The sum over geometries can now be rewritten as
$$\int {\cal D}g\rightarrow \sum_{g=0}^\infty \int\prod_{i=1}^{6g-6}
dm_i\ \int {\cal D}\phi\ \int {\cal D}\xi
 \times\hbox{Jacobian}. \eqno(2.6)$$
A Jacobian arises because of the change of variables from
$g_{\alpha\beta}$ to $\phi,\xi$.${}^{[\poy]}$ It will be discussed next.
In (2.6), it is implied that we do not sum over the $SL(2,C)$--modes
of $\phi$ mentioned above.
Otherwise, (2.6) would be infinite. The integral over diffeomorphisms
$\xi$ now cancels
the volume of the diffeomorphism group in (2.1),
provided that there is no gravitational anomaly, which means that
the matter sector must
not include self-dual spin--$2$ fields.${}^{[\awi]}$
\eject

 {$\underline{\hbox{2.3. The Jacobian}}$}\vskip2mm

When we make a
linear change of variables $y^i\rightarrow
y'^i = A^i_j y^j$
in finite--dimensional
integrals, we pick up a Jacobian $\det A$:
$$\int\prod {dy^i}'\ \sim\ \int\prod dy^i\ \det A.$$
$A$ also appears in the norm in $y$ space:
$$\Vert\delta\vec y\Vert^2=\sum_i(A_i^j\delta  y_j)^2.$$
Let us apply this to infinite--dimensional integrals.
To find the Jacobian in (2.6), consider the natural (covariant) definition of
the
measure, i.e., of the norm in the space of metrics:${}^{[\poy]}$
$$\eqalign{\Vert\delta g\Vert^2 &\equiv \int d^2 \sigma{\sqrt g}
g^{\alpha\beta}g^{\gamma\delta}\delta g_{\alpha\gamma}\delta g_{\beta\delta}
\cr &=\int d^2\sigma{\sqrt{\hat g}}
e^\phi[(\delta\phi
+\hat\nabla^\gamma\xi_\gamma)^2
+(L\xi)_{\alpha\gamma}(L\xi)^{\alpha\gamma}],
}\eqno(2.7)$$
where infinitesimal deformations of the metric have been decomposed as
 $$\delta g_{\sigma\rho}=g_{\sigma\rho} \delta\phi+
\nabla_\sigma\xi_\rho+\nabla_\rho\xi_\sigma.$$
The operator $L$ in (2.7) and its adjoint
$L^\dagger$
are given by
$$\eqalign{
&(L\xi)_{\alpha\beta}\equiv\hat\nabla_\alpha\xi_\beta
+\hat\nabla_\beta\xi_\alpha
-\hat g_{\alpha\beta}\hat\nabla^\gamma\xi_\gamma,\cr
&(L^\dagger h)_\gamma=\hat\nabla^\alpha h_{\alpha\gamma}.}$$
$L^\dagger$ acts on
a traceless, symmetric tensor.
$\hat\nabla_\alpha$ is the covariant derivative with respect to
$\hat g$. From the above, the Jacobian in (2.6) is seen to be
$$\det L\ \equiv\ (\det L^\dagger L)^{1\over 2}.$$
\eject

This determinant is often
represented with the help of anticommuting
``Faddeev-Popov ghost fields''${}^{[\gsw]}$
$c^\alpha,b_{\alpha\beta}$ by the
fermionic functional integral
$$(\det L^\dagger L)^{1\over2}
=\int {\cal D}b\ {\cal D}c\ \exp{\{-\int d^2\sigma{\sqrt g}\ g^{\alpha\gamma}
c^\beta\nabla_\alpha b_{\beta\gamma}}\}.\eqno(2.8)$$
Here, $b_{\alpha\beta}$ is a traceless, symmetric tensor of
conformal dimension 2, and $c_\beta$ is a vector of dimension --1.
We can now write (2.1) as
$$\sum_{g=0}^\infty\ \lambda^{(2g-2)}\int\prod_{i=1}^{6g-6} dm_i
\int {\cal D}\phi\ (\det\ L^\dagger L)^{1\over2}_{\hat g e^\phi}
\ (\det\ \Delta)^{-{D\over2}}_{\hat g e^\phi}\
\exp\{-\mu\int\sh e^\phi\}.\eqno(2.9)$$
The partition function for $x$ has been written as the determinant
of the laplacian.
The subscripts $\hat g e^\phi$ indicate that the determinants
are to be evaluated in the curved background $\g(m_i)$.
In the above, we have not discussed variations of the moduli $m_i$.
Since in general
it is not known how to integrate over the moduli spaces
and sum over topologies in the continuum
theory, let us focus on the $\phi$--integral in the
following.

\vskip1cm

{\leftline{\bf 3. The Trace Anomaly and DDK}}
\vskip.5cm

 {$\underline{\hbox{3.1. The Conformal Anomaly}}$}

The next step is to decouple $\phi$ from the determinants in (2.9).
Classically,
the free scalars $x^i$ in (2.1) and the ghosts in (2.8) are not coupled
to the metric $\g$ at all: their actions are diffeomorphism
invariant (of course) and conformally invariant, because
the corresponding stress tensors are traceless:${}^{[\gsw]}$
$$\eqalign{
T^{(x)}_{\alpha\beta}&=\partial_\alpha x\partial_\beta x
-{1\over2}g_{\alpha\beta}
\partial^\gamma x\partial_\gamma x  \cr
T^{(b,c)}_{\alpha\beta}&=
 {1\over2}c^\gamma\nabla_{\alpha}b_{\beta \gamma}
+ {1\over2}c^\gamma\nabla_{\beta}b_{\alpha\gamma}
+ (\nabla_{\alpha}c^\gamma) b_{\beta\gamma}
+ (\nabla_{\beta}c^\gamma) b_{\alpha\gamma}\cr
&-{1\over2}g_{\alpha\beta}(
c^\gamma\nabla^{\beta}b_{\beta \gamma}
+ 2(\nabla^{\beta}c^\gamma) b_{\beta\gamma}
).}$$
\eject
Quantum mechanically however,
the background metric enters the determinants through one--loop
graphs like
\vskip 1cm
$${- - - - - - -\bullet\hskip15mm\bullet- - - - - - -}\eqno(3.1)$$
\vskip 1cm

As mentioned, there is no diffeomorphism anomaly, so the
determinants are diffeomorphism invariant.
But Weyl invariance is spoiled by the conformal anomaly: generally,
$$\det X_\g=\det X_{\hat g}\ \times \exp\{-S_{\hbox{eff}
}(\hat g,\phi)\}\eqno (3.2)$$
where $X$ represents some conformally invariant operator.
$S_{\hbox{eff}}$ can be obtained by first computing (3.1) in weak
gravitational backgrounds, then using general covariance
to determine from this the effective action and then writing it in conformal
gauge.${}^{[\polov]}$ But it is more straightforward to integrate the
trace anomaly $<T^\alpha_\alpha>$ of the stress tensors of
the fields $x,b$ and $c$, since
$$\delta S_{\hbox{eff}}=-\int d^2\sigma{\sqrt g}
<T_{\alpha\beta}(\sigma)>\delta g^{\alpha\beta}(\sigma),$$
hence
$${{\delta  S_{\hbox{eff}}[\hat g,\phi]}\over{\delta\phi}}=
-{\sqrt g}<T^\alpha_{\alpha}>.\eqno(3.3)$$
In any dimension, $<T_\alpha^\alpha>$
can be found using the Schwinger--de Witt method
by expanding the Green's
function of $X$ in a curved background.${}^{[\bd]}$ In
two dimensions, things are much easier: first we regularize the
determinants by introducing a short distance cutoff $a$, for example by
putting the theories on a lattice.
Since
$X$ is conformally invariant, $T_\alpha^\alpha$ is zero
and $<T^\alpha_\alpha>$
comes only from short distance
quantum effects and is therefore local. It must also be generally covariant,
and is thus a polynomial in the curvature. Dimension counting
then determines $<T^\alpha_\alpha>$
up to a parameter $c$, the ``central charge":
$$<T^\alpha_\alpha>=
1\times O({1\over{a^2}})
+{c\over{48\pi}} R + O(a^2).\eqno(3.4)$$
$c$ can be read off from the most singular term in
the operator product expansion
of the stress tensor with itself:${}^{[\bpz,\fri]}$
$$T(r)T(0)\sim{{c/2}\over{\vert r\vert^4}}+
\hbox{...}\eqno(3.5)$$
Well--known results are $c=1$ for each free scalar field, $c={1\over2}$ for
each free fermion and $c=-26$ for the ghosts $b,c$. The
leading term in (3.4)
is infinite, but this will only renormalize the cosmological constant,
as will be seen.
\vskip.5cm

 {$\underline{\hbox{3.2. The Liouville Action}}$}

We can now integrate (3.3), using (3.4--5). For $d=2$, the curvature is
${\sqrt g}R={\sqrt{\hat g}}(\hat R -\hatbox\phi).$
This yields
the effective action${}^{[\poy]}$
$$\eqalign{S_{\hbox{eff}}(\hat g,\phi)
\equiv S_L&={c\over{48\pi}}S_0[\hat g,\phi]
+\int d^2\sigma {\sqrt{\hat g}}\mu' e^\phi\cr
S_0&\equiv\int d^2\sigma {\sqrt{\hat g}}({1\over2}
\hat g^{\alpha\beta}\hat\partial_\alpha
\phi \hat\partial_\beta \phi+\hat R\phi )
}\eqno(3.6)$$
with induced cosmological constant $\mu'$. $S_L$ is called the
Liouville action.
We see that the conformal anomaly induces a kinetic term
for the conformal factor, even
though the metric did not seem to be a dynamical variable in (2.1).
This is in contrast with the random walk, where
$<T^\alpha_\alpha>\sim 1\times O({1\over a})+O(a)$
by dimension counting, thus resulting only in
a renormalization of the cosmological constant.
In four--dimensional gravity, $<T^\alpha_\alpha>$ and $S_{\hbox{eff}}$
will also include generally
covariant fourth--order derivative terms,${}^{[\bd]}$ leading to a unitarity
problem in Minkowski space (but not in Euclidean space). See part III
for further discussion.

Defining $c_m$ as the combined conformal anomaly of the matter
(i.e., if we include matter other than the scalar fields $x$),
the $\phi$--integral in (2.9) becomes
$$(\det\ L^\dagger L)^{1\over2}_{\hat g}\
(\det\ \Delta)^{-{D\over2}}_{\hat g}\ \int {\cal D}\phi\
\exp\{{{26-c_m}\over{48\pi}}S_0[\hat g,\phi]
-(\mu+\mu')\int{\sqrt{\hat g}}e^\phi\}.
\eqno(3.7)$$
\vskip.5cm

 {$\underline{\hbox{3.3. The Measure for $\phi$}}$}

(3.7) is not yet a field theory as usual, because of the geometric
meaning of $\phi$ as conformal factor.
This shows up in the definition of the measure:
the generally covariant definitions of the norm in the space of metrics
and of the cutoff are (see (2.7); the term $\vec\nabla\cdot\vec\xi$ has been
absorbed in a shift of $\phi$):
$$\Vert\delta \phi\Vert^2
=\int d^2\sigma{\sqrt{\hat g}}\ e^\phi\ (\delta\phi)^2,\ \ \ \ \
e^\phi\ (\delta\sigma)^2\ \ge\ a^2.\eqno(3.8)$$
If $\phi$ were just another field, we would have
$$\Vert\delta \phi\Vert^2
=\int d^2\sigma{\sqrt{\hat g}}\ (\delta\phi)^2,\ \ \ \ \
(\delta\sigma)^2\ \ge\ a^2.\eqno(3.9)$$
The cutoff $a$ can be introduced, e.g., by regularizing the sum over surfaces
as a sum over random triangulations with
triangle side length $a$ (see appendix).
We see that $\phi$
lives on a half--line:${}^{[\polch]}$ for fixed $\delta\sigma$,
$\phi$ must be bounded from below
to have $e^\phi(\delta\sigma)^2\ge a^2$. The bound is set by
the smallest possible $\delta\sigma$ (that is, $\delta\sigma_{min}$ between
two neighboring lattice sites):
$$\phi\ge\phi_0\ , \ \ \ \ e^{\phi_0}\ (\delta\sigma_{min})^2 =
a^2.\eqno(3.10)$$

Unlike the measures discussed in subsection 2.3, the
measures defined by (3.8) and (3.9)
do not correspond to a linear change of variables that can be absorbed
in a simple Jacobian.
One way to circumvent the problem of the unusual
measure for $\phi$ is to
write (3.7) in light--cone gauge rather than
conformal gauge, following Knizhnik,
Polyakov and Zamolodchikov.${}^{[\kpz]}$
An $SL(2,R)$ symmetry of the model can then be
used to solve it with methods of conformal field theory.
\vskip.5cm

 {$\underline{\hbox{3.4. DDK}}$}

It is more convenient, though, to proceed in conformal gauge
following
David,${}^{[\d]}$ Distler and Kawai${}^{[\dk]}$ (DDK):
Their idea was to replace $\phi$ with an ordinary field,
also called $\phi$, whose measure $D_{\hat g}\phi$
is defined by (3.9). Their conjecture, later confirmed in [\dho], was
that this change in the measure can be absorbed by replacing
the action (3.6) for $\phi$ with the
most general local renormalizable action. This is the Liouville
action itself, but with modified coefficients $Q,\alpha,\mu$:
$$S[\hat g,\phi]={1\over{8\pi}}
\int d^2\sigma {\sqrt {\hat g}}\{
\hat g^{\alpha\beta}\partial_\alpha\phi\partial_\beta\phi+Q\ \hat R\phi
+\mu\ e^{\alpha\phi}\}.\eqno(3.11)$$
Here $\phi$ has been normalized so that the kinetic term is standard.
Quantum gravity is now described by three ordinary field theories,
for $b+c$, $x$, and $\phi$. For this to be consistent, the combined
theory must be scale invariant: scale invariance was part of the general
covariance of the original theory.
An arbitrary gauge choice corresponding to the background metric
$\hat g$ has been made in (2.5) to parametrize
metrics in conformal gauge. Now that $\phi$ is just a dummy integration
variable, everything should be invariant under rescaling of $\hat g$.

Scale invariance turns out to specify the action
for $\phi$ completely.
For $\mu=0$, it means that the total conformal anomaly must vanish:
\foot{This can also be seen directly by studying the original
theory in (3.7), by simultaneously shifting $\hat g\rightarrow
\hat g e^\sigma, \phi\rightarrow\phi-\sigma$: The $\phi$ theory
behaves exactly like a conformal field theory with central charge
$c_\phi=26-c_m$.}
$$c_\phi+c_m-26=0\ \ \ \rightarrow\ \ \ 3Q^2=25-c_m,$$
because (3.11) has $c_\phi=1+3Q^2$. This is derived as follows.
Even for $\mu=0$, (3.11) is not quite conformally invariant because
of the term $Q\hat R\phi$. Locally, we can write
$$\hat g_{\alpha\beta}
=\delta_{\alpha\beta}e^{\hat\phi},\ \ \hbox{so}\ \
 {{\sqrt{\hat g}}}\hat R=-\bx\hat\phi.$$

Then,
$$\eqalign{S_0(\hat g,\phi)&\sim\int d^2\sigma {\sqrt {\hat g}}\{
\hat g^{\alpha\beta}\partial_\alpha\phi\partial_\beta\phi+Q\ \hat R\phi\}\cr
&=-\int d^2\sigma\{\phi\box\phi+Q\hat\phi\box\phi\}
=-\int d^2\sigma (\phi+{Q\over2}\hat\phi)\box(\phi+{Q\over2}\hat\phi)-
 {Q^2\over4}S_0(\delta,\hat\phi).}$$
Up to a shift of $\phi$, the first term on the RHS describes an ordinary
scalar field with $c=1$. The second term is just the Liouville action
for $\hat g=\delta$. Comparing with (3.6), we see that this term gives
a ``classical contribution'' $3Q^2$ to the central charge.

It follows from the above that
$$Q={\sqrt{{25-c_m}\over3}}.\eqno(3.12)$$
When $\mu$ is turned on in (3.11), scale invariance
implies that the cosmological constant $e^{\alpha\phi}$
must be a marginal operator, i.e.,
of conformal dimension two (to cancel the two from $\sqrt{ \hat g}$).
To exploit this piece of information, we first shift
$\phi$ by $(Q/2)\hat\phi$, as above. Then the condition on the dimension
becomes
$$\dim(e^{\alpha\phi})=2+Q\alpha\ \ \ \hbox{with action}\ \ \
 {1\over{8\pi}}\int d^2\sigma(\partial \phi)^2.$$
With this action, the propagator $<\phi(r)\phi(0)>$ is $-\log (r^2)$.
\foot{It is sufficient to use the free field action to compute
the dimension.${}^{[\gn]}$ An infrared cutoff, which is required,
is not shown.}
The easiest way to compute the (classical plus anomalous) dimension
of the operator $e^{\alpha\phi}$, which is assumed to be normal ordered,
is to consider the two--point function
$$<e^{\alpha\phi(r)}e^{-\alpha\phi(0)}> = e^{-\alpha^2 <\phi(r)\phi(0)>}
= (r^2)^{\alpha^2},$$
yielding the dimension $-\alpha^2$.

So without
shifting $\phi$,
$$\hbox{dim}(e^{\alpha\phi})=-\alpha(\alpha+Q),\eqno(3.13)$$
$$\rightarrow\ \ \ \ \alpha={1\over{2{\sqrt3}}}
({\sqrt{25-c_m}}+{\sqrt{1-c_m}}).\eqno(3.14)$$
For $\alpha$ (and therefore the action) to be real, we need
$c_m\le1$. That is, the ``matter'' must be a ``minimal model''
\foot{E.g., the Ising model with $c=1/2$} or
a single coordinate $x$.
It is believed that the surfaces are in a
branched polymer phase for $c_m>1$. More about this $c=1$
barrier will be said in the next section.

When $c_m=25$, the $\hat R\phi$--term in (3.11) vanishes and
$\phi$ is an ordinary scalar field. From (3.14), $\alpha$ is
imaginary in this case;
it can be made real by redefining $\phi\rightarrow i\phi$.
Then $\phi$ becomes timelike and
(3.11) is the usual
world sheet action of the critical bosonic string,
obtained by coupling 25 space coordinates $x^i$ and
one time coordinate $\phi$ to 2D gravity and ignoring
the conformal anomaly.
\vskip.5cm

 {$\underline{\hbox{3.5. Gravitational Dressing}}$}

So far, the $x^\mu$ have been free fields in (2.1), but
we can also add interactions
in the form of scaling operators $\Phi_i(x)$
with positive scaling
dimensions, $h_i\ge0$,
and small coupling constants $t^i$. Scaling operators are operators of
definite scaling dimensions. Their two--point functions are
just powers of their distance. Examples in two dimensions are the normal
ordered
operator $\cos px$, or the operator $e^{\alpha\phi}$ of the
last subsection.

Before coupling to gravity,
the perturbed matter action is
$$S=\int d^2\sigma{\sqrt g}\{
(\partial x)^2+t^i\ \Phi_i(x)
\}.\eqno(3.15)$$
Here, summation on $i$ is understood.
After coupling to gravity and replacing the measure for $\phi$
by (3.9), the interaction terms $\Phi_i(x)$ will get ``gravitationally
dressed,'' that is, they will become mixed operators
$\hat V_i(x,\phi)$. So we make the ansatz
$$S=\int d^2\sigma{\sqrt g}\{
(\partial x)^2+(\partial\phi)^2+Q\hat R\phi+t^i\ \hat V_i(x,\phi)
+\hbox{cosmol. const.}\}.\eqno(3.16)$$
Then scale invariance again determines the $\hat V_i$.
It implies that
the $t^i$ do not ``run,''
that is, their beta functions must be zero. The beta functions
are
${}^{[\car]}$
$$0=\beta^i=(\Delta^i_j-2\delta^i_j)t^j+\pi c^i_{jk}t^jt^k+...\eqno(3.17)$$
Here, $\Delta^i_j$ is the dimension matrix
of the operators $\hat V_i$, defined by
$$(L_0+\bar L_0)\hat V_j=\Delta^i_j\hat V_i,$$
where $(L_0+\bar L_0)$ is the generator of scale transformations.
The $c^i_{jk}$
are the operator product coefficients in the short--distance expansion
$$\hat V_j(\vec r)\hat V_k(\vec 0)\sim ({\vec r}^2)^{-1}
\sum_i c^i_{jk}\hat V_i(\vec 0).$$
To lowest order in $t$,
$\Delta^i_j$ must be $2\delta^i_j$. From (3.14), this is obeyed by
$$\hat V_i=\Phi_i(x) \ e^{\gamma_i\phi}\ \ \ \hbox{with}\ \ \
\gamma_i={1\over{2{\sqrt3}}}({\sqrt{25-c_m}}-{\sqrt{1-c_m+24h_i}}).\eqno(3.18)$$
In particular, the cosmological constant is the `dressed' unit operator $1$.
If $c_m$ is such that
$\alpha$ in (3.14) is real, all the $\gamma_i$ will be real, since
the operators $\Phi_i$ have dimensions $h_i\ge0$. Therefore, the
$\hat V_i$ will not lower the $c=1$ barrier.

While background invariance was imposed to first order in $t$ by DDK,
leading to (3.18), the implications of (3.17) at $O(t^2)$ have not
previously been studied. This will be done in part II. We will see
that this requires new terms of $O(t^2)$ in (3.16).

\eject

{\leftline{\bf 4. Applied Liouville Theory}}

Here we use the formalism developed above to discuss briefly
some critical exponents (in subsection 4.1) and the
spectrum of the theory (in subsection 4.2) with its geometric interpretation
(in subsection 4.3).
We are restricted to the case
$c_m\le1$, where (3.11) is well--defined. The case $c=1$ will be
used as an example in subsection 4.4.
More aspects of Liouville theory and its
gravitational interpretation
are discussed in the appendices of Part II.
\vskip.5cm

 {$\underline{\hbox{4.1. Correlation Functions and Critical Exponents}}$}

Unfortunately, there is not yet a satisfactory way to compute
correlation functions
$$<\prod_i\ \Phi_i(x)\ e^{\gamma_i\phi}>$$
in Liouville theory.
The main obstacle is the exponential potential $e^{\alpha\phi}$. In particular,
$\phi$--momentum is not conserved, as it would be in free field theory.
The potential cannot be
treated perturbatively in $\mu$, starting from the free theory
with $\mu=0$, because the cosmological constant diverges in the
infrared ($\alpha\phi\rightarrow\infty$).
Thus, it cannot be made small -- rescaling $\mu$ just
shifts it in $\phi$--space.
So it must be included in the path integral
from the start and dealt with nonperturbatively.
It is not yet clear how to do this integral in general (see however
[\triv,\gou]).

The area--dependence of the correlators can be
extracted quite easily, though, and from this some critical coefficients
can be deduced.
To this end, consider the sum over surfaces of given area $A$. The
fixed--area partition function is defined as
$$Z(A)
\equiv<\delta(\int d^2\sigma\ {\sqrt {\hat g}}e^{\alpha\phi}-A)>\
\ =e^{-\mu A}\ Z_0(A),
\eqno(4.1)$$
where $Z_0$ is the partition function with action $S_0$,
the free part of (3.11).
$Z_0(A)$ can be found up to a proportionality factor
by shifting $\phi$ by a constant:${}^{[\d,\dk]}$

$$\eqalign{\phi\rightarrow\phi+c\ \ &\Rightarrow\ \
S_0\rightarrow S_0+4\pi\chi c\cr
&\Rightarrow\ \ Z_0(A\ e^{\alpha c})
=Z_0(A)\ e^{c(Q(1-g)-\alpha)}\cr &\Rightarrow\ \
Z_0(A)\propto A^{Q(1-g)/\alpha-1}\cr
&\Rightarrow\ \ Z(A)\propto e^{-\mu A}\ A^{\gamma-3},}\eqno(4.2)$$
$$\hbox{with}\ \ \ \gamma=2+(1-g){Q\over\alpha}=2+{{1-g}\over12}
(d-25-{\sqrt{(25-d)(1-d)}}).\eqno(4.3)$$
The coefficient $\gamma$ is called the string susceptibility, and
this formula for $\gamma$ agrees
with the matrix model results.${}^{[\mxm]}$

Similarly, one finds for the fixed--area
correlation functions of the operators (3.18), up to a proportionality factor:
$$\eqalign{<\hat V_1...\hat V_n>_A
&\equiv<\hat V_1...\hat V_n\
\delta(\int d^2\sigma\ {\sqrt {\hat g}}e^{\alpha\phi}-A)>\cr
&\propto e^{-\mu A}\ A^{\gamma-3+
\sum\gamma_i/\alpha}.}$$
This can be integrated over $A$ from some cutoff
$\epsilon$ to $\infty$, assuming
$(2g-2)+\sum\gamma_i>0$ so that the integral converges.
Due to the cutoff on $\phi$,
i.e., on $A$, cutoff-dependent terms will be added otherwise.
${}^{[\sei]}$
The result for the $\mu$--dependence of the correlators is
$$\eqalign{<\hat V_1...\hat V_n>_\mu&\equiv\int_\epsilon^\infty dA\ e^{-\mu A}
<\hat V_1...\hat V_n>_A\cr
&\propto\mu^{-(\gamma-2
+\sum\gamma_i/\alpha)}
.}\eqno(4.4)$$
This also agrees with the matrix model results.${}^{[\mxm]}$
The power of $\mu$ is in general fractional. This
confirms that we must treat the cosmological
constant nonperturbatively, because perturbation theory in $\mu$
could have produced
integer powers of $\mu$ only.
\eject

Another interesting coefficient is the
Haussdorff dimension $d_H$ of random surfaces:
$$<x^2>_A\ \propto e^{-\mu A}\ A^{2/{d_H}}.$$
This measures the mean extension $<x^2>$ of the surface in target space
versus the intrinsic area $A$. The random walk is well-known to have
$d_H=2$, independently of the dimension of the embedding
space. For random surfaces, one can show with the above
methods that:${}^{[\dkaw]}$
$$d_H={24\over{1-D+{{\sqrt{(25-D)(1-D)}}}}}.\eqno(4.5)$$
This, too, agrees with the matrix model results.
Note that $d_H$ depends on the embedding dimension.
For $D=1$, $d_H=\infty$, i.e., $<x^2>\propto \log A$.

The Haussdorff dimension is useful to investigate the relevance
of interactions in the first--quantized formulation. E.g.,
$\phi^4$ theory can be viewed
as a second--quantized
self--interacting random walk. It
becomes free in the renormalization group sense in $D>2d_H=4$ embedding
dimensions,
because then two paths typically do not intersect and the interaction
term is thus irrelevant.
Unfortunately, the analogous statement
for string theory can
presently only be made for $D\le1$, where $2d_H>D$. The statement
is that interactions of the $D\le1$ string are ``relevant.''
\vskip.5cm

 {$\underline{\hbox{4.2. States and Operators}}$}

For further applications, it is important to know
the eigenstates of the Hamiltonian (if we think of one of the
coordinates as time), or equivalently, to know the
operators that create these states when acting on the vacuum.
Those are the scaling operators
(i.e., the operators of definite scaling dimension, like $e^{\alpha\phi}$)
that can be constructed in the theory. The reason is that, up to a constant,
the Hamiltonian can be identified with the generator of scale transformations:
\eject

Consider inserting an operator at a point $P$ of the surface $\Sigma$ on which
our theory lives (fig.3a).
Deform the surface into a cylinder by a conformal transformation
that maps $P$ to infinity, as shown in
fig. 3b.
Translations along
the axis of the cylinder,
generated by the Hamiltonian ${\cal H}$, correspond to scale transformations
on $\Sigma$, whose generator we call
$L_0+\bar L_0$ as before. When the action of the fields that live on $\Sigma$
is conformally invariant, ${\cal H}$ and $L_0+\bar L_0$ would be the same
if it were not for the conformal anomaly. Due to the latter,
${\cal H}$ and $L_0+\bar L_0$ actually differ, but only by a constant,
which is proportional to the central charge $c$ (see e.g., ref.
[\car]).

\vskip 8cm
\line{\hskip1.5cm Fig. 3a: A surface with an\hskip2cm Fig. 3b: The same
surface,
\hfill}
\line{\hskip1.9cm operator inserted at $P$\hskip2.8cm deformed to a
cylinder\hfill}
\vskip.5cm

As explained above, if we think of Liouville theory
as quantum gravity, we should combine operators of the $x$
and the $\phi$ sectors to obtain scaling operators of dimension two,
\foot{More precisely, we should impose the Virasoro constraints.}
as in (3.18). This will be done for the example of the $c=1$ model
in subsection (4.4). We should also interpret
$\phi$ as the conformal factor. This will be
done in subsection (4.3). Here, let us forget about
gravity and just study the
spectrum of Liouville theory as a theory of its own.

What are the scaling operators in the theory? Ignoring for now
the cosmological constant,
\foot{With cosmological constant, these operators will get
modified at large negative $\phi$.${}^{[\polch]}$}
and considering only operators without derivatives,
they are the exponentials
$$\exp\{\epsilon\phi\}\ \ \ \hbox{with dimension}\
\ d_\epsilon= -\epsilon(\epsilon+Q)
=-(\epsilon+{Q\over2})^2+{Q^2\over8},\eqno(4.6)$$
as discussed above.
For the dimension to be real, $\beta\equiv\epsilon+Q/2$ must be real
or imaginary.
For real $\beta$, the dimensions are not bounded from below. Thus
the Hamiltonian of the $\phi$ sector is not bounded from below, but this
will be taken care of by the $x$ part of the operators (3.18).
For imaginary $\beta$, the hermitean combinations of
the operators are actually
$$\exp\{-{Q\over2}\phi\}\sin (\beta\phi+\Theta).\eqno(4.7)$$

Next, what are the eigenstates of the Hamiltonian?
To answer this,
deform a Riemann surface with boundary
to a half--open cylinder, insert all the background curvature (and
possible handles) and an operator $O_i$ in the far past as shown below (fig.
4a),
and consider the wave function $\psi_i (\phi)$ on the boundary.
Let us only consider the quantum mechanics problem of
the constant mode $\phi_c(\sigma,\tau)=\phi_c(\tau)$ ($\sigma$ is here the
coordinate along the circle, and $\tau$ is ``time'').
{}From the corresponding minisuperspace action
$$S(\phi_c,\dot\phi_c)\sim{1\over{8\pi}}\int d\tau(\dot\phi_c^2
+\mu e^{\gamma\phi_c})$$
one derives the Schr\"odinger equation
$$(-{1\over 2}{{\partial^2}\over{\partial
\phi_c^2}}+\mu e^{\gamma\phi_c})\psi(\phi_c)=E\psi(\phi_c).\eqno(4.8)$$
The potential and the solutions${}^{[\sei]}$ are shown in fig. 4b.
\eject

\line{}\vskip8cm
\line{\hskip2mm Fig. 4a: Riemann surface with boundary,\hskip1cm Fig. 4b:
Two types of
wave functions\hfill}
\line{\hskip.5cm two handles and an operator insertion\hskip1.7cm of states
with
$E>0$ and $E<0$\hfill}
\vskip1cm

For $E>0$, there are
oscillating states, behaving like $\sin(\beta\phi_c+\Theta)$
for $\phi_c\rightarrow\infty$.
If the range of $\phi_c$ was $[-\infty,\infty]$, there would be no ground
state: for $E=0$, the
wave function diverges linearly for $\phi_c\rightarrow\infty$
and would thus not be normalizable. But
since there is an upper bound $\phi_0$ on $\phi_c$, the state
exists and its wave function is peaked
at $\phi_c=\phi_0$. Likewise, all other states whose wave functions
diverge as $\phi_c\rightarrow\infty$ should be included in the
spectrum.${}^{[\sei]}$ On the other
hand, states that diverge at $\phi_c\rightarrow-\infty$ do not exist.

Third, how do the states correspond to the operators?
Because of the background charge,
the vacuum wave function
behaves like $e^{-(Q/2) \phi_c}$ and
the operator
$e^{\epsilon\phi}$ in (4.6), when acting on the vacuum,
creates the state with
$\psi(\phi_c)\propto e^{(\epsilon-{Q/2})\phi_c}$ as
$\phi_c\rightarrow -\infty$.
Since no states exist for
$\epsilon<{Q\over 2}$, one concludes
that the operators (4.6) also exist only
for $\epsilon\ge{Q\over2}$.${}^{[\sei,\pch]}$
More precisely, one can argue that the operators
with $\epsilon<{Q\over 2}$
cannot be renormalized (see ref. [\pch]).
Finally, the oscillating states obviously correspond to the operators (4.7).
\vskip5mm

 {$\underline{\hbox{4.3. Geometric Interpretation}}$}

Interpreting $\phi$ as conformal factor means interpreting
$$A=\int_\Sigma d^2\sigma\ e^{\alpha\phi} \ \ \ \ \ \hbox{and}\ \ \ \ \
l=\oint_{\partial\Sigma} d\sigma\ e^{{\alpha\over2}\phi}$$
as the area of the surface $\Sigma$ and the length of its boundary
$\partial\Sigma$. From the previous discussion,
the expectation value $<l>$
is of the order of the cutoff ($e^{(\alpha/2) \phi_0}$)
for the exponentially growing
states.\foot{We see this from the minisuperspace approximation of the last
subsection. From the matrix model results it is known that this approximation
is exact.}
They are thus called
``microscopic.''${}^{[\mxm]}$
$<l>$ is finite for the operators (4.10) and their
states, which are thus called ``macroscopic.''
Although these
operators are local in the background metric, they are not local
in the physical metric. Inserting them into the surface cuts a
macroscopic hole into it: a closed line drawn around the
insertion, no matter how closely, will always have some finite
circumference $<l>$.

This gives a geometric interpretation to
the $c=1$ -- barrier of 2D gravity.
Consider the cosmological
constant operator $e^{\alpha\phi}$ with $\alpha$ given by (3.14).
For $c>1$, $\alpha$ acquires an imaginary part and the cosmological constant
becomes a macroscopic operator like (4.7). While inserting it into
the surface cuts a hole, adding it to the action, i.e.,
inserting its exponential, destroys the surface.${}^{[\sei]}$
We thus expect a phase transition
at $c=1$.

In the language of string theory, the barrier is related to the
presence of target--space
tachyons for embedding dimension $c>1$.
The condition
that the vertex operator
$:e^{i\vec p\vec x+\epsilon\phi}:$ be of dimension two yields
the mass $m$ of the lowest string state:
$${\vec p} {}^2-(\epsilon+{Q\over2})^2={Q^2\over4}-2\equiv m^2.\eqno(4.9)$$
$m^2$ is negative for $Q^2<8$, i.e., $c>1$. In this case there are
plane waves with imaginary $\epsilon$ (macroscopic operators)
and negative $m^2$ (tachyons),
one of them ($\vec p=0$) being the cosmological constant. On the other
hand, for unitary theories
with $c\le1$ ($Q^2\ge8$), all physical states are microscopic
${}^{[\sei]}$ (real $\epsilon$)
and there are no tachyons.
\eject

 {$\underline{\hbox{4.4. 2D String Theory at $c=1$}}$}

The most physical case that can
be disussed in the present framework is that of
gravity coupled to an (uncompactified)
scalar field $x$ with $c=1$. This is the theory of random surfaces
embedded in one dimension.
{}From (3.12), $Q=2{\sqrt2}$. (4.3) and (4.5)
yield the critical coefficients $\gamma=0$ on genus zero and $d_H=\infty$.
The action is
$${1\over{8\pi}}\int d^2\sigma{\sqrt{ \hat g}}
\{\partial_{\alpha}x\partial^{\alpha}x+\partial_\alpha \phi\partial^\alpha\phi
+2{\sqrt2}\hat R^{(2)}\phi+\mu e^{\gamma\phi}
+\hbox{ghosts}\ b,c\}.\eqno(4.10)$$
Examples of matter scaling operators are
$e^{ikx}$,
$\sin kx, \cos kx$
with dimensions $h_k={k^2}.$
{}From (3.18), the dressed operators are
$$\hat V(k)\equiv \ e^{ikx+\epsilon\phi}\
\ \hbox{with}\ \ \epsilon =-{\sqrt 2}\pm k.\eqno(4.11)$$

What makes the $c=1$ model particularly interesting is the
fact that (4.10) can be viewed as the world--sheet action
of a critical string theory in {\it two}
target space dimensions $x$ and $\phi$.${}^{[\polch]}$
The dilaton background $\Phi(x,\phi)=Q\phi$ in (4.10)
is responsible for lowering the critical dimension from 26 to 2.
This is further discussed in part II, appendix A.

{}From the analysis of the $c=1$ matrix model,
the correlation functions of (4.10) are known
to all orders in the string loop expansion, i.e., summed over
all genera, and beyond. This
makes 2D string theory an interesting
toy model for more realistic (26D) string theories.
One might think that it
is a rather boring toy model, because there are no transverse directions
in which the string can oscillate. Thus the
spectrum can contain no target--space gravitons or higher excited modes, only
the
``tachyons'' corresponding to the
operators (4.11). This is not quite
so: the spectrum contains discrete remnants of the graviton
and the higher string modes at special
momenta${}^{[\wtt,\pkl]}$ (see part II, section 2.3).
For a review of the $c=1$ model,
see~ref.~[\kle].
\eject

{\leftline{\bf Appendix: Random Lattices and Matrix Models}}

\def\tr{\hbox{tr}}

In the above we have often referred to numerical results
obtained from
the ``matrix models.''
For completeness, the basic idea of this approach is explained below.
For a complete review, see e.g., [\mxm,\kle].

 {$\underline{\hbox{A.1. Random Triangulations}}$}

The path integral over unembedded ($D=0$) two--dimensional
Euclidean geometries can be regularized in a diffeomorphism invariant
way as a sum over triangulations of a
surface (figure 5a).
The side length $a$ of the triangles is held
fixed while the number of triangles joining at each vertex
is allowed to vary.
The sum runs over all distinct graphs, i.e., all graphs
that cannot be mapped onto each other.

Let us define $F,E$ and $V$ as the
numbers of faces (triangles), edges and vertices
of a graph like the one in figure 3.
$V-E+F$ is well known to be the Euler characteristic
$\chi=2-2g$ of the manifold, $g$ being its genus.
The area is $\sim\ {a^2}\times F$. The discretized partition function
(2.1) (without $x$) is thus
$$W(\lambda_0,\mu_0)=
\sum_{\hbox{graphs}} \lambda_0^{-(V-E+F)}\ \exp\{-\mu_0 F\},\eqno(A.1)$$
with bare ``string coupling constant''
$\lambda_0=e^{4\pi\gamma}$ and
bare cosmological constant $\mu_0$.

Let us first restrict ourselves to genus zero:
$V-E+F=2$. The
sum over genera will be discussed in subsection A.3.
It is known (and suggested by (4.2)) that the
number of distinct triangulations with a fixed number of triangles
grows to leading order
like $e^{\mu_c F} F^{\gamma-3}$ as $F\rightarrow \infty$, with some
coefficients $\mu_c,\gamma$.
By fine--tuning $\mu_0\sim \mu_c$ and simultaneously letting $a$ go to zero,
the continuum limit is reached
where the sum (A.1) just starts to diverge.
\footnote{*}{Actually, for pure gravity $\gamma-3=-{7\over2}$,
so there is no continuum limit. This can be cured by either inserting
operators or adding matter.}
Intuitively one expects that, in this limit and for genus zero, (A.1)
becomes the partition
function of
two--dimensional quantum
gravity on the sphere,
$$Z={1\over{\lambda_0^2}}\int {Dg\over{\hbox{Diff}}}
\ \exp\{-\mu\int d^2\sigma\ {\sqrt g\ }\}\eqno(A.2)$$
with renormalized cosmological
constant $\mu$:
$$\mu={{(\mu_0-\mu_c)}\over{a^2}}.\eqno(A.3)$$

To establish the equivalence of (A.1) and (A.2)
beyond the intuitive level, one has to show that in the
continuum limit the measure for
the discretized sum becomes the Polyakov measure (2.7).
This has not been proven rigorously, but it
can be expected on the grounds that
the definition of the sum is
diffeomorphism invariant (invariant under permutation of the
vertices), and that (2.7) is
the only diffeomorphism invariant measure
for $g_{\alpha\beta}$ that can be constructed.
Further confirmation comes from the agreement of the results
of the continuum approach and the matrix model approach.

\line{}\vskip9cm
\line{\hskip3.7cm {fig. 5a:}\hskip5.5cm {fig 5b:}\hfill}
\line{\hskip2cm A random triangulation\hskip4cm Its dual\hfill}
\eject

 {$\underline{\hbox{A.2. The Matrix Models}}$}

The trick that makes it possible to actually perform the sum (A.1)
is by now well--known:${}^{[\kada]}$ Consider the partition function
of a zero--dimensional
$\phi^3$ theory, where $\phi$ is a hermitean $N\times N$
matrix:
$$e^{W(\alpha)}=\int d^{N\times N} \phi\
\exp\{-N\ \tr({1\over2} \phi^2+\alpha\phi^3)\}.\eqno(A.4)$$
The normalizations have been chosen for later convenience.

There are two ways to compute $W(\alpha)$: (i), perturbatively
by summing the connected
Feynman diagrams, or (ii), by just doing the
integral.${}^{[\itz,\bre]}$
(ii) is much easier and therefore used to actually
do the computation. We refer to the literature for this.${}^{[\mxm]}$

(i), on the other hand,
serves to establish that (A.4) is equivalent to (A.1).
One only has to note that
the connected Feynman graphs of (A.4) and the triangulations
of (A.1) are in one--to--one correspondence: they are dual to each other.
\footnote*{The dual graph of a triangulation is obtained
by replacing each face by a vertex
and each vertex by a face.} The diagrams are gluon--diagrams
as in fig. 5b.
$V$ is the number of loops, $E$ the number of propagators
and $F$ the number of vertices in the Feynman graphs.
With propagators $1/N$, each graph in the perturbation expansion of (A.4)
is weighted by
$$(\alpha N)^F\ N^{-E}\ N^V\ =\ \alpha^F\ N^{V-E+F},
\eqno(A.5)$$ where the factor $N^V$
comes from the $N$ different flavors propagating in each loop.
With $\alpha=e^{-\mu_0}$ and $\lambda_0=1/N$,
this yields precisely (A.1).
In particular, as $N\rightarrow\infty$, only planar diagrams will survive,
corresponding to a sum over triangulations of a genus zero surface.
In the continuum limit
$\mu_0\rightarrow\mu_c$, where the gluon--nets become dense,
this sum becomes the partition function
(A.2) for quantum gravity on a sphere.

Correlation functions can also be represented in terms of matrix
integrals.${}^{[\mxm]}$ Consider (A.4) with insertions of $\tr\ \phi^n$:
$$\int d^{N\times N} \phi\
\exp\{-N\ \tr({1\over2} \phi^2+\alpha\phi^3)\}\ \tr\ \phi^{n_1}
...\tr\ \phi^{n_k}.$$
This corresponds to summing diagrams with external legs. One easily
sees that their duals are triangulations with $k$ holes of sizes
$n_1\cdot a,...,n_k\cdot a$. In the continuum limit,
these holes correspond to insertions
of operators -- of ``microscopic operators,'' if $n_k$ is held fixed
as $a\rightarrow0$, and of ``macroscopic operators'' if $n_k\cdot a$
is held fixed
(compare with subsections 4.2, 4.3).

The matrix model (A.4) can be generalized
by modifying the potential in the exponential or by introducing
more than one matrix.
The resulting matrix integrals have been identified as partition functions
and correlation functions of gravity coupled to matter
with central charge $c\le1$.
The most interesting solvable matrix model is the one--dimensional
one:${}^{[\kmi]}$
The matrix is a one--dimensional field $\phi(t)$, and the partition
function is
$$e^W=\int D^{N\times N} \phi(t)\ \exp\{-N\int dt\
\tr({1\over2}\dot\phi^2+{1\over2}m\phi^2+\alpha\phi^3)\}.\eqno(A.6)$$
In the perturbation expansion,
each diagram is now weighted by
$$\alpha^F\ N^{(V-E+F)}\ \int\prod_k dt_k \prod_{<ij>}
e^{-m\vert t_i-t_j\vert}, \eqno(A.7)$$
instead of (A.5). Here, $<ij>$ are neighboring vertices in
the Feynman diagrams, and
$e^{-m\vert t\vert}/N$ is the propagator of (A.6). In the
continuum limit,
obtained by tuning $\alpha$ to its
critical value,  $t$ turns into a scalar field on a two--dimensional
surface.
It is believed that in the continuum limit universality allows
replacing the propagator by
$e^{-m t^2}/N$.
Then the Gaussian nearest--neighbor interaction becomes
a standard kinetic term for $t$.
So the integral in (A.7) becomes
$$\int Dt(\sigma)\
\exp\{-m\int d^2\sigma\ (\partial t)^2\}.$$
This establishes that, in the continuum limit,
$W$ is
the partition function of
a free boson with $c=1$, coupled to gravity.

The beautiful nonperturbative calculation of integral
(A.6), by interpreting it as the
ground state energy of $N$
free fermions${}^{[\itz]}$ (the matrix eigenvalues),
is not the topic of this review, so we just refer to the extensive
literature, e.g., [\kle].

\vskip5mm

 {$\underline{\hbox{A.3. Sum over Topologies}}$}

Matrix models can also be used to sum over all genera.${}^{[\double]}$
Consider again the sum (A.1), but now with surfaces of arbitrary genus,
$V-E+F=2-2g$. Then
$$W(\lambda_0,\mu_0)=\sum_{g=0}^\infty \lambda_0^{2g-2}\
W_g(\mu_0).\eqno(A.8)$$
{}From (4.3--4) we know how $W_g$ scales with $\mu\propto(\mu_0-\mu_c)$
in the continuum limit:
$$\eqalign{&W_g(\mu_0)\propto
(\mu_0-\mu_c)^{\gamma_0(1-g)}\cr \Rightarrow \ \
&W(\lambda_0,\mu_0)=W(\kappa)=\sum_{g=0}^\infty\kappa^{(2g-2)}\ w_g
\ \ \ \hbox{with}\ \ \
\kappa={{\lambda_0}\over{(\mu_0-\mu_c)^{\gamma_0/2}}},}
\eqno(A.9)$$
with $\gamma_0=5/2$ for pure gravity
\footnote*{More generally, $\gamma_0=-Q/\alpha$ for gravity with matter}
and some constants $w_g$.
In order to obtain a sensible continuum limit, one must therefore also take
$\lambda_0\rightarrow0$, as $\mu_0-\mu_c\rightarrow0$. Recalling (A.3),
one sees that the string coupling constant gets renormalized
\footnote{**}{Note, however, that from (4.3), $\kappa=\lambda_0$ for $c=25$.}
to
$$\lambda={\lambda_0}({a^2})^{-\gamma_0/2}
\ \ \ \hbox{so that}\ \ \ \kappa=\lambda\mu^{-\gamma_0/2}
=\hbox{finite}.$$
\eject

In the matrix
model, we had $\lambda_0=1/N$ and $\mu_0=-\log\alpha$.
The limit where $N\rightarrow\infty$
and simultaneously $\alpha
\rightarrow\alpha_c=e^{-\mu_c}$ so that
$\kappa$ is kept fixed is called the double--scaling limit.
The matrix integral (A.4) in this limit
is the partition function of quantum gravity
with topological expansion parameter $\kappa$.
Quantum gravity on the sphere is recovered
for $\kappa\rightarrow0$, that is,
$N\rightarrow\infty$ for small but fixed $\alpha-\alpha_c$.

Integral (A.4)
in the double--scaling limit can be evaluated.${}^{[\double]}$
It is found that $\tilde W(t)$, defined by
$\tilde W(t)=W(\kappa)$ with
$t=\kappa^{-\gamma_0/2}$,
obeys the Painlev\'e equation
$$t=\tilde W^2-{1\over3}\ddot {\tilde W}.\eqno(A.10)$$
This result has yet to be obtained from the Liouville approach.

\vfill\eject



\line{{\bf PART II:\ \
RUNNING COUPLING CONSTANTS IN 2D GRAVITY}
\foot{based on a paper to be published in Nuclear Physics B}
}
\vskip1cm

{\leftline{\bf 1. Introduction}}

It was pointed out in subsection 3.5 of part I, that interactions
in two--dimensional quantum gravity
must be exactly marginal.
As will be shown in this part of the thesis, this implies that
new terms must be added to the action (3.16). They have not been considered
previously, but
are important for understanding
the renormalization group flow and
can be observed
in recent matrix model results for the phase diagram
of the Sine--Gordon model coupled to gravity.

Let us recall the issue.
Two--dimensional quantum gravity coupled
to $c\le 1$ matter `$x$' is described
in conformal gauge${}^{[\d,\dk]}$
in terms of fields propagating
in a fictitious background metric $\hat g_{\alpha\beta}$.
The action is the appropriate conformally invariant free action
plus interaction terms which are usually assumed to be of the form
$${\cal L}_{int}= \hbox{cosmological constant}
+\sum_i\tau^i\int\Phi_i(x)\ e^{\alpha_i\phi},\eqno(1.1)$$
where $\Phi_i$ are primary fields of the matter theory,
the $\tau^i$ are small coupling constants, $\phi$ is the Liouville mode
and the $\alpha_i$ are adjusted to
make the dimensions of the operators equal to two.
However, (1.1) cannot be the complete interaction, for at least two reasons:

1. The operators in (1.1) are not exactly marginal.
\foot{An operator is marginal if its dimension is two, and exactly
marginal if its beta function is zero to all orders.}
They should be,
because
the Liouville theory must be background
independent as a consequence of general covariance.${}^{[\d,\dk]}$
Therefore the beta functions
of the theory must be zero to all orders in the couplings.
Adjusting the $\alpha_i$ in (1.1) makes them zero to first order,
but the beta functions have quadratic pieces
whenever there are nontrivial OPE's,
\foot{See section 2 for the issue of
renormalization schemes and field redefinitions}
as in formula (3.17) of part I.

2. The renormalization group flow
would be quite trivial with (1.1). As mentioned, there
should be no flow with respect to the fictitious background scale $\sqhat$.
But, as explained in section 3,
a constant shift of $\phi$ should be interpreted as a rescaling
of the physical cutoff,${}^{[\polch,\suss,\poly]}$
and should, in particular, result in a mixing (flow) between
different operators. This does not happen in (1.1).

It is shown in section 2 that the first problem
can be solved by adding a term
$$\propto\ \ \
-c^k_{ij}\tau^i\tau^j\int \Phi_k(x)\ \phi\ e^{\alpha_k\phi}\eqno(1.2)$$
to the interaction (1.1), where $c^k_{ij}$ are the operator
product coefficients.
This, in fact, also resolves the second problem: the modified
interaction displays the expected operator mixing
under shifts of $\phi$ by a constant.
Requiring that there be {\it no} flow with respect to
the background scale
$\sqrt{\hat g}$ determines the
flow with respect to the physical scale ${\sqrt{\hat g}}e^{\alpha\phi}$.
For the case of the Sine--Gordon model coupled to gravity it will be seen
that this
flow qualitatively agrees with recent matrix model results
by Moore.${}^{[\moo]}$

Equation (1.2) should be viewed as a second--order
correction to the gravitational dressing of the
$\Phi_i(x)$.
We conjecture that further modifications of (1.1)+(1.2)
can be made order by order in the $\tau^i$, leading to an infinite
dimensional space of exactly
marginal perturbations. Our calculations serve to verify this conjecture to
second order.
This part of the thesis is organized as follows:

In section 2, the second--order corrections (1.2)
are discussed. First, it is shown in subsection 2.1 that
the interaction (1.1) plus (1.2) is marginal up to second order.
That the correction (1.2) is essentially unique is argued
in appendix A by thinking of the marginality conditions
as equations of motion of string theory.
The $c=1$ model coupled to gravity is
discussed as an example. In subsection 2.2, the interaction term
is taken to be the Sine--Gordon interaction near
the Kosterlitz-Thouless momentum $p={\sqrt2}$
and near $p={1\over2}\sqrt2$. In subsection 2.3, the interaction terms
are taken to be the ``discrete operators.''
The effects of including the cosmological constant
are studied in appendix B.
The conclusions of appendices A and B are summarized in subsection 2.4.

In section 3, running coupling constants are discussed.
They are defined in subsection 3.1 so that they
absorb a constant
shift of $\phi$. In subsections 3.2 and 3.3
this is applied to the Sine-Gordon model and the resulting phase
boundaries are compared with those found with the
nonperturbative matrix model
techniques.${}^{[\moo]}$
It is seen that
the presence of the terms (1.2) is crucial even for qualitative
agreement of the matrix model and the Liouville theory
approaches. A more detailed comparison of both
is left for future work.
The one--loop beta functions for the discrete $c=1$ operators
are also obtained.

In section 4,
possible extensions of this work are pointed out, as well as
implications for black--hole hair and correlation functions.
In particular, it is argued that the relation between
correlation functions in the matrix model and in the Liouville
approach is more complicated than often assumed.

\def\h{h_{\mu\nu}}

\def\t{\cos px\ e^{(p-{\sqrt2})\phi}}
\def\tt{\tau}

\def\stwo{{\sqrt2}}

\vskip1cm
{\leftline{\bf 2. Exactly Marginal Operators}}

$\underline{\hbox{2.1. The Terms of Order $\tau^2$}}$

In the approach of
David, Distler and Kawaii (DDK),
a conformal field theory with central charge
$c$ and Lagrangian ${\cal L}_m(x)$
coupled to 2D gravity is described by the action${}^{[\d,\dk]}$
$$S_0={1\over{8\pi}}\int d^2\sigma\sqhat\{ {\cal L}_m(x)+(\partial\phi)^2+Q
\hat R\phi+\hbox{cosmological constant}+\hbox{ghosts}\}\eqno(2.1)$$
with $Q={\sqrt{(25-c)/3}}$ and conformal factor $\phi$. The
cosmological constant will be neglected at first, but included later.
(See subsection 2.4.)

When ${\cal L}_m(x)$ is
perturbed by operators $t^i\Phi_i(x)$, these operators get
``dressed'' upon coupling to gravity.
As emphasized above, the dressed interaction
must be an {\it exactly}
marginal operator, not only an operator of dimension two.
Exact marginality
is needed, because in DDK's approach the
background metric $\hat g$ corresponds to an arbitrary gauge choice that
nothing physical should depend on.
In particular, coupling constants
should not run with respect to $\hat g$: all beta functions must be
zero to all orders.

In prior work, this condition has been exploited only
to first order.${}^{[\d,\dk]}$
Here it will be investigated in second order.
Generally,${}^{[\car]}$
the beta functions for a perturbed conformally
invariant theory (see sect. 3.5 of part I),
$$S=S_0+\tau^i\int d^2\sigma\ V_i,
\foot{\hbox{Here and below we omit powers of a length scale
$a$, needed to make the $\tau^i$ dimensionless.}}$$
$$\hbox{are}\ \ \ \ \ \beta^i=(\Delta^i_j-2\delta^i_j)\ \tau^j+\pi c^i_{jk}
\tau^j\tau^k+O(\tau^3),\eqno(2.2)$$
if the $V_i$ are primary fields of dimension $\Delta_i$ close to two.
$\Delta^i_j$ is the dimension matrix computed with $S_0$.
If the operators $V_k$ on the RHS of the operator algebra
\foot{keeping only the radial dependence on the RHS;
the rest drops out after
integrating over $\vec r$.}
$$V_i(r)V_j(0)\sim \sum_k \vert r\vert^{-\Delta_i-\Delta_j+\Delta_k}
c^k_{ij} V_k(0)$$
also have dimension
close to two, the coefficients $c^k_{ij}$ are universal constants,
independent of the renormalization scheme used to compute them.
Operators of other dimensions also appear on the RHS.
For them, the
$c^i_{jk}$ are scheme--dependent, that is, not invariant under
coupling constant redefinitions. Let us ignore them here.
\foot{
Presumably the scheme can be chosen so that they vanish.}

We now show that $\beta^i=0+O(t^3)$ for the perturbation (1.1) plus
(1.2):
\foot{The question of the uniqueness
of (1.2) is deferred to subsection 4.4.}
$$\delta S\ =\ \tau^i\int d^2\sigma\ V_i(x,\phi)\ \equiv \ \tau^i\int
d^2\sigma\ \hat V_i(x,\phi)
-\pi\ c^k_{ij}\tau^i\tau^j\int d^2\sigma\ X_k(x,\phi),\eqno(2.3)$$
$$V_i=\hat V_i-\pi c^k_{ij}\tau^jX_k,\ \ \ \ \hat V_i \equiv
\Phi_i(x)\ e^{\alpha_i\phi},\ \ \ \
X_k\equiv -{1\over{Q+2\alpha_k}}\ \Phi_k(x)\ \phi\ e^{\alpha_k\phi}.
\eqno(2.4)$$
$\alpha_i$ is adjusted to make the dimension of $\hat V_i$
exactly two. Without the $O(\tau)$ corrections in $V_i$, we would thus
have
$\Delta^i_j=2\delta^i_j$ and
$\beta=0+O(\tau^2)$ from (2.2). With them,
$$\Delta^i_j=2\delta^i_j-\pi c^i_{kj}\tau^k+O(\tau^2),\eqno(2.5)$$
hence $\beta=0+O(\tau^3)$ in (2.2). (2.5) can be derived by
writing
$$X_k =\ -{1\over{Q+2\alpha_k}}\ \Phi_k(x)\ {\partial\over{\partial\alpha_k}}
\ e^{\alpha_k\phi},$$
defining the generator $L_0+\bar L_0$ of global scale transformations,
and differentiating with respect to $\alpha_k$ the dimension formula
$$\eqalign{
&(L_0+\bar L_0) e^{\alpha_k\phi} = -\alpha_k(\alpha_k+Q)\ e^{\alpha_k\phi},\cr
\Rightarrow\ \ \ &(L_0+\bar L_0) X_k= 2X_k+ \hat V_k\cr
\Rightarrow\ \ \ &(L_0+\bar L_0) V_k= 2V_k-\pi c^i_{jk}\tau^jV_i+O(\tau^2).}$$

As a simple check of all this, one can consider rescaling
$\psi\rightarrow (1+\lambda)\psi$ in
$$S_{\hbox{ toy model}}\ =
\ {1\over{8\pi}}\int d^2\sigma\ ((\partial \psi)^2+\gamma
\cos\stwo \psi)\ \ \ \hbox{with}\ \ \
\lambda\ll \gamma\ (\times a^2).$$
This should keep
the interaction marginal at $O(\gamma\lambda)$ and is equivalent to
adding the terms $2\lambda (\partial \psi)^2
-\stwo\lambda \gamma \psi\sin\stwo \psi$ to
the Lagrangian. Using the above method, one can check
that the second term indeed arises as the correction to the first term.
\foot{Here, $8\pi \hat V_1= \cos \stwo\psi,\ 8\pi\hat V_2
= 2(\partial \psi)^2$,  $c^1_{12}=c^1_{21}=-2/\pi$ and
$8\pi X_1=-1/(2\stwo)\ \psi\sin \stwo\psi$.}

\vskip 5mm

$\underline{\hbox{2.2. The Sine--Gordon model}}$

As an example, consider an uncompactified
scalar field $x$ coupled to gravity. Then $c=1$ and $Q=2\stwo$. First,
we perturb this model
by the Sine--Gordon interaction,
$$S={1\over{8\pi}}\int d^2\sigma\ \sqhat\{ (\partial x)^2+(\partial\phi)^2
+2{\sqrt2}
\hat R\phi+\hbox{ghosts}\}+
m\int d^2\sigma\ \t,\eqno(2.6)$$
and determine the $O(m^2)$ corrections (2.3).
To find the coefficients $c^k_{ij}$, consider
the operator product expansion (OPE), using the propagator $-\log r^2$:
$$\eqalign{
\cos px\ &e^{(p-{\sqrt2})\phi}(r)\ \
\cos px\ e^{(p-{\sqrt2})\phi}(0)\cr
&\sim \vert r \vert
^{-2-4(p-{1\over2}{\sqrt2})^2}\ e^{2(p-{\sqrt2})\phi}
\ \{{1\over2}-\vert r\vert^2
 {{p^2}\over8}\partial x^2
+...\}\cr &+\vert r\vert
^{-2+(4{\sqrt2}p-2)}\ \cos 2px\
e^{2(p-{\sqrt2})\phi}
\ \{{1\over2}-\vert r\vert^2
\ {{p^2}\over8}\partial x^2+...\}
}\eqno(2.7)$$
As mentioned, we must look for nearly quadratic singularities,
so that the $c^i_{jk}$ are universal constants.
The second line in (2.7)
has $\vert r\vert^{-2}$ singularities at the ``discrete momenta''
$p\in\{...,0,{1\over2}\stwo,\stwo,...\}$
\foot{corresponding to the discrete tachyons $\Phi_{j,\pm j}$ of the
next subsection}
and the third line at
$p\in\{...,0,{1\over4}\stwo.\}.$
\foot{However, for $p<{1\over2}\stwo$ the operators
on the RHS do not exist (see sect. 4.2 of part I).
As a consequence, $\cos 2px$ terms are not induced and
no phase transition occurs at those momenta, as argued in
appendix B.}
Let us study the neighborhoods of $p={1\over2}\stwo$
and $p=\stwo$. There the induced operators are:
$$\eqalign{\hbox{at}\ p={1\over2}\stwo+\delta:\ \ \
& \hat V_1=
e^{(2\delta-\stwo)\phi}\ \ \ \hbox{with}\ c^1_{mm}={1\over2}\cr
\hbox{at}\ p=\stwo+\epsilon:\ \ \
& \hat V_2=(\partial x)^2\ e^{2\epsilon\phi}\ \ \ \hbox{with}\ c^2_{mm}
=-{{p^2}\over8}.}$$
{}From (2.3) and (2.4), the leading order corrections to (2.6)
are obtained:
$$\eqalign{
&\hbox{near } p={1\over2}\stwo:\ \ \delta S=
 {{m^2\pi}\over{8\delta}}\
\int d^2\sigma\ \phi\ e^{-\stwo\phi} \cr
&\hbox{near } p=\stwo:\ \ \delta S=
-{m^2\pi\over{8\stwo}}\int d^2\sigma\ \phi\ (\partial x)^2.
}\eqno(2.8)$$
This will be further discussed in section 3. Note the
factor $\delta^{-1}$ in the first line. Note also that from
the string theory
point of view, (2.8) describes the backreaction of the tachyon
onto itself and the graviton.
\eject

$\underline{\hbox{2.3. The Discrete $c=1$ Operators}}$

As a second example, consider perturbing the $c=1$ model with
the nonrenormalizable so--called
(chiral) discrete primaries $\Phi_{jm}(x)$,${}^{[\seg]}$
$$\Phi_{jm}=\ f_{jm}[\partial x,\partial^2 x,...]
e^{im{\sqrt 2}x}\ \equiv (H^-)^{j-m}\ e^{ij{\sqrt 2}x}
\eqno(2.9)$$
with dimension $j^2$ and
$SU(2)$ indices $j,m$, the $SU(2)$ algebra being generated by
$$H^{\pm}\sim\ \oint dz\ e^{\pm i{\sqrt 2}x(z)}\ =\oint dz\ \Phi_{1,\pm1}(z),
\ \ \ H^3\sim
\oint dz\ i\stwo\partial x(z)=\oint dz\ \Phi_{1,0}(z).$$
Here the integrations are along contours in the $z$ plane that encircle
the operators that $H^\pm,H^3$ act upon.
If an interaction
$t^{jm}\Phi_{jm}[x]\bar\Phi_{jm}[\bar x]$
is added to the matter Lagrangian, the dressed interaction is, to
first order in the coupling constants,
$${\cal L}_{int}=
\tau^{jm}\ \hat V_{jm},\ \ \ \hat V_{jm}\equiv  \Phi_{jm}(x)
\bar\Phi_{jm}(\bar x)\ e^{\alpha_j(\phi+\bar\phi)}$$
with $\alpha_j=(j-1){\sqrt 2}$. The $\Phi_{jm}$ can be rescaled
such that the operator algebra of the
$\hat V_{jm}$ has the
$w_\infty$
structure${}^{[\wtt,\pkl]}$
$$c^{jm}_{kn\ k'n'}=(kn'-k'n)^2\ \delta_{j,k+k'-1}\delta_{m,n+n'}.$$
{}From (2.3) and (2.4) one obtains the second--order interaction term
$$\delta {\cal L}=
\sum_{j,m}\ \Phi_{jm}\bar\Phi_{jm}\ \phi\ e^{\alpha_j\phi}\ \
\times\ {\pi\over{2\stwo j}}
\sum_{j'+j''=j+1\atop m'+m''=m}
(j'm''-j''m')^2\tau^{j'm'}\tau^{j''m''}.
\eqno(2.10)$$
${\cal L}_{int}+\delta {\cal L}$ is marginal up
to order $(\tau)^2$.
Again, depending on the renormalization scheme, operators whose dimensions
are not two may also appear in $\delta {\cal L}$.
\eject

$\underline{\hbox{2.4. Uniqueness and the Cosmological Constant}}$

Next, we must ask whether the modifications (1.2) of the operators
(1.1) are
the unique modifications that achieve marginality
up to order $(\tau)^2$.
The situation
is greatly clarified by thinking of the marginality conditions
as equations of motion of string
theory, as in ref. [\polch].
One concludes the following (more details are given in appendix A):

The marginality conditions are second--order differential equations
in $\phi$ and $x$. Their solutions are unique after two boundary
conditions are imposed, namely:
(i): the modifications must
vanish at $\phi=0$, and (ii): the second, more negative of the two possible
Liouville dressings (as e.g., in (A.5)) does not appear.

Boundary condition (i) comes about because
the Liouville mode
$\phi$ lives on a half line:${}^{[\polch]}$
The sum over geometries can be covariantly regularized as a sum over random
lattices. Then no two points can come closer to each other than the
lattice spacing $a$:
$$\hat g_{\mu\nu}\ e^{\alpha\phi}\ d\sigma^\mu d\sigma^\nu\ge a^2\ \ \
\Rightarrow\ \ \ \phi\le\phi_0\ \ \hbox{with}\ \
e^{\alpha\phi_0}\propto a^2\eqno(2.11)$$
(recall that $\alpha<0$.)
After shifting $\phi$ so that $\phi_0=0$, boundary condition
(i) states that the action $S(\phi=0)$ at the cutoff scale is the bare
action (see e.g., (A.4)).${}^{[\polch]}$
Boundary condition (ii) arises because operators with the more
negative Liouville dressing
do not exist (see sect. 4.2 of part I).

The correction terms found above
obey the boundary conditions (i) and (ii) and are therefore
unique.
Of course, there is always an ambiguity due to field redefinitions,
that is, choosing different renormalization schemes when
computing the beta functions.
There is no problem as long as we stick to one scheme.
\foot{
Actually, the scheme used in subsection 2.1
is not the same as the one used for the string equations of motion
in appendix A, but this does not affect the above conclusions.}

Another important question is
how the cosmological constant
modifies our results.
The problem with the cosmological constant operator is that it cannot be made
small in the IR ($\phi\rightarrow -\infty$). It can only be
shifted in the $\phi$ direction.
Thus it cannot be treated perturbatively, rather it should be included
from the start in $S_0$ of (2.1). In its presence
the OPE's used above are modified. Applying
the discussions in refs. [\polch,\sei,\pch],
one tentatively
concludes the following (more details are given in appendix B):

1. The effects of the
cosmological constant on gravitational dressings can be
neglected in the ultraviolet ($\phi\sim0$), but not in the
infrared ($\phi\rightarrow -\infty$).

2. In the Sine--Gordon model coupled to gravity, no unwanted
terms with $\cos 2px$ are induced because the OPE's
are ``softer'' than in free field theory (see (A.4-5)).

These conclusions will be
confirmed in section 3 by observing the agreement with
matrix model results.

\vskip 1cm

{\leftline{\bf 3. Running Coupling Constants}}

$\underline{\hbox{3.1. Renormalization Group Transformations}}$

In subsections 3.1 and 3.2, the cosmological term will be assumed to be
$\mu\ e^{-\stwo\phi}$ to simplify the discussion. This can be
generalized to more complicated forms like $T_\mu(\phi)$ in (B.2).

Consider rescaling the cutoff
$a\rightarrow ae^\rho$
in the path integral of 2D gravity,
$$\int_{\phi\ge\phi_0} D\phi\ Dx\ Db\ Dc\ e^{-S(\phi,x,b,c)}.$$
{}From (2.11) one sees that this induces a shift
of the bound $\phi_0\rightarrow
\phi_0+\lambda,$ in addition to an
ordinary RG transformation. From (2.11),
$$\lambda=\phi_0(a e^\rho)-\phi_0(a)={2\over\alpha}\rho=-\stwo\rho.
\foot{\hbox{More generally,
$\rho={1\over2}\log{{T_\mu(\lambda)}\over{T_\mu(0)}}$
with cosmological constant $T_\mu$ as in (B.2)}}\eqno(3.1)$$
In fact, since ordinary RG transformations
are irrelevant (since all beta functions are zero),
$only$ the shift of the bound remains.
The constant shift of the
bound is equivalent to a constant shift of the Liouville mode,
$\phi\rightarrow \phi+\lambda$.
Let us absorb this shift in
``running coupling constants'' $\vec\tau(\lambda),\vec\tau_0
\equiv \vec\tau(0)$,
defined by:
$$S[\vec\tau(\lambda),x,\phi+\lambda]=S[\vec\tau_0,x,\phi].\eqno(3.2)$$
After expressing $\lambda$
in terms of $\rho$,
\foot{This is more complicated with $T_\mu$,
but to find phase boundaries,
$\vec\tau(\lambda)$ will be good enough.}
one obtains the renormalization group
flow $\vec\tau(\rho)$. (For similar conclusions, see
[\suss,\kut].)

As mentioned above,
the action (3.2) corresponds to a classical solution of string theory
with two--dimensional target space ($x,\phi$).
The equations of motion of classical string theory thus play the role
of the Gell-Mann--Low equations in the presence of gravity.${}^{[\suss]}$
They contain
second-- (and higher--) order derivatives of $\phi$, which we have
just interpreted as ``renormalization group time.''
It has been suggested that those
are due to the contribution of pinched
spheres in the functional integral over metrics.${}^{[\pkov]}$

\vskip5mm
$\underline{\hbox{3.2. Sine--Gordon Model near $p=\stwo$}}$

We now apply the preceding procedure to the examples worked out in section 2,
starting with the Sine--Gordon model.
In flat space, at $p=\stwo$ the Kosterlitz--Thouless phase transition occurs.
With gravity, at $p=\stwo+\epsilon$ the action
is to order $(m,\epsilon)^2$
(see (2.8); we ignore $O(\mu)$--corrections
of the Sine--Gordon interaction):
$$\eqalign{S={1\over{8\pi}}\int d^2\sigma\
\sqhat&\{ (\partial x)^2+(\partial\phi)^2+2{\sqrt2}
\hat R\phi
+\hbox{ghosts}+\mu\ e^{-\stwo\phi}\}\cr
&+m\int d^2\sigma\ \cos(\stwo+\epsilon)x\ e^{\epsilon\phi}
-{\pi\over{8\stwo}}\ m^2\int d^2\sigma\ \phi\ (\partial x)^2.
}\eqno(3.3)$$
To $O(m,\epsilon)^2$, a shift
$\phi\rightarrow\phi+\lambda$ can be absorbed in the
$\lambda$--dependent couplings
$$m(\lambda)=m_0e^{-\epsilon\lambda},\ \ \
\epsilon(\lambda)=\epsilon_0-{\pi^2\over{2}}\lambda m^2,\ \ \
\mu(\lambda)=\mu_0e^{\stwo\lambda}.
$$
In deriving $\epsilon(\lambda)$,
the $\lambda m^2(\partial x)^2$ term has been absorbed
in a redefinition of $x$ and then in a shift of $\epsilon$. Defining
`prime' as ${d\over{d\lambda}}$, we get
$$\epsilon'=-{\pi^2\over{2}}m^2+...,\ \ \
m'=-\epsilon m+...,\ \ \
\mu'=\stwo\mu+...
$$
Defining `dot' as ${d\over{d\rho}}
=-\stwo{d\over{d\lambda}}$ yields the
beta functions
$$\dot\epsilon={\pi^2\over{\stwo}}m^2,
\ \ \ \dot m={\stwo}\epsilon m,\ \ \ \dot\mu=-2\mu.\eqno(3.4)$$
$\dot\mu$ serves as a check: $\mu$ decays in the UV according
to its dimension (two).
The coupling constant flow is qualitatively the same as in flat
space and is given by the Kosterlitz--Thouless diagram
(Figure 1). We see that the $m^2 \phi\ (\partial x)^2$
correction in (3.3), which is an example of the corrections
(1.2) to (1.1), plays a crucial role:
ignoring it would be like forgetting about field renormalization
in the ordinary Sine-Gordon model.
\vfill
\centerline{{\bf fig.1:}}
\centerline{KT--transition with gravity at $O(\epsilon,m)^2$
(Arrows point towards infrared).}
\eject

{}From (3.4), the phase boundary for $p>\stwo$ is linear,
$m \propto \epsilon$.
To this order, this agrees with the matrix model result ${}^{[\moo]}$
$$m\ \propto\ \epsilon\ e^{{1\over2}\stwo\epsilon
\log\epsilon}.\eqno(3.5)$$
With the normalization of $m$ and $\epsilon$ as in (3.3), we obtain
the slope $\stwo/\pi$ for the phase boundary. After comparing
the normalizations, this should also
be checked with the matrix model.
It will also be interesting to see if the logarithm in
(3.5)
follows from the
modifications of higher order in $m$, needed to keep the interaction
near $p=\stwo$ marginal beyond $O(m^2)$.

We can now interpret the phase diagram of [\moo]
(figure 2) near $m,\epsilon=0$:
For $\epsilon<0$ (regions II and V of [\moo]),
$m$ grows exponentially towards the IR.
The model thus flows to (infinitely many copies of) the $c=0$,
pure gravity model.${}^{[\kut,\gkl]}$

For $\epsilon>0$, but $m$ greater than a critical value $m_c(\epsilon)$
(region VI of [\moo]), the flow goes again towards
the $c=0$ model in the IR. For $m<m_c(\epsilon)$ (region III of [\moo])
the flow seems to go to the free $c=1$ model. However,
the domain of small $\epsilon,m$ is now the IR domain. As noted
in subsection 2.4, the cosmological
constant cannot be neglected there and further investigation is needed.

\vfill
\centerline{{\bf fig.2:}}
\centerline{Regions of the Sine--Gordon model with gravity at $O(m^2)$.}
\eject

$\underline{\hbox{3.3. Sine--Gordon Model near $p={1\over2}\stwo$}}$

At $p={1\over2}\stwo+\delta$, the situation is less clear.
{}From (2.8),
instead of the $(\partial x)^2$ term a ``1'' term
is induced. That is,
the cosmological constant
is modified by the induced
operator $\phi\ e^{-\stwo\phi}$. The latter becomes comparable with
the background cosmological constant at $\delta\sim{{m^2}\over\mu}$.
Let us tentatively
\foot{Here we use $\phi\ e^{-\stwo\phi}$ instead of the
simple form $e^{-\stwo\phi}$ for the cosmological
constant (see refs. [\sei,\pch]).}
write the action
to leading order as:
$$\eqalign{S={1\over{8\pi}}&\int d^2\sigma\ \sqhat\{
(\partial x)^2+(\partial\phi)^2+2{\sqrt2}
\hat R\phi
+\hbox{ghosts}\}\cr
&+m\int d^2\sigma\ \cos({1\over2}\stwo+\delta)x\
e^{(-{1\over2}\stwo+\delta)\phi}
+({\mu\over{8\pi}}+{{m^2}\pi\over{8\delta}})\int d^2\sigma\ \phi\
e^{-\stwo\phi}
.}\eqno(3.6)$$
With our normalizations, the effective cosmological constant is now
$\mu+{{m^2}\over{\delta}}\pi^2$. For fixed $\mu$, it blows up
as $\vert\delta\vert\rightarrow 0$.
For $\delta<0$ and
$m\ge{1\over\pi} \sqrt{\vert\mu\delta\vert}$, it is negative.
Indeed, in the matrix model a singularity of the free energy has been found
at
$$\delta<0,\ \ \ \
m\propto{\sqrt{\vert\mu\delta\vert}} e^{{1\over2}\stwo \delta\log\delta}.
\eqno(3.7)$$
Let us therefore identify the region
where $\mu+{{m^2}\over{\delta}}\pi^2$ is negative with
region IV of [\moo].
We leave a further interpretation of the situation
near $p={1\over2}\stwo$ for the future.

\vskip5mm

$\underline{\hbox{3.4. The Discrete Operators}}$

We can also determine the one--loop beta function
for the ``discrete'' interactions (2.9)
of subsection 2.3. From (2.10),
$${\cal L}+\delta {\cal L}=\sum_{j,m}
\Phi_{jm}
\bar\Phi_{jm}\ e^{(j-1)\stwo\ \phi}
\{\tau^{jm}+\phi\
 {\pi\over{2\stwo j}}
\sum_{j'+j''=j+1\atop m'+m''=m}
(j'm''-j''m')^2\tau^{j'm'}\tau^{j''m''}\}.$$
Constant shifts $\phi\rightarrow\phi+\lambda$
are absorbed up to $O(\tau^2)$ in:
$$\tau^{jm}(\vec \tau_0,\lambda)=\{\tau^{jm}_0
-\lambda\times {\pi\over{{2\sqrt 2 }j}}
\sum_{j'+j''=j+1\atop m'+m''=m}
(j'm''-j''m')^2
\tau^{j'm'}_0\tau^{j''m''}_0
\} e^{-(j-1)\stwo \hbox{$\lambda$}}
.$$
{}From this we find the one--loop beta function (using (3.1)):
$$\dot \tau^{jm}
=2(j-1)\ \tau^{jm}
+{\pi\over{2j}}\sum_{{j'+j''=j+1}\atop{m'+m''=m}}
(j'm''-j''m')^2\ \tau^{j'm'}\tau^{j''m''}
\ +O(\tau^3).
\eqno(3.8)$$
Thus, turning on
the operators $\Phi_{jm}\bar\Phi_{j'm'}$ with $j'>1$ will in general
induce an infinite set of higher spin operators $\Phi_{jm}\bar\Phi_{jm}$ at
$O(\tau^2)$, whose couplings
were originally turned off.
This is what one expects from these nonrenormalizable
operators, but it would not happen without
the $O(\tau^2)$ modification $\delta{\cal L}$.

\vskip 1cm

{\leftline{\bf 4. Outlook}}

$\underline{\hbox{4.1. Correlation Functions}}$

The modifications (1.2) are important not only for understanding
the renormalization group flow but also
for computing correlation functions in Liouville theory.
They imply the identification (the notation is as in (2.4)):
$$<\exp\{\int t^i \Phi_i\}>_{G}\ \sim\
<\exp\{\int(\tau^i\ \hat V_i+\kappa_l c^l_{ij}\tau^i\tau^j
\phi \hat V_l+..
)\}>_{L},\eqno(4.1)$$
where $<...>_{G}$ and $<...>_{L}$ denote correlation functions
computed in the matrix model (Gravity)
and in Liouville theory, respectively, and
$\kappa_l=\pi/(Q+2\alpha_l)$.
$\tau^i$ is related to the $t^j$ in some nontrivial way:
$\tau^i=t^i+O((t)^2)$.${}^{[\mss]}$

Geometrically, the extra terms on the RHS can be interpreted as arising from
pinched spheres in the sum over surfaces.
(4.1) has consequences for the correspondence of matrix model and
Liouville correlation functions. Expanding both sides and temporarily
identifying $t$ and $\tau$
\foot{
The nontrivial relation between $\tau$ and $t$ noted in [\mss] corresponds
to the appearance of operators $\hat V_l$ (instead of $\phi \hat V_l$) on the
RHS of (4.1). They are also present, but let us here focus on
the new type of operators $\phi \hat V_l$.}
yields e.g, for the two-point function:
$$<\int\Phi_i\int\Phi_j>_{G}\ =\int d^2z\int d^2w<\hat V_i(z)\hat V_j(w)>_{L}
+\ 2\kappa_l
c^l_{ij}\int d^2w<\phi\ \hat V_l(w)>_{L}.
\eqno(4.2)$$
In fact, the last term is necessary for background invariance: Inserting
a covariant regulator
like $\Theta(\sqphi\vert z-w\vert^2-a^2)$
into the two-point function
induces new background dependence, coming from the
integration region $z\sim w$.${}^{[\car]}$
By construction, the one--point functions added in (4.2) are precisely
what is needed to cancel this dependence.

Additional terms like the ones in (4.2) are
also present in higher point functions. They can be determined by
background invariance. It should be possible to see them in matrix model
computations, e.g., of higher--point functions of tachyons at the
``discrete'' momenta.
It then needs to be better understood
why we can recover some of the matrix model correlators
with the method of
Goulian and Li
from the Liouville correlators,${}^{[\triv,\gou,\kle]}$
without the
extra terms in (4.2).
\vskip .8cm

$\underline{\hbox{4.2. Black Hole Hair}}$

The conjecture that all the discrete operators, in particular the
`static' ones $\Phi_{j,0}$ with zero $x$--momentum can be turned into exactly
marginal ones implies that
each $\Phi_{j,0}$ adds
a new dimension to the space of black hole solutions of
classical 2D string theory, corresponding to higher spin (not
only metric) hair.
It will be very interesting to better understand how
significant this is for the issue of information loss
in black holes.${}^{[\jhs]}$

\vskip .8cm
$\underline{\hbox{4.3. Four Dimensions}}$

It would also be interesting to extend this work to four
dimensions. Four--dimensional Euclidean quantum geometry
is, at the least, an interesting statistical mechanical model.
In part III, we show that at
the ultraviolet fixed point of infinite Weyl coupling,
where the theory is asymptotically free,${}^{[\ft]}$
it can be solved
with the methods of two--dimensional quantum
gravity in conformal gauge.${}^{[\schm]}$
Perturbing away
from this limit is similar to adding perturbations to
the free $c=1$ theory. One might be able to find a phase diagram for
Euclidean quantum gravity
by generalizing the method suggested here.

\vskip.8cm
$\underline{\hbox{4.4. Summary}}$

In the Liouville theory approach to 2D quantum gravity coupled to an
interacting scalar field, new terms
appear in the Lagrangian at higher orders in the coupling constants.
They are required by background independence and
cannot be eliminated by a field redefinition when the interaction
is given by one of the discrete tachyons or higher--spin operators.

The new terms are crucial for obtaining
the correct phase diagram, as found with
the nonperturbative matrix model techniques in the
case of the Sine--Gordon model. We have partly interpreted
this diagram, but the transition below $p=$
${1\over2}$ of the Kosterlitz--Thouless momentum must be clarified more.
The cosmological constant must be treated more rigorously, and
the cubic terms in the beta function (2.2), which are also universal,
should be derived. The new terms have various other implications
and should, in particular, be important for the correct computation
of higher--point correlation
functions.

\vskip2.5cm
\centerline{{\bf APPENDICES}}\vskip1.5cm
{}

$\underline{\hbox{Appendix A:
String Equations of Motion and Boundary Conditions}}$
\vskip .5cm

The question addressed here is whether (1.2) are the unique modifications that
make the interaction (1.1) marginal up to order $\tau^2$.
It is useful to think of 2D quantum gravity as classical
string theory.${}^{[\polch]}$
Let us first discuss the example of
the Sine--Gordon model.
The discussion will be restricted to genus zero.

It is well known that, for genus zero, exactly marginal
perturbations of the world--sheet action
correspond to classical solutions of string theory.
Some of them can be found by expanding
in powers of $m$ the dilaton $\Phi$, the graviton $G_{\mu\nu}$
and the tachyon $T$ in the sigma model
$$S={1\over{8\pi}}\int d^2\sigma\ \sqhat\{
G_{\mu\nu}(x,\phi)\partial x^\mu\partial x^\nu
+\hat R\Phi(x,\phi)+T(x,\phi)\}\hskip1cm\hbox{in m:}
\foot{\hbox{The cosmological constant will be included in the tachyon
in appendix B.}}
\footnote{}{\hbox{
Its presence justifies the expansion in $m$.}}
\eqno(A.1)$$
$$\eqalign{
T(x,\phi)&
=m\t+m^2T^{(1)}(x,\phi)
+..\cr
\Phi(x,\phi)&
=2{\sqrt2}\phi+m^2\Phi^{(1)}(x,\phi)+..\cr
G_{\mu\nu}(x,\phi)&
=\delta_{\mu\nu}+m^2\h(x,\phi)+..
}\eqno(A.2)$$
and by then solving the equations of motion derived from
the low-energy effective action of two-dimensional string
field theory,${}^{[\gsw,\tsey]}$
$$\int dx\ d\phi\ {\sqrt G}e^\Phi
\{R+\nabla\Phi^2+8+\nabla T^2-2T^2+{4\over3}T^3+O(m^4)
\}.$$
The corrections to $G,\Phi$ in (A.2) are of order $m^2$ because $T$ appears
in the Hilbert-Einstein equations only in the tachyon stress tensor, which
is quadratic in $T$. The $T^3$ term is ambiguous,${}^{[\tsey,\ban]}$
but this will not be important here.
It is useful to choose a gauge in which the dilaton is linear,
i.e., $\Phi^{(1)}=0$.
\foot{This is always possible at least at order $m^2$ and $m^3$.}
To $O(m^2)$,
the equations of motion are second--order differential equations:
$$\eqalign{&\nabla_\mu\nabla_\nu\Phi-{1\over2}G_{\mu\nu}
({\vec\nabla}\Phi^2+2\bx\Phi-8)=\Theta_{\mu\nu}\cr
&\bx T+{\vec\nabla}\Phi{\vec\nabla}T+2T-2T^2=0
}\eqno(A.3)$$
with tachyon stress tensor $\Theta_{\mu\nu}$.

To specify a
solution, we need two boundary conditions.
Following ref. [\polch], we will adopt boundary conditions
given (i) in the
ultraviolet
by the bare action and (ii) in the infrared
by the requirement
of regularity.
Let us for now assume the simple form
$e^{\alpha\phi}$ for the cosmological constant. `Infrared' means
$\phi\rightarrow -\infty$
since $\alpha=-\stwo<0$.

(i) UV: As pointed out in (2.11), the Liouville coordinate is bounded:
$$\hat g_{\mu\nu}\ e^{\alpha\phi}\ d\xi^\mu d\xi^\nu\ge a^2\
\Rightarrow\ \ \ \phi\le\phi_0\sim {1\over\alpha}\log a^2.$$
This bound on $\phi$ does not modify the Einstein equations.
\foot{The implicit assumption here is that the term
$\log\sqrt{ \hat g}$ in the definition of $\phi_0$ is absorbed in the
gravitational dressing of the operators. Otherwise $\phi_0$ varies
with $\hat g$ and we can no longer expect that the perturbations
are (1,1), let alone exactly marginal.}
It just requires
specifying the action at the cutoff, $S(\phi=\phi_0)$.
As in [\polch], we identify it with the unperturbed action
$S_0$ plus the bare matter interaction
($\Delta$ is the bare cosmological constant:)
$$S(\phi=\phi_0)=S_0+{1\over{8\pi}}
\int d^2\sigma\ (\Delta+m_B\cos px)\ \Leftrightarrow\ \cases{
&$T(\phi_0)=\Delta+m_B\cos px$\cr
&$G_{\mu\nu}(\phi_0)=\delta_{\mu\nu}$\cr}\eqno(A.4)$$

(ii) IR: It has been pointed out${}^{[\sei,\pch]}$ that
operators that
diverge faster than $e^{-Q/2\ \phi}$ as $\phi\rightarrow -\infty$ do not
exist in the Liouville theory (2.1).
This provides the second boundary condition. Given one solution of (A.3)
for $T^{(1)}$ and $h_{\mu\nu}$, the other solutions are obtained by adding
linear combinations of $O(m^2)$ of
the on-shell tachyons and the
two discrete gravitons
$$\cos px\ e^{(p-{\sqrt2})\phi}
,\ \ \ \cos px\ e^{(-p-{\sqrt2})\phi}
,\ \ (\partial x)^2\ \ \hbox{and}\ \
(\partial x)^2\ e^{-2{\sqrt2}\phi}.\eqno(A.5)$$
Boundary condition (ii)
means essentially
that the operators with the more negative Liouville dressing
must be dropped.
For a more precise statement, see appendix B.

So far, the discussion has been restricted to the tachyon and the graviton.
Including the discrete operators of subsection
2.3 as interactions corresponds to turning on
higher spin backgrounds in the sigma model,
and the same arguments seem to apply.
That two boundary conditions still suffice to specify a solution
is suggested by the fact that there are only two possible
Liouville dressings for each of the discrete operators of the
$c=1$ model.

Setting the bound $\phi_0=0$, we see that the operators found in
section 2 already satisfy the boundary conditions
(i) and (ii), and are thus the unique marginal perturbations.
\vfill\eject

{}$\underline{\hbox{Appendix B: The Cosmological Constant}}$

Gravitational dressings
in the presence of a cosmological constant $\mu$
can in principle be found as follows (see [\polch,\pch] for some details).
One includes the cosmological constant in the tachyon of string theory,
replacing e.g., for the Sine--Gordon model,
the ansatz (A.2) by
$$\eqalign{
&T(x,\phi)=T_\mu(\phi)+m\cos px\ f_\mu(p,\phi)
+m^2 T^{(1)}_\mu(x,\phi)
+..\cr
&G_{\alpha\beta}(x,\phi)
=\delta_{\alpha\beta}^\mu(\phi)+m^2 h_{\alpha\beta}^\mu(x,\phi)
+..}\eqno(B.1)$$
where $T_\mu$ is the cosmological constant and
$f_\mu$, ${T}_\mu^{(1)}$
and ${h}^\mu$ are the modified dressings, exact in $\mu$
order by order in $m$.
\foot{Although we cannot expand in
$\mu$, for $\mu>0$ we can expand in $m$.}
Let us assume that $m\ll\mu$, but that both are small.

First, one must find $T_\mu$ and $\delta^\mu$ exactly.
$T_\mu$
has the form of a kink centered at a free parameter $\bar\phi$,
related to $\mu$ by $\mu=e^{{\sqrt2}\bar\phi}$ and to the bare
cosmological constant $\Delta$ by $\Delta\propto\bar\phi
e^{{\sqrt2}\bar\phi}$:
${}^{[\polch]}$
$$T_\mu(\phi)=T_0(\phi-\bar\phi),\ \ \ T_0(\phi)=\cases{
1 &for $\phi\rightarrow-\infty$\cr
\propto\ \phi e^{-{\sqrt2}\phi} &for $\phi\rightarrow\infty$\cr
 }\eqno(B.2)$$
(B.2) satisfies
the boundary conditions (i), $T_\mu(0)=
\Delta$ (by definition of $\Delta$)
and (ii), $T_\mu$ does not diverge as $\phi\rightarrow-\infty$
($T\rightarrow1$). $\delta^\mu$
differs from $\delta$ because of the backreaction of the tachyon $T_\mu$ on
the metric. Like $T_\mu$, this difference
decays exponentially in the UV.

Next, one must find the dressings
$f_\mu$, ${T}_\mu^{(1)}$
and ${h}^\mu$
by solving the string equations of motion (A.3) order by order in $m$.
E.g., the tachyon equation of motion,
linearized around the
background $T_\mu$, determines
$f_\mu$:${}^{[\polch]}$
$$\{\partial_\phi^2+2\stwo
\partial_\phi+2-p^2-4T_\mu\}f_\mu=0.
\eqno(B.3)$$
Since $T_\mu$ and
$\delta^\mu$ are very small in the UV
$(\bar\phi\ll\phi<0)$,
the equations of
motion for $f_\mu$, ${T}_\mu^{(1)}$
and ${h}^\mu$
are the same as for $\mu=0$ in this regime
and the only role of the cosmological constant is
to set the second boundary condition (ii) of appendix A.
E.g., the solutions
of (B.3) in the UV are${}^{[\polch]}$
$$c_1e^{(-p-\stwo)\phi}
+ c_2e^{(+p-\stwo)\phi}
\ \propto\  e^{-\stwo\phi}\sinh(p\phi-\Theta).$$
In the IR region $\phi\ll\bar\phi$, where $T_\mu\sim$ constant,
the solution of (B.3)
that is regular
at $\phi\rightarrow -\infty$ grows exponentially.
The other, divergent solution
does not exist as an operator.
To match the solutions for
$\phi<\bar\phi$ and $\phi>\bar\phi$, one needs roughly
$\Theta\sim p\bar\phi$. In the UV, $\phi-\bar\phi$ is large,
of order $\vert\log a^2\vert$. So unless
$p$ is close to zero, $f_\mu$ is just
$e^{(p-\stwo)\phi}$ there.
Boundary condition (ii) then simply means dropping the term with
the second Liouville dressing, as without cosmological constant.
At $O(m^2)$, the same arguments can be repeated for
${T}_\mu^{(1)}$
and ${h}^\mu$.

Next, let us discuss OPE's in the presence of the cosmological constant.
In free field theory, the OPE of two operators
with Liouville momenta $\alpha,
\beta$ would produce an operator with Liouville momentum
$\alpha+\beta$. But in Liouville theory
momentum is not conserved because of the
exponential potential. Also, if $\alpha+\beta<-Q/2$, the operator
$e^{(\alpha+\beta)\phi}$ does not exist.
Instead, new primary fields $V_\sigma=e^{-Q/2\ \phi}\sin(\sigma\phi+\Theta)$
will be produced, with some weight $f(\sigma)$ and
less singular coefficients:${}^{[\pch]}$
$$\eqalign{
e^{\alpha\phi(r)}\ e^{\beta\phi(0)}\ \sim\int^\infty_0{d\sigma}
&\vert r\vert^{-2\alpha\beta+(\alpha+\beta+Q/2)^2+\sigma^2}
f(\sigma)\ V_\sigma\cr \hbox{instead of\hskip 2cm}
&\vert r\vert^{-2\alpha\beta}\ e^{(\alpha+\beta)\phi}.}\eqno(B.4)$$
For the Sine--Gordon model, this modification of the OPE's seems to cure
the problem of new $\vert r\vert^{-2}$
singularities that would naively appear in (2.7)
below $p={1\over2}\stwo$. They would give rise to unwanted
counterterms like $\cos 2px$ at $p={1\over4}\stwo$ and
$(\partial x)^2 \cos 2px$ at $p=0$. The
modified OPE of $\cos px\ e^{\epsilon\phi}$ with itself produces
$$
\int {{d\sigma}\over{2\pi}}f(\sigma)
\vert r\vert^{-2+4p^2+\sigma^2}\cos 2px\
V_\sigma(\phi)\ +...
\eqno(B.5)$$
Except for the (negligible) case $p=\sigma=0$,
all singularities are milder than quadratic.

\vfill\eject


\leftline{{\bf PART III:
A 4D ANALOG OF 2D GRAVITY}
\foot{based on a paper published in Nuclear Physics B 390, 188 (1993)}}
\vskip1cm

{\leftline{\bf 1. Introduction}}

In parts I and II, a theory of two--dimensional quantum gravity in
conformal gauge has been developed.
It is natural to ask how much can be learned from this
about four--dimensional quantum
gravity. Here the following answer will be given:
A natural analog of two--dimensional gravity is four--dimensional
gravity with the action
$$S=\int_M d^4x\ {\sqrt g}\{\lambda+\gamma R + \eta R^2 + \rho
W^2\}\eqno(1.1)$$
in the limit of infinite Weyl coupling, $$\rho\rightarrow\infty.\eqno(1.2)$$
Here, $M$ is a manifold of fixed topology,
$\lambda$ and $\gamma$ are the cosmological and the inverse
Newtonian constants, $R$
is the Ricci--scalar, and $W$ is
the Weyl tensor, the traceless part of the Riemann tensor:
$$W_{\mu\nu\sigma\tau}=R_{\mu\nu\sigma\tau}
-{1\over 2}(g_{\mu\sigma}R_{\nu\tau}+g_{\nu\tau}R_{\mu\sigma}
-g_{\mu\tau}R_{\nu\sigma}-g_{\nu\sigma}R_{\mu\tau})
+{1\over6}(g_{\mu\sigma}g_{\nu\tau}-g_{\mu\tau}g_{\nu\sigma})R.$$

It will be seen that in the limit (1.2) the path
integral over the metric reduces to an integral over the conformal
factor and a moduli space, as in two dimensions.
As a consequence, most of
the developments described in part I and II have four--dimensional analogs.
In particular, the analog of the $c=1$ barrier of two--dimensional gravity
will be derived below, as well as scaling
laws that can be compared with computer simulations.
On the other hand, all the important
features in four--dimensional gravity that go
beyond those present in the limit
$\rho\rightarrow\infty$
do not seem to have two--dimensional analogs.

Apart from the fact that the theory (1.1) in the limit
$\rho\rightarrow\infty$ can be studied
with the methods of part I and II, is it of any interest otherwise?
(1.1) is the most general local, renormalizable action of four--dimensional
gravity, up to topological invariants.
One
interesting aspect of the limit (1.2), first
pointed out by Fradkin and Tseytlin,${}^{[\ft]}$ is that, at least for
$\eta=0$, it is an ultraviolet fixed point of (1.1).
It could thus be viewed as a ``short--distance phase'' of
fourth--order derivative gravity.

But of course there is a well--known ghost problem common to all
fourth--order derivative actions like (1.1):
we can rewrite them in terms of new fields with two
derivatives only, but some of them will have the wrong
sign in the kinetic term. With Minkowskian signature this leads
to nonunitarity.
For this reason, let us consider
(1.1) only in Euclidean space, as a (still interesting)
statistical mechanical model of quantum geometry, or ``three--branes.''
In the future it will hopefully be possible to extract information
about the physically interesting case in Minkowski space,
$\rho=\eta=0$, by means of a $1/\rho$--expansion.

In section 2 we define (1.1) in the limit (1.2)
rigorously by introducing a Lagrange multiplier $p$. That is, we begin
by studying
the path integral
$$\int Dg\ Dp\ \exp\{- \int_{M} d^4 x{\sqrt g}(\lambda +
 \gamma R +\eta R^2 +ip W_+)\}.\eqno(1.3)$$
In section 6 we show that this theory is the limit (1.2) of (1.1).
Here, $W_{\pm}$ is the (anti-) self-dual part of the Weyl tensor
$$W_{\pm\ \mu\nu\sigma\tau}\equiv {1\over 2}
( W_{\mu\nu\sigma\tau} \pm
 {1\over 2}\epsilon_{\mu\nu}^{\ \ \ \alpha\beta} W_{\alpha\beta\sigma\tau}).$$
The Lagrange multiplier $p$ is a $4^{th}$ rank self-dual tensor
field which (like $W_+$) transforms as a (2,0) representation of
the Euclideanized Lorentz group $SO(4)\sim SU(2)\times SU(2)$,
i.e., like a spin 2 field.

In section 2, (1.3) will be
rewritten as an integral over a moduli space
and over the conformal factor
$\phi$, with a few determinants
in this gravitational background. The moduli space is that of conformally
self--dual metrics and plays a role analogous to the moduli space of
Riemann surfaces in two--dimensional gravity.

As in two dimensions,
the determinants can be decoupled from $\phi$
by introducing a 4D analog of the Liouville action.
Its form has recently been found in [\ma]. It consists of a free $4^{th}$ order
derivative piece (essentially $\phi\box^2\phi$) plus pieces that renormalize
$\lambda, \gamma$ and $\eta$ in (1.3), as explained in section 3.
The proposal of DDK, explained in part I, is generalized to four dimensions
in section 4.
The cosmological constant, the Hilbert-Einstein term and the $R^2$
term each become exactly marginal operators of the new theory,
but so far I have explored this only to lowest order.

In section 5
the fixed volume and fixed average curvature partition functions
and the correlation
functions of local operators in their dependence on the cosmological constant
are derived, as has been done in two dimensions. It would be very interesting
to explore whether the condition $W_+=0$ can be imposed in
computer simulations of random triangulations, or whether -- equivalently --
the limit (1.2) can be taken. Then the predictions (5.11)
could be compared with ``experiment." The analog of the $c=1$ barrier
is also given, in (5.3). In contrast with two dimensions, it is not
crossed by adding too much matter.

In section 6, the theory, which we call ``conformally self--dual
gravity,'' is discussed as the limit (1.2) of (1.1).
It is also suggested that conformally self-dual gravity is
connected with four dimensional topological gravity,${}^{[\wiw]}$
as in the two dimensional case.${}^{[\wv]}$

Part of our analysis is concerned with the four
dimensional analog of the Liouville action and of DDK.
In a different context the induced action for the conformal factor
and its renormalization have also been studied recently
by Antoniadis
and Mottola.${}^{[\ma]}$
I have used some of their calculations.
However, when I discuss the four--dimensional analog of DDK's
method of decoupling the conformal factor from its measure, my treatment
and my conclusions will differ from those of [\ma]. I will state the main
differences.
\eject

{\leftline{\bf 2. Conformal Gauge}}

The Lagrange multiplier $p$ in (1.3) restricts
the path integral over $g$
to conformally self-dual metrics, i.e., metrics with $W_+=0$.
$W_+$ has five independent components and the condition $W_+=0$ is Weyl
and diffeomorphism invariant. So, up to a finite number of moduli,
the five surviving components of the metric are the
conformal factor and the diffeomorphisms.
Let $m_i$ parametrize the moduli space of conformally self-dual
metrics modulo diffeomorphisms $x \rightarrow x+\xi$ and Weyl
transformations $g \rightarrow g e^\phi$. Let us
fix a representative $\hat g (m_i)$ via, say, the condition
$\hat R=0$ and Lorentz gauge
$\partial^\mu \hat g_{\mu\nu}=0$,
and let us pick a conformally self dual metric
$$g_0 = (\hat g(m_i) e^\phi )^\xi,\eqno(2.1)$$ where $\xi$ indicates
the action of a diffeomorphism.
At $g_0$ we can split up a fluctuation of
$g$: $$\delta g_{\mu\nu} = g_{0\mu\nu}\delta \phi
+ \nabla_{(\mu}\delta\xi_{\nu)}+\delta \bar h_{\mu\nu}.$$
The four $\delta\xi$'s generate infinitesimal diffeomorphisms and the
five $\bar h_{\mu\nu}$ parametrize the space of metrics
perpendicular to $\xi$,  $\phi$ and the moduli, i.e., perpendicular
to the conformally self--dual ones.
The measure for $g$ is defined, in analogy to two dimensions${}^{[\poy]}$, by
$$\Vert \delta g \Vert ^2 \equiv \int d^4 x {\sqrt g}(4
(\delta \phi+{1\over 2}\nabla^\mu \delta\xi_\mu)^2
 + (L\delta\xi)^2 + (\delta \bar h)^2)\eqno (2.2)$$ with
$$(L\delta\xi)_{\mu\nu} \equiv \nabla_{(\mu} \delta\xi_{\nu )} -
 {1 \over 2}g_{\mu\nu} \nabla^\rho \delta\xi_\rho.\eqno(2.3)$$
Apart from restricting the path integral, integrating out $p$ and $\bar h$
in (1.3) will contribute the determinant
$$\det(O^\dagger O)_
 {g_0}^{-{1\over 2}}\eqno(2.4) $$
where $O^\dagger$ is the linearized $W_+$-term
$$(O^\dagger_{g_0} \bar h)_{\mu\nu\sigma\tau} \equiv\lim_{\epsilon
\rightarrow 0} {1\over \epsilon}( W_{+\ \mu\nu\sigma\tau}
[g_0 +\epsilon \bar h] -W_{+\ \mu\nu\sigma\tau}[g_0]),\eqno(2.5)$$
$O$ is its adjoint and $O^\dagger O$ is a $4^{th}$
order, conformally invariant,
linear differential operator in the curved background $g_0$, acting on $p$
of (1.3).

We are left with an integral over the conformal
equivalence class of each $\hat g$.
 {}From (2.2) it is seen that changing variables from $g$ to $\phi$
and $\xi$ in this equivalence class leads to a Jacobian
$$(\det L^\dagger L)_g^{1\over 2},$$
where the zero modes of the operator $L$, defined in
(2.3), have to be projected out.
After dropping the integral over the diffeomorphism group $D\xi$ (since,
in the absence of other gauge field backgrounds,
gravitational anomalies can occur only in $4k+2$ dimensions${}^{[\awi]}$),
the path integral (1.3) reduces
to an integral over the
moduli space of conformally self-dual metrics and $\phi$:
$$ \int \prod_i dm_i\ D\phi\
(\det O^\dagger O)^{-{1\over 2}}_{\hat g e^\phi}\
(\det L^\dagger L)^{1\over 2}_{\hat g e^\phi}\
(\det \triangle)^{-{1\over 2}}_{\hat g e^\phi}\
e^{-\int d^4 x {\sqrt g}(\lambda+\gamma R+\eta R^2)}, \eqno (2.6)$$
where free, conformally invariant matter fields have been added to the
theory for generality, and their partition function has been
denoted by $\det(\triangle)^{-{1\over 2}}$. Despite the notation, let us
allow the matter to be fermions,
Yang-Mills fields, etc., as well as conformally coupled scalars.

The moduli space of conformally self-dual metrics
is a very interesting subject by itself
which will not be discussed here. On the four sphere
its dimension is zero: all conformally self-dual metrics on $S^4$
are conformally flat.
On $K^3$, its dimension is~57.${}^{[\egu]}$

\eject

{\leftline{\bf 3. Liouville in 4D}}

Let us now decouple the determinants in (2.6)
from $\phi$. For conformally invariant differential operators $X$:
\footnote*{More precisely, if $X\equiv M^\dagger M$,
$M$ has to transform as $e^{p\phi}Me^{q\phi}$ under
$g\rightarrow ge^\phi$.}
$$\det X_{\hat g e^\phi} = \det X_{\hat g} e^{-S_i[\hat g,\phi]}\eqno(3.1)$$
where the induced action $S_i$ is obtained
by integrating the trace anomaly of the stress tensor ${}^{[\bd]}$
$$-2{{\delta S_i[\hat g,\phi]} \over{ \delta \phi}} ={\sqrt g}
<T^\mu _\mu> = {1\over{16\pi^2}}{\sqrt g}[a(F+{2 \over 3} \box R) + bG
] \ - 4\lambda'{\sqrt g} - 2\gamma'{\sqrt g} R \eqno (3.2)$$
where $$F=W_+^2+W_-^2$$ is the square of the Weyl tensor. (3.2) has,
apart from the divergent parameters $\lambda'$ and $ \gamma'$,
two finite parameters $a,b$.
${\sqrt g}G$ is the Gauss-Bonnet
density,
$$G=R^{\mu\nu\sigma\tau}R_{\mu\nu\sigma\tau}-4R^{\mu\nu}R_{\mu\nu}+R^2 ,$$
whose integral over the manifold
is proportional to the Euler characteristic.
Following Antoniadis and Mottola,${}^{[\ma]}$ (3.2) can
easily be integrated by noting that with $g=\hat ge^\phi$ the combination
$${\sqrt g}(G-{2\over 3}\box R)={\sqrt {\hat g}}\hat M
 \phi + {\sqrt{\hat g}}(\hat G-{2\over 3}\hatbox\hat R)$$
is only linear in $\phi$ with the fourth--order differential operator
$$\eqalign{\hat M &\equiv 2\hatbox^2 + 4\hat R^{\mu\nu}\hat\nabla_\mu\hat
\nabla_\nu-{4\over 3}\hat R\hatbox+{2\over 3}
(\hat\nabla^\mu\hat R)\hat\nabla_\mu\cr
&= 2\hatbox^2 + 4\hat R^{\mu\nu}
\hat\nabla_\mu\hat\nabla_\nu\ \ \hbox{if}\ \ \hat R=0\cr
&= 2\hatbox^2 \ \ \hbox{if}\ \ \hat g_{\mu\nu}
=\delta_{\mu\nu} e^{\phi_0}.}\eqno(3.3)$$
${\sqrt g} F$ is independent of $\phi$ and ${\sqrt g}\box R$
is just the variation of the $R^2$ action with respect to $\phi$.
So the four--dimensional analog
of the Liouville action consists of a free part plus
a cosmological constant term, a Hilbert-Einstein term and an $R^2$ term:

$$S_i[\hat g,\phi] =
 {-b\over{32\pi^2}}S_0[\hat g,\phi]+{-a\over{32\pi^2}}S_1[\hat g,\phi]
 +{{a+b}\over{72\pi^2}}S_{R^2}+\gamma'S_R+\lambda'S_{c.c.},\eqno(3.4)$$ where
$$\eqalign{
S_0[\hat g,\phi] &= \int d^4 x {\sqrt{ \hat g}}
[ {1 \over 2} \phi \hat M \phi+(\hat G -{2\over 3}\hatbox\hat R)\phi]\cr
S_1[\hat g,\phi] &= \int d^4 x {\sqrt{ \hat g}} \hat F \phi\cr
S_{c.c} &= \int d^4x \sqrt{ \hat g} e^{2\phi}\cr
S_{R}&= \int d^4 x \sqrt{ \hat g} e^{\phi} [\hat R -{3\over 2}
(\hat \nabla \phi)^2 -3 \hatbox \phi]\cr
S_{R^2}&= \int d^4 x \sqrt{ \hat g}
[\hat R -{3\over 2}(\hat \nabla \phi)^2 -3 \hatbox \phi]^2.}\eqno(3.5)$$
$b$ will turn out to be negative for `normal' operators $X$.
$\gamma', \lambda' $ and ${{a+b}\over{72\pi^2}}$
just renormalize $\gamma, \lambda $ and $\eta$ in (2.6).
A $\phi$-independent local term
$${-\int d^4 x\sqrt {\hat g}}({{a+b}\over{72\pi^2}}\hat R^2 +\gamma'
 \hat R+\lambda')\eqno(3.6)$$
has been omitted in (3.4) and will frequently be omitted in the following.
If it is included, we see from (3.1) that for some action $S_j$:
$$S_i[\hat g,\phi]=S_j[\hat g e^\phi]-S_j[\hat g].\eqno(3.7)$$
${-b\over{32\pi^2}}S_0$ is the 4D analog of the 2D action
$$S_{2D}={c\over {48\pi}}\int d^4 x
 {\sqrt {\hat g}}({1\over 2}\phi\hatbox\phi-\hat R\phi).$$
If $\hat g = \tilde g e^{\phi_0}$ for any $\tilde g$, $S_0$ can be written:
$$S_0[\hat g,\phi]=\int d^4x {\sqrt {\tilde g}}{1\over 2}
[(\phi+\phi_0)\tilde M(\phi+\phi_0)-\phi_0\tilde M\phi_0].\eqno(3.8)$$
Adding up the anomaly coefficients in (3.2) for
$(\det O^\dagger O)^{-{1\over2}},
(\det L^\dagger L)^{+{1\over2}},(\det\Delta)^{-{1\over2}}$,
$$A_0\equiv a_O +a_L+a_{mat}\hskip 1in
 B_0\equiv b_O+b_L+b_{mat},\eqno(3.9)$$
(2.6) can now be rewritten as

$$\eqalign{\int \prod_i dm_i\ &\chi(m_i)
\int D\phi\ e^{{B_0\over{32\pi^2}}S_0[\hat g ,\phi]+{A_0\over{32\pi^2}}
[\hat g, \phi]-\eta_1 S_{R^2}-\gamma_1S_R-\lambda_1S_{c.c.}}\cr
&\chi(m_i)\equiv
(\det O^\dagger O)^{-{1\over 2}}_{\hat g(m_i) }\
(\det L^\dagger L)^{1\over 2}_{\hat g(m_i)}\
(\det \triangle)^{-{1\over 2}}_{\hat g(m_i) }} \eqno(3.10)$$
$\chi(m_i)$ is now purely a function of the moduli $m_i$, once
we have fixed a representative $\hat g(m_i)$ for each point in moduli space.

The coefficients $a$ and $b$ in (3.2) are:
${}^{[\bd,\ft,\bvi,\amm]}$

\line{\hfill $120\ a$\hskip1.5cm $-360\ b$\hskip2cm}
\vskip 1mm\line{\hskip5mm conformally coupled
scalars ($\triangle \sim\box -{1\over 6}R$):\hfill
1\hskip 2.3cm 1\hskip 2.5cm}
\line{\hskip.5cm spin ${1\over 2}$ (four component) fermions:\hfill
6\hskip 2.2cm 11\hskip 2.4cm}
\line{\hskip .5cm massless gauge fields:\hfill
12\hskip 2.1cm 62\hskip 1.35cm(3.11)}
\line{\hskip.5cm $(\det O^\dagger O)^{-1/2}(\det L^\dagger L)^{1/2}$:\hfill
796\hskip 1.7cm 1566\hskip 2.25cm}
\line{\hskip.5cm $M\sim 2\box^2+..$ of (3.3):\hfill
--8\hskip 1.9cm --28\hskip 2.4cm}
\vskip 5mm

Note that the fourth--order derivative induced action
makes the theory power counting renormalizable, and also bounded if
$b<0$.
The price is the existence of a ghost,
the general problem of fourth--order derivative actions mentioned in
the introduction.
It has been suggested in [\ma], where the theory was studied
in Minkowski space, that the reparametrization
constraints $T_{\mu\nu}\sim 0$ eliminate these ghosts from the physical
spectrum, as they do in two dimensions.${}^{[\gsw]}$
This would be very interesting to verify.
\vskip1cm

{\leftline{\bf 4. DDK in 4D}}

Let us now focus
on the $\phi$ integral over the conformal equivalence class of $\hat g$:
$$Z[\hat g]\equiv\int D_{\hat ge^\phi}\ \phi
\ e^{{B_0\over{32\pi^2}}S_0[\hat g ,\phi]+{A_0\over{32\pi^2}}
S_1[\hat g, \phi]-\eta_1 S_{R^2}-\gamma_1S_R-\lambda_1S_{c.c.}}\eqno(4.1)$$
where the dependence of
$Z$ on $\eta_1,\gamma_1$ and $\lambda_1$ has been suppressed.
In $D_{\hat ge^\phi}\phi$ it is indicated that the measure
for $\phi$ depends on $\phi$ itself, namely in two ways:
First, the metric itself must be used to define a norm in the space of metrics:
$$\Vert\delta\phi\Vert^2\equiv\int d^4 x{\sqrt g}
\ (\delta\phi(x))^2=\int d^4x{\sqrt{\hat g}}e^{2\phi(x)}(\delta\phi(x))^2.$$
Second, in order to define a short distance cutoff one should also
use the metric $\hat ge^\phi$ itself: the cutoff fluctuates with the field.
Let us follow David, Distler and Kawai ${}^{[\d],[\dk]}$ and assume
that the $\phi$ dependence of the measure in (4.1) can be absorbed
in a local renormalizable action:
$$D_{\hat g e^\phi}\ \phi\ e^{-S_i[\hat g,\phi]} =
D_{\hat g }\ \phi\ e^{-S_{{loc}}[\hat g,\phi]},\eqno(4.2)$$
where now on the right-hand side
$$\Vert\delta\phi\Vert^2\equiv\int d^4 x{\sqrt{\hat g}}(\delta\phi(x))^2$$
and the cutoff no longer fluctuates.

What is $S_{{loc}}$? Although the $\phi$ dependence of the measure
in (4.1) looks inconvenient, we do learn something important
from (4.1): simultaneously changing
$$\hat g_{\mu\nu}\rightarrow\hat g_{\mu\nu}e^{\phi_0},
\hskip 1in \phi\rightarrow\phi-\phi_0$$
does not change the measure or $S_{R^2},S_R,S_{c.c.}$.
It does change the induced action. From (3.7) we see (reinstating the $\phi$
independent terms (3.6) into (4.1)):
$$S_i[\hat g,\phi]\rightarrow S_i[\hat g,\phi]-S_i[\hat g,\phi_0].$$
We conclude that
$$Z[\hat ge^{\phi_0}]=Z[\hat g]\ e^{S_i[\hat g,\phi_0]}\eqno(4.3)$$
and the $\phi$ theory behaves as if it
were a conformal field theory with conformal anomaly (3.2) given by
$a=-A_0,\ b=-B_0$. This is, of course, precisely what is needed
in order to insure that the background metric is really a fake:
if we vary it,
$\hat g_{\mu\nu}\rightarrow\hat g_{\mu\nu}e^{\phi_0}$, the variation
of the determinants in (3.10) is determined by their total conformal
anomalies $+A_0,\ +B_0$, defined in (3.9), and that just
cancels the $-A_0,\ -B_0$ from the $\phi$ theory.

So let us replace (4.1) as in (4.2) by a four--dimensional
conformal field theory with conformal anomaly given by
$$a=-a_O-a_L-a_{{mat}},\hskip 1in b=-b_O-b_L-b_{{mat}}.\eqno(4.4)$$
I will propose -- and justify in a moment --
that, as in two dimensions, $S_{loc}$ in (4.2) is again the induced action
with modified coefficients $A,B$ and modified interactions:
$$Z[\hat g]\sim\int D_{\hat g}\ \phi\ e^{{B\over{32\pi^2}}
S_0[\hat g ,\phi]+{A\over{32\pi^2}}S_1[\hat g, \phi]-\eta_2\hat S_{R^2}
-\gamma_2\hat S_R-\lambda_2\hat S_{c.c.}},\eqno(4.5)$$
where $\hat S_{R^2}, \hat S_R$, and $ \hat S_{c.c.}$ are marginal
operators of the free theory given by $S_0$ and $S_1$ and will be discussed
below.

The free theory ($\eta_2, \gamma_2, \lambda_2 =0$) of (4.5) has conformal
anomaly $$a=-A+a_M,\hskip 1in b=-B+b_M,\eqno(4.6)$$
where $M$ is the operator (3.3). This can be seen as follows: Setting
$\hat g_{\mu\nu}=\tilde g_{\mu\nu}e^{\phi_0}$ we see from (3.8):
$$\eqalign{\int D_{\hat g}\ \phi\ &e^{{B\over{32\pi^2}}
S_0[\hat g ,\phi]+{A\over{32\pi^2}}S_1[\hat g, \phi]}\cr
&=\int D_{\hat g}\ \phi\ e^{\int d^4x {\sqrt{\tilde g}}
\{{B\over{64\pi^2}}[(\phi+\phi_0)\tilde M(\phi+\phi_0)-\phi_0\tilde M
\phi_0] +{A\over{32\pi^2}}[\tilde F(\phi+\phi_0)-\tilde F\phi_0]\}}\cr
&=e^{-{B\over{32\pi^2}}S_0[\tilde g ,\phi_0]-
 {A\over{32\pi^2}}S_1[\tilde g, \phi_0]}
\int D_{\hat g}\ \phi\ e^{\int d^4x ({\sqrt{\hat g}}
 {B\over{64\pi^2}}\phi\hat M\phi +{A\over{32\pi^2}}\hat F\phi)}
}\eqno(4.7)$$
by shifting $\phi\rightarrow\phi+\phi_0$ and using the fact
that ${\sqrt g}M$ and ${\sqrt g}F$ are conformally invariant.
So $-A,-B$ are the ``classical" contributions\footnote*{Here
and below I will call these contributions ``anomalies,"
although they actually arise from the fact that the action is classically
not quite conformally invariant.} to (4.6) and $a_M,b_M$ are
the quantum contributions from $M$. Therefore we see from (4.4)
that the
ansatz (4.5) is consistent if
$$
A=a_O+a_L+a_{{mat}}+a_M\hskip 1in
B=b_O+b_L+b_{{mat}}+b_M.
\eqno(4.8)$$

How do we know that $a_M,b_M$ do not depend on the moduli $m_i$?
The only local scale invariant quantity they could depend on
is $\int d^4x{\sqrt{\hat g}}\hat F$,
which is a topological invariant in the case of $W_+=0$.

Why does the free part of $S_{{loc}}$ in (4.2) have to be of the
form of the free part of the induced action $S_i$ again? One can plausibly,
though not rigorously, argue as follows:
there are two ways to obtain the right effective action (4.3) via (4.2);
(a) $S_{{loc}}$ is classically conformally invariant
and $a,b$ come purely from the quantum anomaly or
(b) the ``classical"
variation of $S_{{loc}}$ is of the form of the induced action $S_i$.
In case (a) $a$ and $b$ would be just numbers
that will in general not cancel the anomalies as needed
in (4.4) (multiplying $S_{{loc}}$ by a factor would then not change
the conformal anomaly). Only in case (b) there
are parameters like $A,B$ in $a,b$ that can be adjusted to satisfy (4.4). But
the only local free action whose ``classical"
variation is the induced action is the induced action itself.

Let us now turn to the operators $\hat S_{R^2},
 \hat S_R$ and $\hat S_{c.c.}$ in (4.5). The consistency conditions
of invariance under rescaling of the background metric
(in particular that the theory is at a renormalization group fixed
point) mean that the integrands of
$\hat S_{R^2}$, the ``dressed" Hilbert-Einstein
action $\hat S_R$, and the ``dressed" cosmological constant
$\hat S_{c.c.}$ must be locally scale invariant operators.
Let us try the ansatz
$$\eqalign{\hat S_{c.c}&=\int d^4 x{\sqrt{\hat g}} e^{2\alpha\phi}\cr
\hat S_R&=\int d^4 x{\sqrt{\hat g}} e^{\beta\phi}(\hat\nabla\phi)^2+...\cr
\hat S_{R^2}&=\int d^4 x{\sqrt{\hat g}} (\hat\nabla\phi)^4+...\cr
}\eqno(4.9)$$
with $\alpha, \beta, $ and ``..."
determined so that the integrands of (4.9) are scaling operators of
conformal dimension $4$,
to cancel the $-4$ from ${\sqrt{\hat g}}$.
In the language of string theory, they
are vertex operators of our theory of
noncritical three branes. All of them should be moduli deformations,
if the background metric $\hat g$ is really fictitious.
So far I have verified this
only for $\hat S_{c.c.}$.
The ``..." includes possible corrections of order
$\eta_2,\gamma_2,\lambda_2$ that
may be needed in order to keep the other operators marginal
as we move away from $\eta_2,\gamma_2,\eta_2=0$.
Some calculations with
$\hat S_{R^2}, \hat S_R$ and $\hat S_{c.c.}$
can also be found in [\ma] (however $\alpha=\beta$ there).

To calculate the (classical plus anomalous)
dimension of $e^{2\alpha\phi}$ with action (4.5) at $\eta_2,\gamma_2,
\lambda_2=0$ one may go to conformally flat $\hat
 g_{\mu\nu}=e^{\phi_0}\delta_{\mu\nu}$ where $\hat M=2\hatbox^2$ and $S_1=0$.
Because of the shift $\phi+\phi_0\rightarrow\phi$ in (4.7), the condition
$$\dim(e^{2\alpha\phi})=4 \ \ \hbox{with action}\ \sim\ S_0 \ \hbox{(given
in (3.5))}$$
is equivalent to the condition
$$\dim(e^{2\alpha\phi})=4-4\alpha\ \ \hbox{with action}\ \sim
\ \int d^4x\ \phi\box^2\phi.\eqno(4.10)$$
Due to the quartic propagator, this four--dimensional theory is formally
very similar to an ordinary free scalar field theory in two dimensions.
In particular, $:e^{2\alpha\phi}:$ will be a scaling operator.
Its dimension in (4.10) is now purely anomalous.
It is found from the two-point function
$$<e^{2\alpha\phi(x)}e^{-2\alpha\phi(y)}>\sim e^{-4\alpha^2 \Delta(\vert
x-y\vert)}\sim\vert x-y\vert^{-{8\alpha^2}\over B},\eqno(4.11)$$
where the propagator
$$\Delta(r)={2\over B}\log r\ \ \hbox{with}\ \
-{B\over{16\pi^2}}\box^2 \Delta(r)=\delta(r)$$
of the free theory
has been used. Thus, $\dim(e^{2\alpha\phi})={{4\alpha^2}\over B}$, and
(4.10) becomes:
$$4-4\alpha={{4\alpha^2}\over B}
.\eqno(4.12)$$
This determines $\alpha$ once $B$ is known. See section 5 for the
numerical discussion.
Similarly, $\beta$ in $\hat S_R=\int{\sqrt{\hat g}}e^{\beta\phi}
(\hat\nabla\phi)^2+...$ is determined by requiring $e^{\beta\phi}$ to have
dimension 2:
$$2-2\beta={{\beta^2}\over B}
.\eqno(4.13)$$
$\alpha$ and $\beta$ are independent of the moduli $m_i$,
for the same reason that
$a_M,b_M$ are.

The result for the dimension of the operator $e^{p\phi}$ agrees with the
result of ref. [\ma] ($Q^2$ of [\ma] is $-2B$).
Let me note two points of disagreement with
ref. [\ma].

First, in [\ma] the theory of the conformal factor was studied as
a `minisuperspace' theory,
rather than as gravity with a self--duality constraint. Further,
it was suggested that this was the relevant description of gravity
in the IR. But we will note below, following ref. [\ft],
that the theory is a UV--, not an IR--fixed
point of Weyl gravity. This justifies the use of the
minisuperspace approximation
in~the~UV, but~not~in~the~IR.

Second, we have found independent values of $\alpha$ and $\beta$ in (4.9).
In ref. [\ma]
it was assumed that $\alpha$ had to be equal to $\beta$,
with the conclusion that the operators $\hat S_R$ and
$\hat S_{c.c.}$ could not both be present at the fixed point.
This led to a suggestion about the cosmological constant problem,
and it implies different critical coefficients and a different
value of the analog of the `$c=1$
barrier' of two--dimensional gravity.

As in two
dimensions, if $\Phi_i$ is a scaling operator
of the matter theory with conformal dimension $\Delta_i$, the operator
$$O_i \equiv \int d^4 x \sqrt {\hat g} e^{\gamma_i \phi} \Phi_i$$
with $\gamma_i$ determined analogously to (4.12) by
$$4-2\gamma_i={{\gamma_i^2}\over B}+\Delta_i
\eqno(4.14)$$
is a marginal operator that can be added to the action, at least
infinitesimally.

Provided that truly marginal operators $\hat S_{R^2},\hat S_R$
can also be found, we can now rewrite (3.10) as
$$\eqalign{\int \prod_i dm_i\
&(\det O^\dagger O)^{-{1\over 2}}_{\hat g }\
(\det L^\dagger L)^{1\over 2}_{\hat g}\
\int D_{\hat g} x\ D_{\hat g}\phi\
e^{-S_{mat}[\hat g,x]-S[\hat g, \phi]},\cr
S[\hat g, \phi]&= {-B\over{32\pi^2}}S_0[\hat g ,\phi]+{-A\over{32\pi^2}}
S_1[\hat g, \phi]\cr &+\eta\hat S_{R^2}[\hat g,\phi]+\gamma\hat
S_R[\hat g,\phi]+\lambda\hat S_{c.c.}[\hat g,\phi]
.}\eqno(4.15)$$
The subscripts on $\eta,\gamma,\lambda$ have been dropped.
(4.15) describes free fields plus marginal interactions
in a gravitational instanton background. $A,B$ are given by (4.8) and (3.11)
and $S_0,S_1,\hat S_{R^2},\hat S_R,\hat S_{c.c.}$ by (3.5),
(4.9), (4.12) and (4.13).
\vskip1cm

{}\hskip-7mm{\bf 5. Results}\footnote*{The
earlier version of this work did not contain
numerical results since I did not know the conformal anomaly
coefficients of the O--L--determinant (fourth line in (3.11)).
After it appeared it was pointed out in reference [\amm] that
they had been computed in [\ft] and they were independently
confirmed. Subsequently part of this section was added.}

We can now make some numerical
predictions: From (3.11) and (4.8),
$$A={1\over {120}}(N_0+6N_{1\over2}+12N_1+788),\hskip .4in
B=-{1\over{360}}(N_0+11N_{1\over2}+62N_1+1538)\eqno(5.1)$$
where $N_0,N_{1\over2},N_1$ are the number of conformally coupled
scalars, spin ${1\over2}$ fermions and massless gauge fields.
(4.12), (4.13) and (4.14) become
$$\eqalign{2\alpha&=-B-\sqrt{B^2+4B},\ \cr
\beta&=-B-\sqrt{B^2+2B}\ ,\ \cr
\gamma_i&=-B-\sqrt{B^2+(4-\Delta_i)B}.}\eqno(5.2)$$
Thus $\alpha$ will be real if $B\ge 0$ or $B\le -4$.
The second constraint is the relevant one since $B$ is negative.
The reality constraint $B\le \Delta_i-4$ on $\gamma_i$
is weaker than the one for $\alpha$ in (5.2) as long as we allow
only operators with positive dimension $\Delta_i$.
The signs in front of the square roots have been picked to give the correct
results $\alpha=\beta=1,\gamma_i=2-{\Delta_i\over 2}$
in the classical limit $B\rightarrow -\infty$.

To compare with two dimensional gravity, define the anomaly coefficient
$\tilde c\equiv -360\ b$ as in (3.11).
$B\le-4$ becomes
$$\tilde c_{mat}+\tilde c_L+\tilde c_O+\tilde c_M\ge 1440
\rightarrow \tilde c_{mat}\ge -98.\eqno(5.3)$$
The analogous restriction in two dimensions is $c_{mat}\le 1$, where
$c_{mat}$ is the matter central charge.
If the cosmological
constant term is absent, the barrier for $\tilde c$, rather than being --98,
is determined by the lowest dimension
operator.
We see that in pure gravity $\alpha$ is
real.
In contrast with two dimensions, the situation is improved by
adding conventional matter like conformally
coupled scalar fields, families of fermions or gauge fields.
The $\tilde c=-98$ barrier would only be crossed by adding exotic matter
with
positive anomaly coefficient $b$ in (3.2).

Let us now derive scaling laws by studying
the integral over the constant mode of $\phi$, as is done in two
dimensions.${}^{{[\dk,\sei]}}$
The fixed volume partition
function at the critical point $\eta,\gamma,\lambda\sim 0$ is defined as
$$Z(V) \equiv \int \prod_i dm_i\ \chi(m_i) \int D_{\hat g} \phi
\ e^{{B\over{32\pi^2}}S_0[\hat g,\phi]+{A\over{32\pi^2}}
S_1[\hat g,\phi]}\ \delta(\int \sqrt{\hat g} e^{2\alpha \phi} - V)\eqno(5.4)$$
with $\chi(m_i)$ as in (3.10).
Under the constant shift $\phi\rightarrow \phi+c$ we see from (3.5):
$$\eqalign{ \delta S_0&=c\int d^4x{\sqrt{\hat g}}
\hat G =32\pi^2c\chi\cr
\delta S_1&=c\int d^4x{\sqrt{\hat g}}\hat F =-48\pi^2c\tau,}\eqno(5.5)$$
where the topological invariants $\chi$
and $\tau$ are the Euler characteristic and  signature of the manifold
($W_+=0$ here):${}^{[\ati,\bd]}$
$$\tau= {1\over {48\pi^2}}\int d^4 x {\sqrt g} (W_+^2 - W_-^2)
\qquad\hbox{and}\qquad \chi={1\over{32\pi^2}}\int d^4 x {\sqrt g} G.
$$
 {}From this it follows that
$$\eqalign{Z(V)&=e^{(-2\alpha+B\chi-{3\over 2}A\tau)c}Z(e^{-2\alpha c}V)\cr
\rightarrow\ \
Z(V) &\sim V^{-1+{1\over{4\alpha}}
(2B\chi-3A\tau)}.
}\eqno(5.6)$$
E.g., for the four sphere ($\chi=2,\tau=0$),
$$Z(V)\sim V^{-1+{B\over\alpha}}.\eqno(5.7)$$
$\alpha$ is given in terms of $B$ by (4.12).

Inserting operators into (5.4) yields
$$<O_1...O_n>(V)\sim V^{-1+{1\over{4\alpha}}(2B\chi-3A\tau+2\sum\gamma_i)}$$
with $\gamma_i$ determined by (4.14).
For nonzero cosmological constant $\lambda$ one finds from
$$<O_1...O_n>_\lambda=\int dV e^{-\lambda V}<O_1...O_n>_0(V)\eqno(5.8)$$
the scaling behavior
$$<O_1...O_n>_\lambda\sim\lambda^{-{1\over{4\alpha}}(2B\chi-3A\tau
+2\sum\gamma_i)},\eqno(5.9)$$
provided the integral (5.8) converges, i.e.
${1\over{4\alpha}}(2B\chi-3A\tau+2\sum\gamma_i)>0.$
Otherwise there will be additional cutoff-dependent terms in (5.9).
${}^{[\sei]}$

Replacing in (5.4)
$$\delta(\int \sqrt{\hat g} e^{2\alpha \phi} - V)\ \ \ \rightarrow\ \ \
\delta({{\int \sqrt{\hat g}e^{\beta \phi}[(\nabla\phi)^2+..]}\over
 {\int \sqrt{\hat g} e^{2\alpha \phi}}
} - \bar R)$$
one obtains the partition function for fixed curvature per volume at
$\eta,\gamma,\lambda=0$:
$$Z(\bar R)\sim \bar R^{-1+{B\chi\over{\beta-2\alpha}}-{3\over2}
 {A\tau\over{\beta-2\alpha}}}.\eqno(5.10)$$

Scaling laws (5.6), (5.8) and (5.9) are similar to the two--dimensional ones,
formulas (4.2), (4.3) and (4.4) of part I. $A,B$ play the role of $Q$ and the
operators $O_i$ were called $\hat V_i$ in part I.
Using the values (5.1), we conclude that for conventional matter,
on the sphere and at the critical point, $Z(V)$ always diverges
(faster than $V^{-1}$) at small
volumes and $Z(\bar R)$ at
large curvature per volume. E.g,
for pure gravity on the sphere,
(5.1), (5.2), (5.7) and (5.10) lead to the predictions:
$$Z(V)\sim V^{ -3.675..} \hskip 2cm Z(\bar R)\sim\bar R^{\ +3.194..}
\eqno(5.11)$$
These quantities should be the easiest ones to check with computer simulations.

\vskip 1cm

{\leftline{\bf 6. Weyl Gravity at Short Distances}}

 {$\underline{\hbox{6.1. Fixed Points Of Gravity With A Weyl Term}}$}

It was claimed in the introduction that
conformally self--dual gravity can also be understood as quantum gravity
with the action
$$\int_{M} d^4 x{\sqrt g}(\lambda +
 \gamma R +\eta R^2 +\rho W_+^2)\eqno(6.1)$$
in the limit $$\rho\rightarrow\infty.\eqno(6.2)$$
Here, $\int W^2$ in (1.1) has been replaced by $\int W_+^2$. They differ
only by the topological invariant $\tau$ of the previous section.

In section 2, the metric was
split into $\phi$, diffeomorphisms $\xi$, moduli $m_i$ and five $\bar h$
components. (6.2) can be understood as the ``classical limit" for the
$\bar h$ components, in which
only the linearized $W_+$ term
$O^\dagger \bar h$ of (2.5)
is important for the $\bar h$ integral. This Gaussian integral
can be performed at each point $g_0(\phi,\xi,m_i)$,
$$\int D \bar h e^{-\rho\int d^4x {\sqrt g} W_+^2}\sim
(\det \rho O O^\dagger)_{g_0}^{-{1\over 2}}=
(\det \rho O^\dagger O)_
 {\hat g e^\phi}^{-{1\over 2}}.$$
This leads again to the integral (2.6), our starting point,
so the two theories are equivalent. (The extra
factor $\rho$ only renormalizes the cosmological constant and does
not influence the anomaly coefficients of section~3.)

As pointed out in the introduction,
the $R^2$ and $W_+^2$ terms would give rise to negative norm states in
Minkowski space. Note however that this need not necessarily
bother us in the
limit $\rho\rightarrow\infty$, because then the $W_+^2$ term decouples
and we can fine--tune away the $R^2$ term in the end.
In any case, there is no unitarity problem in Euclidean space.

One might worry that the renormalization group flow will
take us from $\rho\sim\infty$ to finite $\rho$ so that the limit
(6.2) does not
make sense as an effective theory. However, since at $\rho\sim\infty$
the five $\bar h$ components decouple from the other five components of the
metric, $\rho\sim\infty$ corresponds to a renormalization group fixed
point. More precisely,
defining $\epsilon\equiv{1\over{\sqrt{\rho}}}$,
rescaling $\bar h\rightarrow
\epsilon\bar h$ and expanding the action in $\epsilon$,
one obtains:
$$L_0[\phi,x]+\bar h OO^\dagger\bar h+\epsilon L_i[\bar h,\phi,x]+
O(\epsilon^2)\eqno(6.3)$$
where $L_0$ is the $\bar h$-independent part, and $L_i$ are interaction terms
of $\bar h$ with itself, $\phi$ and $x$, $x$ representing matter
fields that might be present. Thus
the beta function for $\epsilon\sim\rho^{-{1\over2}}$
will receive contributions only from diagrams
that couple $\bar h$ and $\phi$, so it will be at least of order $\epsilon$
and vanish as $\epsilon\rightarrow 0$.
If $(\bar\lambda,\bar\gamma,\bar\eta)$ is a fixed point
of $L_0$, then $(\bar\lambda,\bar\gamma,\bar\eta,\rho
=\infty)$ will be a
fixed point of (6.1).

It has been pointed out in ref. [\ft], that for
$\lambda=\gamma=\eta=0,\ \rho=\infty$ is an
ultraviolet fixed point. Since the cosmological constant and the
Hilbert--Einstein action are relevant operators, this
fixed point is UV--stable in the $\lambda$ and $\gamma$ directions,
but I presently do not know whether it is stable in the $\eta$ direction.
If so,
conformally self--dual gravity can be viewed a short--distance phase of
Euclidean gravity. Of course, in this case ``short distance" means
``distance much shorter than the Planck length,'' a notion that might
be meaningless in the real world (but not in statistical mechanics).

Hopefully the results found in section 5
will be the starting point for finding similar results for $\rho$ finite
or zero, by means of an expansion in $1/\rho$.
It would be very interesting to investigate if the barrier in (5.3)
then becomes positive.
\vskip1cm\eject

 {$\underline{\hbox{6.2. Topological Gravity}}$}

In two--dimensional quantum gravity the correlation functions of
local operators
are related to the correlation functions of topological gravity
${}^{[\wv]}$ which are intersection numbers of submanifolds
on the moduli space of Riemann surfaces with punctures.
Given the similarity of conformally self-dual quantum gravity
to two--dimensional quantum gravity,
it would be very interesting to see if there is a similar relation between
it and four--dimensional topological gravity.${}^{[\wiw]}$

This is suggested by the fact that the moduli
space of the latter theory seems to be precisely the moduli space of
conformally self-dual metrics that arose here.
One might be able to
find a matter system, analoguous to the $c=-2$ system in two
dimensions${}^{[\dis]}$
that, coupled to gravity,
reproduces the BRST multiplet of 4D topological gravity. In this sense,
Euclidean
quantum geometry might have a topological description at short distances.
\vskip1cm

{\leftline{\bf Conclusion of Part III}}

Surprisingly enough, methods of two--dimensional
quantum gravity can be applied to four--dimensional quantum gravity
at least in the limit of infinite Weyl coupling.
The scaling predictions (5.11)
can hopefully
be compared with numerical simulations based on random triangulations.
It will be interesting to investigate how the $\tilde c=-98$ barrier
moves, as $\rho$ in (6.1) moves away from $\infty$.

Many other interesting questions could now be asked,
but this will be left for future work.

\vfill\eject

\par\penalty-400\vskip\chapterskip\spacecheck\referenceminspace
   \ifreferenceopen \Closeout\referencewrite \referenceopenfalse \fi
   \line{\fourteenrm\hfil REFERENCES\hfil}\vskip\headskip
   \input referenc.txa
   
\end